\documentclass[twoside,11pt,final]{entics} 

\usepackage{graphicx}
\usepackage[all]{xy}
\usepackage{amsmath}
\usepackage{mathtools}
\usepackage{xspace}
\usepackage{hyperref}
\usepackage{nicefrac}
\usepackage{listings}
\usepackage{wrapfig}
\usepackage{cutwin}

\usepackage{xypic}
\usepackage[all]{xy}
\xyoption{all}
\xyoption{2cell}
\UseTwocells
\xyoption{tips}
\SelectTips{cm}{}  

\newenvironment{myproof}[1][{\sc\quad Proof.}]%
{ \begin{trivlist}%
		\item[\hskip \labelsep {\bfseries #1}]%
	}%
	{ \end{trivlist}%
}

\newif\ifignore 
\ignorefalse

\newcommand*{\fatten}[1][.4pt]{%
	\textpdfrender{
		TextRenderingMode=FillStroke,
		LineWidth={\dimexpr(#1)\relax},
	}%
}

\ifdefined\mathsl\else
\DeclareMathAlphabet{\mathsl}{\encodingdefault}{\rmdefault}{\mddefault}{\sldefault}
\SetMathAlphabet{\mathsl}{bold}{\encodingdefault}{\rmdefault}{\bfdefault}{\sldefault}
\fi

\makeatletter
\newcommand{\mathoverlap}[2]{\mathpalette\mathoverlap@{{#1}{#2}}}
\newcommand{\mathoverlap@}[2]{\mathoverlap@@{#1}#2}
\newcommand{\mathoverlap@@}[3]{\ooalign{$\m@th#1#2$\crcr\hidewidth$\m@th#1#3$\hidewidth}}
\makeatother

\newcommand{\QEDbox}{\textcolor{darkgray}{\ensuremath{\square}}}

\newcommand{\setin}[3]{\{#1\in#2\;|\;#3\}}

\newcommand{\idmap}[1][]{\ensuremath{\mathsl{id}_{#1}}}
\newcommand{\after}{\mathrel{\circ}}

\newcommand{\nnR}{\mathbb{R}_{\geq 0}}
\newcommand{\shortplus}{\ensuremath{{\kern-2pt}+{\kern-2pt}}}
\newcommand{\shortminus}{\ensuremath{{\kern-1.5pt}-{\kern-1.5pt}}}
\newcommand{\finset}[1]{\ensuremath{\boldsymbol{#1}}}
\newcommand{\supp}{\mathsl{supp}}
\newcommand{\normal}{\mathsl{nrm}}
\newcommand{\disintegration}{\mathsl{disint}}
\newcommand{\domain}{\mathsl{dom}}
\newcommand{\flip}{\mathsl{flip}}
\newcommand{\no}[1]{#1^{\scriptscriptstyle \bot}} 

\newcommand{\one}{\ensuremath{\mathbf{1}}}
\newcommand{\zero}{\ensuremath{\mathbf{0}}}

\newcommand{\uniform}[1][]{\ensuremath{\mathsl{uf}_{{\kern-.3ex}#1}\xspace}}
\newcommand{\andchan}{\mathsl{ac}}
\newcommand{\orchan}{\mathsl{oc}}
\newcommand{\Dst}{\mathcal{D}}
\newcommand{\subDst}{\mathcal{D}_{\leq 1}}
\newcommand{\fullDst}{\ensuremath{\mathcal{D}_{{\kern-0.4pt}\mathsl{fs}}}}
\newcommand{\FracDst}{\ensuremath{\mathcal{F}{\kern-2.0pt}\mathcal{D}}}
\newcommand{\Mlt}{\mathcal{M}}
\newcommand{\fullMlt}{\ensuremath{\Mlt_{{\kern-0.2pt}\mathsl{fs}}}}

\newcommand{\intd}{{\kern.2em}\mathrm{d}{\kern.03em}}
\newcommand{\Kl}{\mathcal{K}{\kern-.4ex}\ell}

\newcommand{\ket}[1]{\ensuremath{|{\kern.1em}#1{\kern.1em}\rangle}}
\newcommand{\ketsl}[1]{\ensuremath{|{\kern.1em}\mathsl{#1}{\kern.1em}\rangle}}
\newcommand{\bigket}[1]{\ensuremath{\big|{\kern.1em}#1{\kern.1em}\big\rangle}}
\newcommand{\ketstrut}{\vrule height 10pt depth 5pt width 0pt}
\newcommand{\Bigket}[1]{\ensuremath{\left|\left.\ketstrut{\kern.1em}#1{\kern.005em}\right>\right.}}
\newcommand{\facto}[1]{\ensuremath{#1{\kern-2.5pt}\raisebox{-2.5pt}{\includegraphics[width=0.9em]{exclamation}}}}
\newcommand{\coefm}[1]{\ensuremath{\fatten[0.6pt]{(}{\kern1pt}#1{\kern1pt}\fatten[0.6pt]{)}}}
\newcommand{\inprod}[2]{\ensuremath{\fatten[0.6pt]{\langle}{\kern1pt}#1,#2{\kern1pt}\fatten[0.6pt]{\rangle}}}
\newcommand{\bibinom}[2]{\left({\kern-.2em}\binom{#1}{#2}{\kern-.2em}\right)}
\newcommand{\setsize}[1]{|{\kern.1em}#1{\kern.1em}|}
\newcommand{\bigsetsize}[1]{\big|{\kern.1em}#1{\kern.1em}\big|}
\newcommand{\Bigsetsize}[1]{\Big|{\kern.1em}#1{\kern.1em}\Big|}
\newcommand{\push}{\mathrel{\mathchoice%
		{\scalebox{-0.5}[1]{$=\ll$}}
		{\scalebox{-0.5}[1]{$={\kern-1.5ex}\ll$}}
		{\scalebox{-0.5}[1]{${\kern.5ex}\scriptstyle={\kern-0.2ex}\ll{\kern.5ex}$}}
		{\scalebox{-0.5}[1]{$\scriptscriptstyle=\ll$}}}}
\newcommand{\pull}{\mathrel{\mathchoice%
		{\scalebox{-0.5}[1]{$\gg=$}}
		{\scalebox{-0.5}[1]{$\gg{\kern-1.5ex}=$}}
		{\scalebox{-0.5}[1]{${\kern.5ex}\scriptstyle\gg{\kern-0.2ex}={\kern.5ex}$}}
		{\scalebox{-0.5}[1]{$\scriptscriptstyle\gg=$}}}}

\newcommand{\Category}[1]{\ensuremath{\mathbf{#1}}\xspace}
\newcommand{\Sets}{\Category{Sets}}

\newcommand{\xeq}[1]{\mathrel{\overset{\makebox[0pt]{\tiny #1}}{=}} }

\newcommand{\ba}{\mathsl{ba}}
\newcommand{\pfc}{\mathsl{pfc}}
\newcommand{\tga}{\mathsl{tga}}
\newcommand{\flt}{\mathsl{flt}}
\newcommand{\pis}{\mathsl{pis}}
\newcommand{\tpd}{\mathsl{tpd}}
\newcommand{\lng}{\mathsl{lng}}
\newcommand{\lt}{\mathsl{lt}}
\newcommand{\DFno}{\mathsl{no}}
\newcommand{\rt}{\mathsl{rt}}
\newcommand{\CMno}{\mathsl{no}}
\newcommand{\mi}{\mathsl{mi}}
\newcommand{\co}{\mathsl{co}}
\newcommand{\tr}{\mathsl{tr}}
\newcommand{\nr}{\mathsl{nr}}
\newcommand{\cg}{\mathsl{cg}}
\newcommand{\ab}{\mathsl{ab}}
\newcommand{\lo}{\mathsl{lo}}
\newcommand{\hi}{\mathsl{hi}}
\newcommand{\eq}{\mathsl{eq}}
\newcommand{\mo}{\mathsl{mo}}
\newcommand{\se}{\mathsl{se}}
\newcommand{\ltfi}{$\textless{\kern-0.1em}$\mathsl{5}}
\newcommand{\fitw}{\mathsl{5}$-$\mathsl{12}}
\newcommand{\gttw}{\mathsl{12+}}

\usepackage{tikzit}
\usetikzlibrary{circuits.ee.IEC}

\tikzstyle{white dot}=[inner sep=0mm, minimum size=1.5mm, draw=black, shape=circle, text depth=-0.2mm, draw=black, fill=white, tikzit category=nodes]
\tikzstyle{black dot}=[inner sep=0mm, minimum size=1.5mm, draw=black, shape=circle, draw=black, fill=black, tikzit category=nodes]
\tikzstyle{observed}=[inner sep=0mm, minimum size=5mm, draw=black, shape=circle, text depth=-0.2mm, draw=white, tikzit draw=gray, fill=white, tikzit category=dag]
\tikzstyle{latent}=[inner sep=0mm, minimum size=5mm, draw=black, shape=circle, text depth=-0.2mm, draw=black, fill=white, tikzit category=dag]
\tikzstyle{small box}=[shape=rectangle, text height=1.5ex, text depth=0.25ex, yshift=0.5mm, fill=white, draw=black, minimum height=6mm, yshift=-0.5mm, minimum width=6mm, font={\small}, tikzit category=boxes]
\tikzstyle{medium box}=[shape=rectangle, draw=black, fill=white, small box, minimum width=8mm, tikzit category=boxes]
\tikzstyle{semilarge box}=[shape=rectangle, draw=black, fill=white, small box, minimum width=12.5mm, tikzit category=boxes]
\tikzstyle{large box}=[shape=rectangle, draw=black, fill=white, small box, minimum width=15mm, tikzit category=boxes]
\tikzstyle{upground}=[circuit ee IEC, thick, ground, rotate=90, scale=1.5, inner sep=-2mm, tikzit shape=circle, tikzit fill=blue, tikzit category=points]
\tikzstyle{downground}=[circuit ee IEC, thick, ground, rotate=-90, scale=1.5, inner sep=-2mm, tikzit shape=circle, tikzit fill=green, tikzit category=points]
\tikzstyle{point}=[regular polygon, regular polygon sides=3, draw, scale=0.75, inner sep=-0.5pt, minimum width=9mm, fill=white, regular polygon rotate=180, tikzit category=points]
\tikzstyle{copoint}=[regular polygon, regular polygon sides=3, draw, scale=0.75, inner sep=-0.5pt, minimum width=9mm, fill=white, tikzit category=points]
\tikzstyle{uniform}=[point, fill=gray, tikzit shape=circle, scale=0.5]
\tikzstyle{label}=[font={\footnotesize}, text height=1.5ex, text depth=0.25ex, tikzit draw=blue, tikzit fill=white, tikzit category=labels]
\tikzstyle{left label}=[label, anchor=east, xshift=2mm, tikzit draw=green, tikzit fill=white, tikzit category=labels]
\tikzstyle{right label}=[label, anchor=west, xshift=-2mm, tikzit draw=purple, tikzit fill=white, tikzit category=labels]
\tikzstyle{disintegration}=[draw=black, fill={gray!50}, tikzit fill=gray, shape=rectangle, minimum width=1.6cm, minimum height=1.2cm, opacity=0.3]
\tikzstyle{empty diag}=[shape=rectangle, draw=darkgray, dashed, minimum width=8mm, minimum height=8mm, yshift=0.5mm]

\tikzstyle{diredge}=[->, >=latex]
\tikzstyle{dashed edge}=[-, dashed, fill=none]

\newif\ifexternalizetikz

\usetikzlibrary{shapes,shapes.geometric}
\usetikzlibrary{shapes.gates.ee,shapes.gates.ee.IEC}
\usetikzlibrary{calc}
\usetikzlibrary{positioning}
\usetikzlibrary{cd}
\usetikzlibrary{decorations.pathmorphing}

\ifexternalizetikz
\usepackage{tikzcdx}
\usetikzlibrary{external}
\makeatletter
\newcommand{\tikzextname}[1]{%
	\tikzset{external/figure name={\tikzexternal@realjob-#1-}}}
\makeatother
\fi

\newsavebox\sbpto
\savebox\sbpto{\begin{tikzpicture}[baseline=-2.5pt]
		\filldraw[fill=white,draw=white] circle (1.4pt);
		\filldraw[fill=white,draw=black,line width=0.2pt]circle(2pt);
\end{tikzpicture}}

\newsavebox\sbground
\savebox\sbground{\begin{tikzpicture}[circuit ee 
IEC,yscale=1,xscale=1]
		\draw (0,-2ex) to (0,0) node[ground,rotate=90,xshift=.65ex] 
		{};
\end{tikzpicture}}
\newcommand\ground{\mathbin{\text{\raisebox{0.2ex}{\usebox\sbground}}}}
\newsavebox\sbunif
\savebox\sbunif{\begin{tikzpicture}[circuit ee IEC,yscale=1,xscale=1]
		\draw (0,0) to (0,2ex) node[ground,rotate=270,xshift=2.5ex] 
		{};
\end{tikzpicture}}
\newcommand\unif{\mathbin{\text{\raisebox{-0.1ex}{\usebox\sbunif}}}}

\tikzset{dot/.style =
	{inner sep=0mm,minimum width=1mm,minimum height=1mm,
		draw,shape=circle}}
\tikzset{minicopy/.style = {dot,fill,text depth=-0.2mm}}
\newsavebox\sbcopier
\savebox\sbcopier{%
	\begin{tikzpicture}[baseline=0pt]
		\node[minicopy,scale=.7] (a) at (0,3.6pt) {};
		\draw (a) -- +(-90:.30);
		\draw (a) -- +(45:.35);
		\draw (a) -- +(135:.35);
\end{tikzpicture}}
\newcommand{\minicopy}{\mathord{\usebox\sbcopier}}
\newsavebox\sbcocopier
\savebox\sbcocopier{%
	\begin{tikzpicture}[baseline=0pt]
		\node[minicopy,scale=.7] (a) at (0,3.6pt) {};
		\draw (a) -- +(90:.30);
		\draw (a) -- +(-45:.35);
		\draw (a) -- +(-135:.35);
\end{tikzpicture}}
\newcommand{\minicocopy}{\mathord{\usebox\sbcocopier}}
\newsavebox\sbcup
\savebox\sbcup{%
	\begin{tikzpicture}[baseline=0pt]
		\node (a) at (-10pt,18pt) {};
		\node (b) at (10pt,18pt) {};
		\draw (a) to[out=-90,in=-90,looseness=2] (b);
\end{tikzpicture}}

\newsavebox\sbcap
\savebox\sbcap{%
	\begin{tikzpicture}[baseline=0pt]
		\node (a) at (-10pt,-6pt) {};
		\node (b) at (10pt,-6pt) {};
		\draw (a) to[out=90,in=90,looseness=2] (b);
\end{tikzpicture}}

\ifexternalizetikz\tikzexternalenable\fi

\newcommand{\alice}[1]{\mathrm{alice}_{#1}}
\newcommand{\bob}[1]{\mathrm{bob}_{#1}}
\newcommand{\location}{\mathrm{location}}

\usepackage{enticsmacro}

\def\conf{MFPS 2025} 	
\volume{5}			
\def\volu{5}			
\def\papno{11}			
\def\mydoi{17029}
\def\lastname{Jacobs, Sz\'eles and Stein} 
\def\runtitle{Compositional inference} 
\def\copynames{AB. Jacobs, M. Sz\'eles, D. Stein}    
\begin{document}
\begin{frontmatter}
  \title{Compositional Inference for\\ Bayesian Networks and 
  Causality} 						
  \author{Bart Jacobs\thanksref{a}\thanksref{emaila}}	
   \author{M\'ark Sz\'eles\thanksref{a}\thanksref{emailb}}
   \author{Dario Stein\thanksref{a}\thanksref{emailc}}		
   \address[a]{Radboud University\\				
    Nijmegen, The Netherlands}  							
   \thanks[emaila]{Email: \href{mailto:bart@cs.ru.nl} 
   {\texttt{\normalshape
        bart@cs.ru.nl}}} 

  \thanks[emailb]{Email: \href{mailto:mark.szeles@ru.nl} 
  	{\texttt{\normalshape
  			mark.szeles@ru.nl}}}
  \thanks[emailc]{Email: \href{mailto:dario.stein@ru.nl} 
  	{\texttt{\normalshape
  			dario.stein@ru.nl}}}

\begin{abstract} 
  Inference is a fundamental reasoning technique in probability 
  theory.
  When applied to a large joint distribution, it involves updating 
  with
  evidence (conditioning) in one or more components (variables) and
  computing the outcome in other components. When the joint 
  distribution
  is represented by a Bayesian network, the network structure may be
  exploited to proceed in a compositional manner --- with great
  benefits.  However, the main challenge is that updating involves
  (re)normalisation, making it an operation that interacts badly with
  other operations.
  
  String diagrams are becoming popular as a graphical technique for
  probabilistic (and quantum) reasoning. Conditioning has appeared in
  string diagrams, in terms of a disintegration, using bent wires and
  shaded (or dashed) normalisation boxes. It has become clear that 
  such
  normalisation boxes do satisfy certain compositional rules. This 
  paper
  takes a decisive step in this development by adding a removal rule 
  to
  the formalism, for the deletion of shaded boxes. Via this removal 
  rule
  one can get rid of shaded boxes and terminate an inference argument.
  This paper illustrates via many (graphical) examples how the 
  resulting
  compositional inference technique can be used for Bayesian networks,
  causal reasoning and counterfactuals.
\end{abstract}
\end{frontmatter}

\section{Introduction}

Disintegration is a technique in probability theory that allows one to
factorise a joint probability distribution as a product of a marginal
distribution and a conditional probability. In traditional probability
notation, one writes $P(X) \cdot P(Y\mid X) = P(X,Y) = P(Y) \cdot
P(X\mid Y)$. Computing such conditional probabilities is an essential
ingredient of Bayesian reasoning.

Because of their basic role, disintegrations (also called
conditionals) have been extensively studied in categorical probability
theory~\cite{ClercDDG17,ChoJ19,Fritz20}. The usual setting for the
axiomatic study of disintegration is the theory of Markov
categories. These categories model key aspects of probabilistic
computation: they can be composed in sequence and in parallel,
technically via a symmetric monoidal structure.  Crucially, there is
also structure for copying and discarding.

Recently, there has been progress in formulating categorical
probability theory in terms of so-called CD-categories
\cite{LavoreR23}.  CD-categories are just like Markov-categories, but
morphisms represent possibly partial computation. For discrete
probability theory, this means that a map $X \rightarrow Y$ is an
$X$-indexed family of subdistributions --- with probabilities adding
up to \emph{at most} one instead of to \emph{precisely} one. This
partial perspective is more natural for disintegration than the
Markov-category approach for two reasons.
\begin{enumerate}
	\item Disintegration $P(X,Y) = P(X) \cdot P(Y\mid X)$ may not
          be well-defined, typically in presence of joint
          distributions $P(X,Y)$ that do not have full support, that
          is, when certain elements of the product $X\times Y$ are not
          in the (support of) the distribution. In the Markov-category
          approach, the conditional distribution $P(Y\mid X)$ is then
          allowed to take an arbitrary value. The partial approach
          involves a least disintegration without any arbitrary, junk
          information. This order-theoretic perspective will be
          elaborated elsewhere, but it does play a role here in the
          background.

	\item Closely related to the previous point is that Bayesian
	updating is an inherently partial operation. If one is
	presented with evidence incompatible with the prior belief,
	then the update cannot be performed. The formalism of
	CD-categories can handle such partiality better than Markov
	categories.
\end{enumerate}	

One advantage of categorical probability theory is that it comes with
the intuitive graphical calculus of string diagrams. In (an early
version of) \cite{Jacobs24a} a notation was introduced
for disintegration using bent wires and shaded boxes. These shaded
areas describe the part of the diagram that is normalised: the
enclosed subdistributions are turned into proper
distributions. Normalisation is a peculiar operation that is typically
highly non-compositional. Hence these shaded boxes did not seem to
match well with the formalism of string diagrams. However, gradually
it became clear that there are useful compositional rules for these
shaded (normalisation) boxes, see Definition~\ref{def:normalisation}
below. This was realised by several authors, see for
instance~\cite{LorenzT23,FritzKMSW24,TullLCKC24,Jacobs24a} (and also
implicitly in~\cite{Fritz20}).

The main contribution of the current paper is to add the final step,
pushing the compositionality of these shaded boxes to a new level, so
that they become actually useful for graphical reasoning with
conditioning.  The crucial new rule here is about the removal of
shaded boxes, see Proposition~\ref{prop:boxremoval} below.  This new
removal rule is combined with already existing rules (in
Definition~\ref{def:normalisation} and
Lemma~\ref{lem:nestedbox}). This makes it possible to apply
disintegration to a Bayesian network by first introducing bent wires
and shaded boxes for conditioning, then performing a number of graph
rewrite steps, finally ending up with a new network from which shaded
boxes are removed. This technique will be illustrated in many examples
below, copied from various places in the literature. While the new
removal rule is not difficult to prove, it introduces a new feature,
namely to return to the world of proper (`non-sub') distributions.
This streamlines graphical reasoning about disintegration.  For
instance, different kinds of interventions treated in \cite{LorenzT23}
can be unified.

We briefly speculate about possible practical advantages of
disintegration in Bayesian networks.
\begin{enumerate}
	\item Disintegration turns a `closed' network into an `open' one, 
	in
	the terminology of~\cite{LorenzT23}. This means that updating does
	not happen pointwise, but yields a single function that performs 
	the
	update for every point at once, see the remarks about the derived
	update function at the very end of
	Section~\ref{sec:bayesiannetwork}.
	
	\item Moreover, these functions (arising via disintegration)
          can be pre-computed for specific, often occurring scenarios,
          e.g.\ in a medical setting, so that the usual point-updates
          do not have to be performed real-time. This may improve the
          usability of inference in Bayesian networks.
\end{enumerate}

This paper is structured as follows. In Section
\ref{sec:distributions} we recall some basics about discrete
probability distributions and subdistributions. Section
\ref{sec:normalisation_comparison} is about the rules of comparators
and normalisation. In Section \ref{sec:disintegration} we recall how
well-behaved comparators and normalisation give rise to
disintegrations. We also prove the derived rule for the disappearence
of shaded boxes. The rest of the paper is dedicated to demonstrating
the versatility of the extended graphical calculus via a zoo of
examples. Section~\ref{sec:bayesiannetwork} shows examples involving
Bayesian networks. Sections~\ref{sec:independence}
and~\ref{sec:programming} contain probabilistic programming examples
on conditional independence and on nested ``reasoning about
reasoning''. Finally, Sections~\ref{sec:causality}
and~\ref{sec:counterfactual} treat causal and counterfactual
reasoning.

\subsection*{Related work}

To put this paper's contributions in context, we briefly summarise the
main developments that precede our work. The string diagrammatic
treatment of Bayesian networks originates in~\cite{Fong12}. It was
extended with updating in~\cite{JacobsZ21}, for forward and backward
inference. The graphical notation for disintegration using shaded
boxes and bent wires was introduced in an earlier version (from 2021)
of the book draft~\cite{Jacobs24a}. Shaded boxes were used to depict
disintegrations in the Kleisli-category of the finitary
subdistribution monad $\Dst$ on the category of sets and functions.
Some compositional rules for the shaded box were identified, but a
systematic formal treatment was missing.

The authors of the (as of yet unpublished) manuscript~\cite{LorenzT23}
took this further by decomposing the notation into cap morphisms and
normalisation boxes. Their choice of semantic universe is the category
$\mathsf{Mat}_{\nnR}$ of matrices with non-negative entries; it is
very similar to our choice of $\Kl(\subDst)$, see
Section~\ref{sec:distributions}. The disappearence rule of
Proposition~\ref{prop:boxremoval} below does not occur
in~\cite{LorenzT23}. The authors of~\cite{LorenzT23} define the notion
of full support for states (distributions), but not for channels as in
Definition~\ref{def:fullsupport} below.

Examples of causal inference and counterfactual reasoning appeared in
string diagrammatic terms in~\cite{JacobsKZ21}.  There, the problem of
the current Section~\ref{sec:causality} is solved using so-called
``comb disintegrations''. That approach is different from the one
adopted in this paper, as it is not internal to the category in which
the model lives, but instead relies on an embedding into a larger
compact closed category. A category theoretic treatment of Pearl's
do-calculus for causal inference appeared in \cite{YinZ22}. The
related topic of conditional independence and causal compatibility is
investigated diagramatically
in~\cite{FritzK23}. In~\cite[Section~8]{LorenzT23}, a diagrammatic
algorithm is presented to identify solutions of a certain class of
counterfactual problems.

\section{Distributions and subdistributions}\label{sec:distributions}

In this paper we shall work with finite discrete (sub)distributions.
We briefly fix notation and recall the essentials. A
\emph{subdistribution} on a set $X$ is a finite formal sum of the form
$\sum_{i} r_{i}\ket{x_i}$, with elements $x_{i}\in X$ occurring with
associated probabilities $r_{i} \in [0,1]$ satisfying $\sum_{i} r_{i}
\leq 1$. It is called a \emph{distribution} when $\sum_{i} r_{i} =
1$. One can identify a (sub)distribution with a function $\omega\colon
X \rightarrow [0,1]$ whose support $\supp(\omega) \coloneqq
\setin{x}{X}{\omega(x)\neq 0}$ is finite and whose sum $\sum_{x}
\omega(x)$ is below or equal to~$1$. We write $\subDst(X)$ for the set
of subdistributions on $X$, with subset $\Dst(X) \subseteq \subDst(X)$
of `proper' distributions. Writing $1$ for a one-element set, say $1 =
\{0\}$, then $\Dst(1) \cong 1$ and $\subDst(1) \cong [0,1]$. Thus,
functions $X \rightarrow \subDst(1)$ can be identified with (fuzzy)
predicates. We shall write $\one \colon X \rightarrow \subDst(1)$ for
the `truth' function/predicate that maps every element $x\in X$ to the
top element $1\in [0,1]$.

These operations $\subDst$ and $\Dst$ are both monads on the category
$\Sets$ of sets and functions. We shall write the associated Kleisli
categories as $\Kl(\Dst) \hookrightarrow \Kl(\subDst)$. Both
categories have arbitrary sets $X$ as objects. A morphism $f$ from $X$
to $Y$ in $\Kl(\subDst)$ is a function $f\colon X
\rightarrow\subDst(Y)$. Composition with $g\colon Y \rightarrow
\subDst(Z)$ is written as $g\after f \colon X \rightarrow \subDst(Z)$
and is defined as $(g\after f)(x) = \sum_{z} \big(\sum_{y}
f(x)(y)\cdot g(y)(z)\big)\ket{z}$. The unit map $X \rightarrow
\subDst(X)$ sends $x$ to the singleton distribution
$1\ket{x}$. Morphisms in $\Kl(\subDst)$ are called \emph{subchannels},
whereas morphisms in $\Kl(\Dst)$ are called \emph{channels}. The
latter can be recognised inside $\Kl(\subDst)$ as those $f\colon X
\rightarrow \subDst(Y)$ that satisfy $\one \after f = \one$. This
means that each $f(x)$ is a proper distribution and that $f$ restricts
to $X \rightarrow \Dst(Y)$.  Distributions and subdistributions can be
recognised within $\Kl(\Dst)$ and $\Kl(\subDst)$ as the morphisms with
a one-element set $1$ as domain.

For two subdistributions $\omega\in\subDst(X)$ and $\rho\in\subDst(Y)$
one can form their parallel / tensor product $\omega\otimes\rho \in
\subDst(X\times Y)$, given by $\big(\omega\otimes\rho\big)(x,y) =
\omega(x)\cdot\rho(y)$. This tensor restricts to $\otimes \colon
\Dst(X)\times\Dst(Y) \rightarrow \Dst(X\times Y)$. Via pointwise
definition it turns the Kleisli categories $\Kl(\subDst)$ and
$\Kl(\Dst)$ into symmetric monoidal categories. The tensor unit is the
one-element set $1$, which is final in $\Kl(\Dst)$.

For each set $X$ there is a copy map $\Delta \colon X \rightarrow
\Dst(X\times X)$, namely $\Delta(x) = 1\ket{x,x}$. There are first and
second projection maps $\Dst(X) \leftarrow X\times Y \rightarrow
\Dst(Y)$, namely $\pi_{1}(x,y) = 1\ket{x}$ and $\pi_{2}(x,y) =
1\ket{y}$. These copiers and projections turn both $\Kl(\subDst)$ and
$\Kl(\Dst)$ into so-called Copy-Delete (CD) categories,
see~\cite{ChoJ19}. The Kleisli category $\Kl(\Dst)$ is not only a
CD-category but also a Markov category, see~\cite{Fritz20}: all its
maps are channels; $\Kl(\subDst)$ is called a \emph{partial} Markov
category in~\cite{LavoreR23}.

\begin{wrapfigure}{l}{0.4\textwidth}
	\label{fig:unital-deterministic}
	\[
	\vcenter{\hbox{\begin{tikzpicture}
	\begin{pgfonlayer}{nodelayer}
		\node [style=small box] (0) at (-6.75, -0.5) {$f$};
		\node [style=small box] (1) at (0, -1) {$g$};
		\node [style=none] (2) at (-6.75, -1.75) {};
		\node [style=none] (3) at (-5.25, -0.5) {$=$};
		\node [style=none] (4) at (-4.25, -1.25) {};
		\node [style=none] (5) at (0, -2) {};
		\node [style=none] (6) at (-1, 1) {};
		\node [style=none] (7) at (1, 1) {};
		\node [style=none] (8) at (3.25, 1) {};
		\node [style=none] (9) at (4.75, 1) {};
		\node [style=none] (10) at (4, -2) {};
		\node [style=none] (11) at (1.75, -0.5) {$\neq$};
		\node [style=upground] (12) at (-6.75, 0.75) {};
		\node [style=upground] (13) at (-4.25, 0.25) {};
		\node [style=black dot] (14) at (0, 0) {};
		\node [style=black dot] (15) at (4, -1.25) {};
		\node [style=small box] (16) at (3.25, 0) {$g$};
		\node [style=small box] (17) at (4.75, 0) {$g$};
	\end{pgfonlayer}
	\begin{pgfonlayer}{edgelayer}
		\draw (12) to (0);
		\draw (0) to (2.center);
		\draw (13) to (4.center);
		\draw [in=165, out=-90, looseness=1.25] (6.center) to (14);
		\draw [in=-90, out=15, looseness=1.25] (14) to (7.center);
		\draw (14) to (1);
		\draw (1) to (5.center);
		\draw (15) to (10.center);
		\draw [in=-90, out=165, looseness=1.25] (15) to (16);
		\draw (16) to (8.center);
		\draw (17) to (9.center);
		\draw [in=-90, out=30, looseness=1.25] (15) to (17);
	\end{pgfonlayer}
\end{tikzpicture}}}
	\]
	\vspace{-1em}
\end{wrapfigure}

These CD and Markov categories have become the standard universes for
categorical probability theory, see
\textit{e.g.}~\cite{Fritz20,ChoJ19}.  They are standardly used with
the convenient language of string diagrams. We shall use the basics of
this language without further explanation --- except that we write
copying and discarding as $\Delta = \minicopy$ and $\one =
\ground$. We refer to~\cite{ChoJ19,Fritz20,PiedeleuZ25} for further
details. We will explicitly describe how to do the less standard
operations of normalisation, comparison and disintegration
(conditioning) within the setting of string diagrams.  The only thing
we wish to emphasise is that a map / box $f$ is called a
\emph{channel} if the equation $\one \after f = \one$ on the left
holds. Further, the copy equation on the right fails in general. The
maps $g$ for which it does hold are called \emph{deterministic}.  A
box without incoming wires is called a \emph{state}.

\section{Normalisation and 
comparison}\label{sec:normalisation_comparison}

\begin{wrapfigure}{r}{0.38\textwidth}
	\label{fig:domain-normal}
	\[
	\vcenter{\hbox{\begin{tikzpicture}
	\begin{pgfonlayer}{nodelayer}
		\node [style=none] (0) at (-6.75, 0) {$\domain(f)$};
		\node [style=none] (1) at (-4.75, 0) {$\coloneqq$};
		\node [style=small box] (2) at (-3.25, 0) {$f$};
		\node [style=upground] (3) at (-3.25, 1.25) {};
		\node [style=none] (4) at (-3.25, -1.25) {};
		\node [style=small box] (5) at (0, 0) {$f$};
		\node [style=none] (6) at (0, 1.75) {};
		\node [style=none] (7) at (0, -1.75) {};
		\node [style=none] (8) at (4, -1.75) {};
		\node [style=black dot] (9) at (4, -1) {};
		\node [style=small box] (10) at (5, 0.5) {$g$};
		\node [style=none] (11) at (5, 1.75) {};
		\node [style=small box] (12) at (3, 0.5) {$f$};
		\node [style=upground] (13) at (3, 1.5) {};
		\node [style=none] (14) at (1.5, 0) {$=$};
	\end{pgfonlayer}
	\begin{pgfonlayer}{edgelayer}
		\draw (3) to (2);
		\draw (4.center) to (2);
		\draw (13) to (12);
		\draw (6.center) to (5);
		\draw (7.center) to (5);
		\draw (8.center) to (9);
		\draw [in=-90, out=150, looseness=1.25] (9) to (12);
		\draw [in=-90, out=30, looseness=1.25] (9) to (10);
		\draw (10) to (11.center);
	\end{pgfonlayer}
\end{tikzpicture}}}
	\]
\end{wrapfigure}
Any subdistribution $\omega\in\subDst(X)$ that is not the (everywhere)
zero subdistribution $\zero$ can be normalised to a proper
distribution $\frac{1}{\|\omega\|}\cdot \omega$ in $\Dst(X)$, where
$\|\omega\| = \sum_{x} \omega(x)$ is the \emph{weight} of $\omega$.
In this section we will first introduce normalisation and then
comparators in string diagrams, following the approach
of~\cite{LorenzT23} and~\cite{LavoreR23}. For an arbitrary map
$f\colon X \rightarrow Y$, written diagrammatically as a box, we
define its domain as $\domain(f) \coloneqq \one \after f$. We say that
a map $g \colon X \rightarrow Y$ is a \emph{normalisation} of $f$ if
$f$ can be written as the tuple of its domain with $g$, as described
on the side. The map
$f$ is called \emph{normalised} if it is a normalisation of itself.

\begin{definition}{(\cite[Def. 8]{LorenzT23})}
  \label{def:normalisation}
  A normalisation structure on a CD-category assigns to every map
  $f\colon X \rightarrow Y$ a normalised map $\normal(f) \colon X
  \rightarrow Y$, written as a shaded box on the left below,
  satisfying the equation on the right, making $\normal(f)$ a
  normalisation of $f$.
  \begin{equation}
    \label{eqn:normalisation}
    \vcenter{\hbox{\begin{tikzpicture}
	\begin{pgfonlayer}{nodelayer}
		\node [style=small box] (0) at (-0.05, 0) {$f$};
		\node [style=none] (1) at (-0.05, 1.25) {};
		\node [style=none] (2) at (-0.05, -1.25) {};
		\node [style=none] (3) at (-1.75, 0) {$\coloneqq$};
		\node [style=disintegration, minimum height=0.9cm, minimum width=0.9cm] (4) at (-0.05, 0) {};
		\node [style=none] (5) at (-3.75, 0) {$\normal(f)$};
		\node [style=small box] (6) at (6.45, 0) {$f$};
		\node [style=none] (7) at (6.45, 1.75) {};
		\node [style=none] (8) at (6.45, -1.75) {};
		\node [style=none] (9) at (10.45, -1.75) {};
		\node [style=black dot] (10) at (10.45, -1) {};
		\node [style=small box] (11) at (11.45, 0.5) {$f$};
		\node [style=none] (12) at (11.45, 1.75) {};
		\node [style=small box] (13) at (9.45, 0.5) {$f$};
		\node [style=upground] (14) at (9.45, 1.5) {};
		\node [style=none] (15) at (7.95, 0) {$=$};
		\node [style=disintegration, minimum height=0.9cm, minimum width=0.9cm] (16) at (11.45, 0.5) {};
	\end{pgfonlayer}
	\begin{pgfonlayer}{edgelayer}
		\draw (1.center) to (0);
		\draw (2.center) to (0);
		\draw (14) to (13);
		\draw (7.center) to (6);
		\draw (8.center) to (6);
		\draw (9.center) to (10);
		\draw [in=-90, out=150, looseness=1.25] (10) to (13);
		\draw [in=-90, out=30, looseness=1.25] (10) to (11);
		\draw (11) to (12.center);
	\end{pgfonlayer}
\end{tikzpicture}}}
  \end{equation}	
			
\noindent The following requirements should hold for this
normalisation.
	\begin{enumerate}
		\item \label{def:normalisation:comp} Normalisation of parallel
		composition can be done separately, and a channel $h$ can be 
		pulled out of sequential composition; discarders 
		$\ground$ and copiers
		$\minicopy$ can be pulled out of normalisation boxes: 
		\[
		\vcenter{\hbox{\begin{tikzpicture}
	\begin{pgfonlayer}{nodelayer}
		\node [style=small box] (0) at (-4, 0) {$f$};
		\node [style=small box] (1) at (-2.25, 0) {$g$};
		\node [style=none] (2) at (-4, 1.5) {};
		\node [style=none] (3) at (-2.25, 1.5) {};
		\node [style=none] (4) at (-2.25, -1.5) {};
		\node [style=none] (5) at (-4, -1.5) {};
		\node [style=small box] (6) at (0.5, 0) {$f$};
		\node [style=small box] (7) at (2.75, 0) {$g$};
		\node [style=none] (8) at (0.5, 1.5) {};
		\node [style=none] (9) at (2.75, 1.5) {};
		\node [style=none] (10) at (2.75, -1.5) {};
		\node [style=none] (11) at (0.5, -1.5) {};
		\node [style=none] (12) at (-0.95, 0) {$=$};
		\node [style=disintegration, minimum height=0.9cm, minimum width=1.7cm] (26) at (-3.15, 0) {};
		\node [style=disintegration, minimum height=0.9cm, minimum width=0.9cm] (27) at (0.5, 0) {};
		\node [style=disintegration, minimum height=0.9cm, minimum width=0.9cm] (30) at (2.75, 0) {};
		\node [style=small box] (31) at (9.5, -0.75) {$g$};
		\node [style=none] (33) at (9.5, -2) {};
		\node [style=small box] (35) at (9.5, 1.25) {$h$};
		\node [style=none] (36) at (9.5, 2.25) {};
		\node [style=none] (37) at (8, 0) {$=$};
		\node [style=small box] (38) at (6.5, -0.75) {$g$};
		\node [style=none] (39) at (6.5, -2) {};
		\node [style=small box] (40) at (6.5, 1) {$h$};
		\node [style=none] (41) at (6.5, 2.25) {};
		\node [style=disintegration, minimum height=1.65cm, minimum width=0.9cm] (42) at (6.5, 0.15) {};
		\node [style=disintegration, minimum height=0.9cm, minimum width=0.9cm] (43) at (9.5, -0.75) {};
	\end{pgfonlayer}
	\begin{pgfonlayer}{edgelayer}
		\draw (2.center) to (0);
		\draw (3.center) to (1);
		\draw (1) to (4.center);
		\draw (5.center) to (0);
		\draw (8.center) to (6);
		\draw (9.center) to (7);
		\draw (7) to (10.center);
		\draw (11.center) to (6);
		\draw (31) to (33.center);
		\draw (36.center) to (35);
		\draw (38) to (39.center);
		\draw (41.center) to (40);
		\draw (35) to (31);
		\draw (40) to (38);
	\end{pgfonlayer}
\end{tikzpicture} 
		\quad\quad \begin{tikzpicture}
	\begin{pgfonlayer}{nodelayer}
		\node [style=medium box] (0) at (-4.75, 0) {};
		\node [style=none] (2) at (-4.25, 0.5) {};
		\node [style=none] (3) at (-5.25, 0.5) {};
		\node [style=none] (4) at (-5.25, 2.25) {};
		\node [style=none] (5) at (-4.75, -1.5) {};
		\node [style=upground] (6) at (-4.25, 1.25) {};
		\node [style=medium box] (7) at (-1.25, 0) {};
		\node [style=none] (8) at (-0.75, 0.5) {};
		\node [style=none] (9) at (-1.75, 0.5) {};
		\node [style=none] (10) at (-1.75, 2.25) {};
		\node [style=none] (11) at (-1.25, -1.5) {};
		\node [style=upground] (12) at (-0.75, 1.5) {};
		\node [style=none] (13) at (-3, 0.25) {$=$};
		\node [style=semilarge box] (14) at (3.25, 0.75) {};
		\node [style=none] (15) at (2.25, 0.25) {};
		\node [style=none] (16) at (3.25, 0.25) {};
		\node [style=none] (17) at (4.25, 0.25) {};
		\node [style=none] (18) at (4.25, -1.5) {};
		\node [style=none] (19) at (2.75, -1.5) {};
		\node [style=black dot] (20) at (2.75, -0.5) {};
		\node [style=none] (21) at (3.25, 2.25) {};
		\node [style=semilarge box] (22) at (7.25, 0.75) {};
		\node [style=none] (23) at (6.25, 0.25) {};
		\node [style=none] (24) at (7.25, 0.25) {};
		\node [style=none] (25) at (8.25, 0.25) {};
		\node [style=none] (26) at (8.25, -1.5) {};
		\node [style=none] (27) at (6.75, -1.5) {};
		\node [style=black dot] (28) at (6.75, -0.75) {};
		\node [style=none] (29) at (7.25, 2.25) {};
		\node [style=none] (30) at (5.25, 0.25) {$=$};
		\node [style=disintegration, minimum height=1.3cm, minimum width=1.2cm] (31) at (-4.75, 0.3) {};
		\node [style=disintegration, minimum height=0.9cm, minimum width=1.2cm] (32) at (-1.25, 0) {};
		\node [style=disintegration, minimum height=1.3cm, minimum width=1.5cm] (33) at (3.25, 0.35) {};
		\node [style=disintegration, minimum height=0.9cm, minimum width=1.5cm] (34) at (7.25, 0.75) {};
	\end{pgfonlayer}
	\begin{pgfonlayer}{edgelayer}
		\draw (6) to (2.center);
		\draw (3.center) to (4.center);
		\draw (5.center) to (0);
		\draw (12) to (8.center);
		\draw (9.center) to (10.center);
		\draw (11.center) to (7);
		\draw (21.center) to (14);
		\draw (17.center) to (18.center);
		\draw [in=150, out=-90, looseness=1.25] (15.center) to (20);
		\draw [in=-90, out=30, looseness=1.25] (20) to (16.center);
		\draw (20) to (19.center);
		\draw (29.center) to (22);
		\draw (25.center) to (26.center);
		\draw [in=150, out=-90, looseness=1.25] (23.center) to (28);
		\draw [in=-90, out=30, looseness=1.25] (28) to (24.center);
		\draw (28) to (27.center);
	\end{pgfonlayer}
\end{tikzpicture}}}
		\]
%
		
		\item \label{def:normalisation:norm} If $f$ is
                  normalised, that is, if $f = \big((\one \after
                  f)\otimes f\big) \after \Delta$, then $\normal(f) =
                  f$. This implies $\normal\big(\normal(g)\big) =
                  \normal(g)$, so that two nested shaded boxes around
                  $g$ can be reduced to a single shaded box.
	\end{enumerate}
\end{definition}

The second equation in point~\eqref{def:normalisation:comp} can
actually be derived, with some effort, see~\cite[Lem.~9]{LorenzT23},
but we include it as a requirement, for simplicity. One can show that
$\normal(f)$ is the least normalisation $f$ in the sense that if a map
$g$ is a normalisation of $f$, then $g$ also normalises $\normal(f)$
\cite[Prop.~100]{LorenzT23}.

\begin{definition}{\cite[Def. 3.23]{LavoreR23}}
	\label{def:comparator}
	A comparator structure on a CD-category assigns to each object $X$
	a map of the form $\nabla : X\otimes X \rightarrow X$, written as
	$\minicocopy$, satisfying commutativity, associativity,
	tensor-compatibility, and the Frobenius equations:
		\begin{equation}
			\label{eqn:frobenius}
			\vcenter{\hbox{\begin{tikzpicture}
	\begin{pgfonlayer}{nodelayer}
		\node [style=black dot] (6) at (5.25, -0.25) {};
		\node [style=black dot] (7) at (5.25, 0.5) {};
		\node [style=none] (8) at (4, 0) {$=$};
		\node [style=none] (9) at (4.75, 1) {};
		\node [style=none] (10) at (5.75, 1) {};
		\node [style=none] (11) at (4.75, -1) {};
		\node [style=none] (12) at (5.75, -1) {};
		\node [style=none] (13) at (7.5, -1) {};
		\node [style=none] (14) at (8.5, -1) {};
		\node [style=none] (15) at (7.75, 1) {};
		\node [style=none] (16) at (8.75, 1) {};
		\node [style=black dot] (17) at (8.5, -0.5) {};
		\node [style=none] (18) at (8.75, 0.25) {};
		\node [style=black dot] (19) at (7.75, 0.5) {};
		\node [style=none] (20) at (7.5, -0.25) {};
		\node [style=black dot] (21) at (-2.5, 0.5) {};
		\node [style=black dot] (22) at (-2.5, -0.5) {};
		\node [style=none] (23) at (-2.5, 1) {};
		\node [style=none] (24) at (-2.5, -1) {};
		\node [style=none] (25) at (-0.75, -1) {};
		\node [style=none] (26) at (-0.75, 1) {};
		\node [style=none] (27) at (-1.5, 0) {$=$};
		\node [style=none] (28) at (3, -1) {};
		\node [style=none] (29) at (2, -1) {};
		\node [style=none] (30) at (2.75, 1) {};
		\node [style=none] (31) at (1.75, 1) {};
		\node [style=black dot] (32) at (2, -0.5) {};
		\node [style=none] (33) at (1.75, 0.25) {};
		\node [style=black dot] (34) at (2.75, 0.5) {};
		\node [style=none] (35) at (3, -0.25) {};
		\node [style=none] (36) at (6.5, 0) {$=$};
	\end{pgfonlayer}
	\begin{pgfonlayer}{edgelayer}
		\draw (7) to (6);
		\draw [in=90, out=-135, looseness=1.25] (6) to (11.center);
		\draw [in=90, out=-45, looseness=1.25] (6) to (12.center);
		\draw [in=-90, out=30, looseness=1.25] (7) to (10.center);
		\draw [in=-90, out=150, looseness=1.25] (7) to (9.center);
		\draw [in=-90, out=30, looseness=1.25] (17) to (18.center);
		\draw [in=90, out=-135, looseness=1.25] (19) to (20.center);
		\draw (15.center) to (19);
		\draw (16.center) to (18.center);
		\draw (14.center) to (17);
		\draw (13.center) to (20.center);
		\draw [in=150, out=-45, looseness=1.25] (19) to (17);
		\draw (21) to (23.center);
		\draw [bend right, looseness=1.25] (22) to (21);
		\draw (24.center) to (22);
		\draw (25.center) to (26.center);
		\draw [bend right, looseness=1.25] (21) to (22);
		\draw [in=-90, out=135, looseness=1.25] (32) to (33.center);
		\draw [in=90, out=-45, looseness=1.25] (34) to (35.center);
		\draw (30.center) to (34);
		\draw (31.center) to (33.center);
		\draw (29.center) to (32);
		\draw (28.center) to (35.center);
		\draw [in=-135, out=30] (32) to (34);
	\end{pgfonlayer}
\end{tikzpicture}
                            \hspace*{1em}
			    \begin{tikzpicture}
	\begin{pgfonlayer}{nodelayer}
		\node [style=black dot] (1) at (-2.5, 0) {};
		\node [style=none] (2) at (-1.25, -0.25) {$=$};
		\node [style=none] (3) at (3.5, -1) {};
		\node [style=none] (4) at (-2.5, 0.5) {};
		\node [style=none] (5) at (-3, -0.5) {};
		\node [style=none] (6) at (-2, -0.5) {};
		\node [style=none] (7) at (-3, -1.25) {};
		\node [style=none] (8) at (-2, -1.25) {};
		\node [style=black dot] (9) at (0, 0) {};
		\node [style=none] (10) at (0, 0.5) {};
		\node [style=none] (11) at (-0.5, -0.5) {};
		\node [style=none] (12) at (0.5, -0.5) {};
		\node [style=none] (13) at (-0.5, -1) {};
		\node [style=none] (14) at (0.5, -1) {};
		\node [style=black dot] (15) at (3, 0.25) {};
		\node [style=none] (16) at (3, 0.75) {};
		\node [style=none] (18) at (3.5, -0.5) {};
		\node [style=black dot] (19) at (5.5, 0.25) {};
		\node [style=none] (20) at (5.5, 0.75) {};
		\node [style=none] (21) at (5, -0.5) {};
		\node [style=black dot] (23) at (2.5, -0.5) {};
		\node [style=none] (25) at (2.25, -1) {};
		\node [style=none] (26) at (2.75, -1) {};
		\node [style=black dot] (27) at (6, -0.475) {};
		\node [style=none] (29) at (5.75, -0.975) {};
		\node [style=none] (30) at (6.25, -0.975) {};
		\node [style=none] (31) at (5, -0.975) {};
		\node [style=none] (32) at (4.25, -0.25) {$=$};
		\node [style=black dot] (33) at (8.75, 0.25) {};
		\node [style=none] (34) at (8.75, 0.75) {};
		\node [style=none] (35) at (8, -0.5) {};
		\node [style=none] (36) at (9.5, -0.5) {};
		\node [style=none] (37) at (8, -1) {};
		\node [style=none] (38) at (9.5, -1) {};
		\node [style=black dot] (39) at (9.5, 0.25) {};
		\node [style=none] (40) at (9.5, 0.75) {};
		\node [style=none] (41) at (8.75, -0.5) {};
		\node [style=none] (42) at (10.25, -0.5) {};
		\node [style=none] (43) at (8.75, -1) {};
		\node [style=none] (44) at (10.25, -1) {};
		\node [style=black dot] (45) at (13.5, 0.5) {};
		\node [style=none] (46) at (13.5, 1) {};
		\node [style=none] (47) at (12.5, -0.25) {};
		\node [style=none] (48) at (14.5, -0.25) {};
		\node [style=none] (49) at (12.5, -1) {};
		\node [style=none] (50) at (14.5, -1) {};
		\node [style=none] (51) at (11.25, -0.25) {$=$};
		\node [style=none] (52) at (8, -1.5) {$X$};
		\node [style=none] (53) at (9.5, -1.5) {$X$};
		\node [style=none] (54) at (8.75, -1.5) {$Y$};
		\node [style=none] (55) at (8.75, 1.25) {$X$};
		\node [style=none] (56) at (10.25, -1.5) {$Y$};
		\node [style=none] (57) at (9.5, 1.25) {$Y$};
		\node [style=none] (58) at (12.25, -1.5) {$X\otimes Y$};
		\node [style=none] (59) at (14.75, -1.5) {$X\otimes Y$};
		\node [style=none] (60) at (13.5, 1.5) {$X\otimes Y$};
	\end{pgfonlayer}
	\begin{pgfonlayer}{edgelayer}
		\draw (4.center) to (1);
		\draw [in=90, out=-150] (1) to (5.center);
		\draw [in=90, out=-30] (1) to (6.center);
		\draw (10.center) to (9);
		\draw [in=90, out=-150] (9) to (11.center);
		\draw [in=90, out=-30] (9) to (12.center);
		\draw [in=90, out=-90] (5.center) to (8.center);
		\draw [in=-90, out=90] (7.center) to (6.center);
		\draw (11.center) to (13.center);
		\draw (12.center) to (14.center);
		\draw (16.center) to (15);
		\draw [in=90, out=-30] (15) to (18.center);
		\draw (20.center) to (19);
		\draw [in=90, out=-150] (19) to (21.center);
		\draw [in=90, out=-150] (23) to (25.center);
		\draw [in=90, out=-30] (23) to (26.center);
		\draw [in=90, out=-150] (27) to (29.center);
		\draw [in=90, out=-30] (27) to (30.center);
		\draw [in=-150, out=90, looseness=1.25] (23) to (15);
		\draw (18.center) to (3.center);
		\draw (31.center) to (21.center);
		\draw [in=90, out=-30] (19) to (27);
		\draw (34.center) to (33);
		\draw [in=90, out=-150] (33) to (35.center);
		\draw [in=90, out=-30] (33) to (36.center);
		\draw (35.center) to (37.center);
		\draw (36.center) to (38.center);
		\draw (40.center) to (39);
		\draw [in=90, out=-150] (39) to (41.center);
		\draw [in=90, out=-30] (39) to (42.center);
		\draw (41.center) to (43.center);
		\draw (42.center) to (44.center);
		\draw (46.center) to (45);
		\draw [in=90, out=-150] (45) to (47.center);
		\draw [in=90, out=-30] (45) to (48.center);
		\draw (47.center) to (49.center);
		\draw (48.center) to (50.center);
	\end{pgfonlayer}
\end{tikzpicture}}}
		\end{equation}
		
		

	We then define a \emph{cap} as $\one \after \nabla \colon 
	X\otimes X
	\rightarrow I$, see on the left below. We further formulate the
	requirement that these comparators are \emph{cancellative} in 
	terms of
	the property below on the right.
	\[ \quad\vcenter{\hbox{\begin{tikzpicture}
	\begin{pgfonlayer}{nodelayer}
		\node [style=black dot] (0) at (-4.25, 0.25) {};
		\node [style=none] (2) at (-4.75, -0.25) {};
		\node [style=none] (3) at (-3.75, -0.25) {};
		\node [style=none] (4) at (-4.75, -0.75) {};
		\node [style=none] (5) at (-3.75, -0.75) {};
		\node [style=none] (6) at (-5.75, 0) {$\coloneqq$};
		\node [style=none] (7) at (-7.75, 0) {};
		\node [style=none] (8) at (-6.75, 0) {};
		\node [style=none] (9) at (-7.75, -0.75) {};
		\node [style=none] (10) at (-6.75, -0.75) {};
		\node [style=upground] (11) at (-4.25, 0.75) {};
		\node [style=small box] (12) at (3.5, 0) {$f$};
		\node [style=small box] (13) at (10.5, 0) {$f$};
		\node [style=none] (16) at (2.5, 0.5) {};
		\node [style=none] (17) at (3.25, 0.5) {};
		\node [style=none] (18) at (2.5, -1) {};
		\node [style=none] (19) at (3.5, -1) {};
		\node [style=none] (20) at (10.5, -1) {};
		\node [style=none] (21) at (4.75, 0.25) {$=$};
		\node [style=none] (22) at (10.25, 1.25) {};
		\node [style=small box] (23) at (13, 0) {$g$};
		\node [style=none] (24) at (13, -1) {};
		\node [style=none] (26) at (11.75, 0) {$=$};
		\node [style=small box] (27) at (6.5, 0) {$g$};
		\node [style=none] (28) at (5.5, 0.5) {};
		\node [style=none] (29) at (6.25, 0.5) {};
		\node [style=none] (30) at (5.5, -1) {};
		\node [style=none] (31) at (6.5, -1) {};
		\node [style=none] (32) at (8.5, 0) {$\Longrightarrow$};
		\node [style=none] (33) at (6.75, 0.5) {};
		\node [style=none] (34) at (3.75, 0.5) {};
		\node [style=none] (35) at (3.75, 1.25) {};
		\node [style=none] (36) at (6.75, 1.25) {};
		\node [style=none] (37) at (10.25, 0.5) {};
		\node [style=none] (38) at (10.75, 0.5) {};
		\node [style=none] (39) at (10.75, 1.25) {};
		\node [style=none] (40) at (12.75, 1.25) {};
		\node [style=none] (41) at (12.75, 0.5) {};
		\node [style=none] (42) at (13.25, 0.5) {};
		\node [style=none] (43) at (13.25, 1.25) {};
	\end{pgfonlayer}
	\begin{pgfonlayer}{edgelayer}
		\draw [in=90, out=-150] (0) to (2.center);
		\draw [in=90, out=-30] (0) to (3.center);
		\draw (2.center) to (4.center);
		\draw (3.center) to (5.center);
		\draw (7.center) to (9.center);
		\draw (8.center) to (10.center);
		\draw [bend left=90, looseness=2.00] (7.center) to (8.center);
		\draw (11) to (0);
		\draw (16.center) to (18.center);
		\draw [bend left=90, looseness=2.00] (16.center) to (17.center);
		\draw (12) to (19.center);
		\draw (13) to (20.center);
		\draw (23) to (24.center);
		\draw (28.center) to (30.center);
		\draw [bend left=90, looseness=2.00] (28.center) to (29.center);
		\draw (27) to (31.center);
		\draw (35.center) to (34.center);
		\draw (36.center) to (33.center);
		\draw (22.center) to (37.center);
		\draw (39.center) to (38.center);
		\draw (40.center) to (41.center);
		\draw (43.center) to (42.center);
	\end{pgfonlayer}
\end{tikzpicture}}} \]
\end{definition}

The second equation below comes 
from~\cite[Lem.~11.11]{Fritz20}.
It can be derived from the first equation.

\begin{lemma}
	\label{lem:nestedbox}
	In a CD-category with normalisation and cancellative comparators,
	nested boxes can be reduced to a single box, as on the left below:
\[ \vcenter{\hbox{\begin{tikzpicture}
	\begin{pgfonlayer}{nodelayer}
		\node [style=small box] (0) at (-11.25, 1) {$g$};
		\node [style=small box] (1) at (-11.25, -1) {$f$};
		\node [style=none] (2) at (-11.25, -2.25) {};
		\node [style=none] (3) at (-11.25, 2.25) {};
		\node [style=small box] (4) at (-8.25, 1) {$g$};
		\node [style=small box] (5) at (-8.25, -1) {$f$};
		\node [style=none] (6) at (-8.25, -2.25) {};
		\node [style=none] (7) at (-8.25, 2.25) {};
		\node [style=none] (8) at (-9.75, 0) {$=$};
		\node [style=disintegration, minimum height=1.8cm, minimum width=0.9cm] (9) at (-8.25, 0) {};
		\node [style=disintegration, minimum height=1.9cm, minimum width=1.0cm] (10) at (-11.25, -0.1) {};
		\node [style=disintegration, minimum height=0.8cm, minimum width=0.8cm] (11) at (-11.25, -1) {};
		\node [style=none] (12) at (7.25, 0) {};
		\node [style=none] (13) at (7.75, -0.25) {};
		\node [style=none] (14) at (6.25, 0) {};
		\node [style=none] (15) at (6.25, -2) {};
		\node [style=none] (16) at (7.75, -1.25) {};
		\node [style=none] (17) at (7.75, -2) {};
		\node [style=none] (18) at (7.75, 0.25) {};
		\node [style=none] (19) at (5.5, 0) {};
		\node [style=none] (20) at (5.5, -2) {};
		\node [style=none] (21) at (7.25, -0.25) {};
		\node [style=medium box] (23) at (7.75, -0.75) {};
		\node [style=none] (24) at (8.25, -0.25) {};
		\node [style=none] (25) at (8.25, 2.25) {};
		\node [style=none] (26) at (2.25, 0) {};
		\node [style=none] (27) at (2.75, -0.25) {};
		\node [style=none] (28) at (1.25, 0) {};
		\node [style=none] (29) at (1.25, -2) {};
		\node [style=none] (30) at (2.75, -1.25) {};
		\node [style=none] (31) at (2.75, -2) {};
		\node [style=none] (32) at (2.75, 0.25) {};
		\node [style=none] (33) at (0.5, 0.25) {};
		\node [style=none] (34) at (0.5, -2) {};
		\node [style=none] (35) at (2.25, -0.25) {};
		\node [style=medium box] (36) at (2.75, -0.75) {};
		\node [style=none] (37) at (3.25, -0.25) {};
		\node [style=none] (38) at (3.25, 2.25) {};
		\node [style=disintegration, minimum height=1.6cm, minimum width=1.8cm] (39) at (7, 0.05) {};
		\node [style=disintegration, minimum height=1.7cm, minimum width=1.8cm] (40) at (2.1, 0) {};
		\node [style=disintegration, minimum height=1.15cm, minimum width=1.3cm] (41) at (2.4, -0.4) {};
		\node [style=none] (42) at (4.5, 0) {$=$};
	\end{pgfonlayer}
	\begin{pgfonlayer}{edgelayer}
		\draw (3.center) to (0);
		\draw (0) to (1);
		\draw (1) to (2.center);
		\draw (7.center) to (4);
		\draw (4) to (5);
		\draw (5) to (6.center);
		\draw (17.center) to (16.center);
		\draw [in=90, out=90, looseness=2.25] (12.center) to (14.center);
		\draw (14.center) to (15.center);
		\draw (13.center) to (18.center);
		\draw (20.center) to (19.center);
		\draw (21.center) to (12.center);
		\draw (24.center) to (25.center);
		\draw [bend right=90, looseness=2.00] (18.center) to (19.center);
		\draw (31.center) to (30.center);
		\draw [in=90, out=90, looseness=2.25] (26.center) to (28.center);
		\draw (28.center) to (29.center);
		\draw (27.center) to (32.center);
		\draw (34.center) to (33.center);
		\draw (35.center) to (26.center);
		\draw (37.center) to (38.center);
		\draw [bend right=90, looseness=2.00] (32.center) to (33.center);
	\end{pgfonlayer}
\end{tikzpicture}}} \]
\end{lemma}

We can now define what `full support' means for a distribution and a
channel, in diagrammatic terms.

\begin{definition}
  \label{def:fullsupport}
  For a state $\omega$ and, more generally, for a map $f$, in a
  CD-category, we say that it has \emph{full support} when:
\[ \vcenter{\hbox{\begin{tikzpicture}
	\begin{pgfonlayer}{nodelayer}
		\node [style=small box] (0) at (5, 0) {$f$};
		\node [style=none] (1) at (4, 0.5) {};
		\node [style=none] (2) at (5, 0.5) {};
		\node [style=none] (3) at (4, -1.25) {};
		\node [style=none] (4) at (5, -1.25) {};
		\node [style=none] (5) at (7.5, -1) {};
		\node [style=none] (6) at (6.5, 0) {$=$};
		\node [style=upground] (7) at (7.5, 0.75) {};
		\node [style=none] (8) at (8.5, -1) {};
		\node [style=upground] (9) at (8.5, 0.75) {};
		\node [style=disintegration, minimum height=1.0cm, minimum width=1.0cm] (10) at (4.8, 0.2) {};
		\node [style=small box] (11) at (-7.25, 0) {$\omega$};
		\node [style=none] (12) at (-8.25, 0.5) {};
		\node [style=none] (13) at (-7.25, 0.5) {};
		\node [style=none] (14) at (-8.25, -1.25) {};
		\node [style=none] (16) at (-4.75, -1) {};
		\node [style=none] (17) at (-5.75, 0) {$=$};
		\node [style=upground] (18) at (-4.75, 0.75) {};
		\node [style=disintegration, minimum height=1.0cm, minimum width=1.0cm] (21) at (-7.45, 0.2) {};
	\end{pgfonlayer}
	\begin{pgfonlayer}{edgelayer}
		\draw (1.center) to (3.center);
		\draw [bend left=90, looseness=2.00] (1.center) to (2.center);
		\draw (0) to (4.center);
		\draw (2.center) to (0);
		\draw (7) to (5.center);
		\draw (9) to (8.center);
		\draw (12.center) to (14.center);
		\draw [bend left=90, looseness=2.00] (12.center) to (13.center);
		\draw (13.center) to (11);
		\draw (18) to (16.center);
	\end{pgfonlayer}
\end{tikzpicture}}} \]
\end{definition}

The normalisation $\normal(\omega)$ of a non-zero subdistribution 
$\omega \in \subDst(X)$ is defined to be $\normal(\omega) = 
\frac{1}{\|\omega\|} \cdot \omega$. We set $\normal(\zero) = 
\zero \in \subDst(X)$. The normalisation structure on $\Kl(\subDst)$ 
is then defined pointwise on subchannels $f : X \to \subDst(Y)$ by 
$\normal(f)(x) = \normal(f(x))$. It satisfies the axioms of Definition
\ref{def:normalisation}.


For each set $X$ there is a comparator map $\nabla : X \times X \to
\subDst(X)$ in $\Kl(\subDst)$, namely $\nabla(x,x') = 1\ket{x}$ if
$x=x'$, and $\nabla(x,x') = \zero$ if $x \ne x'$. Comparators in
$\Kl(\subDst)$ are cancellative. Concretely, this means the following
for parallel subchannels $f,g : X \to \subDst(Y \times Z)$. The
equation $f = g$ holds if $f(x)(y,z) = g(x)(y,z)$ for all $x \in X$,
$y \in Y$, and $z \in Z$.

A subchannel $f : X \to \subDst(Y)$ has full support, according to
Definition~\ref{def:fullsupport}, if and only if $f(x)(y) > 0$ for all
$x \in X$ and $y \in Y$. This can hold only if the set $Y$ is
finite. This notion differs from what is sometimes meant by $f$ having
full support, which is that for all $y \in Y$ there exists an
$x \in X$ such that $f(x)(y) > 0$, see e.g. \cite{FritzGLPS23}.  The 
notion of full support in
Definition~\ref{def:fullsupport} for channels is more restrictive than
this, since it requires that every distribution $f(x)$ has full
support.

\section{Disintegration and daggers}\label{sec:disintegration}

Disintegration is the technique of extracting a conditional
probability $P(z\mid y)$ from a joint probability $P(y,z)$, such that
$P(z\mid y) \cdot P(y) = P(y,z)$. More generally, this may be done in
parameterised form, by extracting $P(z\mid y,x)$ from a joint
probability $P(y,z\mid x)$, such that $P(z\mid y,x) \cdot P(y\mid x) =
P(y,z\mid x)$. Crucially, disintegration involves normalisation. Its
diagrammatic description below is now standard, see
\textit{e.g.}~\cite{LorenzT23,LavoreR23,FritzKMSW24,Fritz20,ChoJ19},
building on earlier categorical formulations, for instance
in~\cite{CulbertsonS14,ClercDDG17}. The separation of normalisation
and comparison for bending gives an intuitive description.

\begin{wrapfigure}{r}{0.25\textwidth}
	\vspace{-2em}
	\[ \quad\vcenter{\hbox{\begin{tikzpicture}
	\begin{pgfonlayer}{nodelayer}
		\node [style=small box] (0) at (0, 0) {$f$};
		\node [style=none] (1) at (-0.25, 0.5) {};
		\node [style=none] (2) at (0.25, 0.5) {};
		\node [style=none] (3) at (0, -1.25) {};
		\node [style=none] (4) at (-1.5, 0.5) {};
		\node [style=none] (5) at (-1.5, -1.25) {};
		\node [style=none] (6) at (0.25, 1.75) {};
		\node [style=disintegration, minimum height=1.0cm, minimum width=1.3cm] (7) at (-0.5, 0.25) {};
		\node [style=none] (8) at (-3, 0) {$\coloneqq$};
		\node [style=none] (9) at (-5, 0) {$\disintegration(f)$};
	\end{pgfonlayer}
	\begin{pgfonlayer}{edgelayer}
		\draw (4.center) to (5.center);
		\draw [bend left=90, looseness=1.75] (4.center) to (1.center);
		\draw (0) to (3.center);
		\draw (2.center) to (6.center);
	\end{pgfonlayer}
\end{tikzpicture}}} \]
	\vspace{-2em}	
\end{wrapfigure}

Thus, in a general CD-category with comparators and normalisation,
disintegration involves the passage from a map $f\colon X\rightarrow
Y\otimes Z$ with two outgoing wires into a map $\disintegration(f)
\colon Y\otimes X \rightarrow Z$ as on the right. We make the crucial
property of disintegration explicit, following~\cite{LorenzT23}.

\begin{theorem}{(\cite[Prop. 109]{LorenzT23})}
	\label{thm:disintegration}
	In a CD-category with normalisation and cancellative comparators 
	one
	can recover a map $f$ from its disintegration in the following 
	manner.
	\begin{equation}
		\label{eqn:disintegration}
		\vcenter{\hbox{\begin{tikzpicture}
	\begin{pgfonlayer}{nodelayer}
		\node [style=small box] (0) at (-3.75, 0) {$f$};
		\node [style=none] (1) at (-4, 0.5) {};
		\node [style=none] (2) at (-3.5, 0.5) {};
		\node [style=none] (3) at (-3.75, -1.75) {};
		\node [style=none] (4) at (-4, 2) {};
		\node [style=none] (5) at (-3.5, 2) {};
		\node [style=none] (6) at (-2.25, 0) {$=$};
		\node [style=medium box] (7) at (1, 1.75) {$\disintegration(f)$};
		\node [style=none] (9) at (1.75, -0.25) {};
		\node [style=none] (10) at (0.25, 1.25) {};
		\node [style=none] (11) at (1.75, 1.25) {};
		\node [style=none] (12) at (1, 2.75) {};
		\node [style=small box] (13) at (-0.25, -0.75) {$f$};
		\node [style=none] (14) at (0, -0.25) {};
		\node [style=none] (15) at (-0.5, -0.25) {};
		\node [style=none] (16) at (-0.25, -2.25) {};
		\node [style=upground] (17) at (0, 0.25) {};
		\node [style=black dot] (18) at (-0.25, -1.75) {};
		\node [style=black dot] (19) at (-0.5, 0.5) {};
		\node [style=none] (20) at (-1.25, 1.25) {};
		\node [style=none] (21) at (-1.25, 2.75) {};
		\node [style=none] (22) at (-4.25, -1.5) {$X$};
		\node [style=none] (23) at (-0.75, -2) {$X$};
		\node [style=none] (24) at (-4.5, 1.75) {$Y$};
		\node [style=none] (25) at (-1.75, 2.5) {$Y$};
		\node [style=none] (26) at (-3, 1.75) {$Z$};
		\node [style=none] (27) at (1.5, 2.75) {$Z$};
	\end{pgfonlayer}
	\begin{pgfonlayer}{edgelayer}
		\draw (4.center) to (1.center);
		\draw (5.center) to (2.center);
		\draw (0) to (3.center);
		\draw (11.center) to (9.center);
		\draw (12.center) to (7);
		\draw (13) to (18);
		\draw (18) to (16.center);
		\draw (14.center) to (17);
		\draw [in=-90, out=15, looseness=1.25] (18) to (9.center);
		\draw [in=165, out=-90, looseness=1.50] (20.center) to (19);
		\draw (19) to (15.center);
		\draw [in=-90, out=15, looseness=1.25] (19) to (10.center);
		\draw (21.center) to (20.center);
	\end{pgfonlayer}
\end{tikzpicture}}} 
		\hspace*{3em}\mbox{\textit{i.e.}}\hspace*{3.2em}
		\begin{array}{rcl}
			P(y,z\mid x)
			& = &
			P(z\mid y,x) \cdot P(y\mid x).
		\end{array}
	\end{equation}
\end{theorem}

\begin{myproof}
  By applying the cancellativity of caps to the next line, where the
  first equation is the normalisation property on the right
  in~\eqref{eqn:normalisation} and the second equation is
  Frobenius~\eqref{eqn:frobenius}.
  \[ \vcenter{\hbox{\begin{tikzpicture}
	\begin{pgfonlayer}{nodelayer}
		\node [style=small box] (0) at (-5, 0) {$f$};
		\node [style=none] (1) at (-5.25, 0.5) {};
		\node [style=none] (2) at (-4.75, 0.5) {};
		\node [style=none] (3) at (-5, -1.75) {};
		\node [style=none] (4) at (-5.25, 1) {};
		\node [style=none] (5) at (-4.75, 2) {};
		\node [style=none] (6) at (-3, 0) {$=$};
		\node [style=medium box] (7) at (9.5, 1.75) {$\disintegration(f)$};
		\node [style=none] (9) at (10.25, -0.25) {};
		\node [style=none] (10) at (8.75, 1.25) {};
		\node [style=none] (11) at (10.25, 1.25) {};
		\node [style=none] (12) at (9.5, 2.75) {};
		\node [style=small box] (13) at (8.25, -0.75) {$f$};
		\node [style=none] (14) at (8.5, -0.25) {};
		\node [style=none] (15) at (8, -0.25) {};
		\node [style=none] (16) at (8.25, -2.25) {};
		\node [style=upground] (17) at (8.5, 0.25) {};
		\node [style=black dot] (18) at (8.25, -1.75) {};
		\node [style=black dot] (19) at (8, 0.5) {};
		\node [style=none] (20) at (7.25, 1.25) {};
		\node [style=none] (22) at (-6, 1) {};
		\node [style=none] (23) at (-6, -1.75) {};
		\node [style=small box] (24) at (0, 0.25) {$f$};
		\node [style=none] (25) at (0.25, 0.75) {};
		\node [style=upground] (27) at (0.25, 1.5) {};
		\node [style=medium box] (29) at (2.25, 0.25) {$\disintegration(f)$};
		\node [style=none] (30) at (1.5, -0.25) {};
		\node [style=none] (31) at (3, -0.25) {};
		\node [style=none] (32) at (-0.25, 0.75) {};
		\node [style=none] (33) at (-0.25, 1.25) {};
		\node [style=none] (34) at (-1.25, 1.25) {};
		\node [style=none] (35) at (-1.25, -0.25) {};
		\node [style=none] (36) at (0, -0.25) {};
		\node [style=none] (37) at (2.25, 0.75) {};
		\node [style=none] (38) at (2.25, 2.5) {};
		\node [style=none] (39) at (-1.75, -2.25) {};
		\node [style=none] (40) at (1.25, -2.25) {};
		\node [style=black dot] (41) at (-1.75, -1.5) {};
		\node [style=black dot] (42) at (1.25, -1.5) {};
		\node [style=none] (43) at (5, 0) {$=$};
		\node [style=none] (45) at (7.25, 1.75) {};
		\node [style=none] (46) at (6.5, 1.75) {};
		\node [style=none] (47) at (6.5, -2.25) {};
		\node [style=black dot] (48) at (-0.75, 1.75) {};
		\node [style=upground] (49) at (-0.75, 2.25) {};
	\end{pgfonlayer}
	\begin{pgfonlayer}{edgelayer}
		\draw (4.center) to (1.center);
		\draw (5.center) to (2.center);
		\draw (0) to (3.center);
		\draw (11.center) to (9.center);
		\draw (12.center) to (7);
		\draw (13) to (18);
		\draw (18) to (16.center);
		\draw (14.center) to (17);
		\draw [in=-90, out=15, looseness=1.25] (18) to (9.center);
		\draw [in=165, out=-90, looseness=1.50] (20.center) to (19);
		\draw (19) to (15.center);
		\draw [in=-90, out=15, looseness=1.25] (19) to (10.center);
		\draw [bend left=90, looseness=1.75] (22.center) to (4.center);
		\draw (22.center) to (23.center);
		\draw (25.center) to (27);
		\draw (33.center) to (32.center);
		\draw [bend left=90, looseness=1.75] (34.center) to (33.center);
		\draw (34.center) to (35.center);
		\draw (37.center) to (38.center);
		\draw [in=-90, out=150, looseness=1.25] (42) to (36.center);
		\draw [in=-90, out=30] (42) to (31.center);
		\draw [in=-90, out=30, looseness=1.25] (41) to (35.center);
		\draw [in=-90, out=135] (41) to (30.center);
		\draw (41) to (39.center);
		\draw (42) to (40.center);
		\draw [bend left=90, looseness=1.75] (46.center) to (45.center);
		\draw (46.center) to (47.center);
		\draw (45.center) to (20.center);
		\draw (49) to (48);
	\end{pgfonlayer}
\end{tikzpicture}}} 
  \eqno{\QEDbox} \]
\end{myproof}

We turn to `daggers', that is, to reversal of maps. This reversal
corresponds to turning a conditional probability $P(y\mid x)$ upside
down into $P(x\mid y)$. Such daggers are standardly defined with
respect to a `prior' distribution $\omega$,
see~\cite{ClercDDG17,ChoJ19,Fritz20}.  Here we define them also in
parametrised form, with a channel as prior, on the right
in~\eqref{eqn:daggerdef}.

\begin{definition}
  \label{def:dagger}
  Let $f\colon Y \rightarrow Z$ be a map in a CD-category. For a state
  $\omega$ on $Y$ and channel $c\colon X \rightarrow Y$ we define the
  dagger $f^{\dag}_{\omega} \colon Z \rightarrow Y$ and the
  parametrised dagger $f^{\dag}_{c} \colon Z\otimes X \rightarrow Y$
  as disintegrations in~\eqref{eqn:daggerdef}.

	\hspace*{-2em}\begin{minipage}{0.45\textwidth}
		\begin{equation}
			\label{eqn:daggerdef}
			\vcenter{\hbox{\begin{tikzpicture}
	\begin{pgfonlayer}{nodelayer}
		\node [style=small box] (0) at (-2.75, -2) {};
		\node [style=small box] (1) at (-2.75, -2) {$\omega$};
		\node [style=black dot] (2) at (-2.75, -1) {};
		\node [style=none] (3) at (-2.75, -1.75) {};
		\node [style=none] (4) at (-2, 0) {};
		\node [style=small box] (6) at (-3.5, 0) {};
		\node [style=small box] (7) at (-3.5, 0) {$f$};
		\node [style=none] (9) at (-3.5, 0.75) {};
		\node [style=none] (10) at (-4.5, 0.75) {};
		\node [style=none] (11) at (-2, 2.25) {};
		\node [style=none] (12) at (-4.5, -3.5) {};
		\node [style=disintegration, minimum height=2.1cm, minimum width=1.5cm] (13) at (-3.25, -0.7) {};
		\node [style=none] (14) at (-1.5, 2) {$Y$};
		\node [style=none] (15) at (-5, -3.25) {$Z$};
		\node [style=none] (29) at (-6, -0.75) {$\coloneqq$};
		\node [style=none] (30) at (-7, -0.75) {$f^{\dag}_{\omega}$};
		\node [style=none] (31) at (1.25, -0.75) {$\coloneqq$};
		\node [style=none] (32) at (0.25, -0.75) {$f^{\dag}_{c}$};
		\node [style=small box] (33) at (4.5, -2) {};
		\node [style=small box] (34) at (4.5, -2) {$c$};
		\node [style=black dot] (35) at (4.5, -1) {};
		\node [style=none] (36) at (4.5, -1.75) {};
		\node [style=none] (37) at (5.25, 0) {};
		\node [style=small box] (38) at (3.75, 0) {};
		\node [style=small box] (39) at (3.75, 0) {$f$};
		\node [style=none] (40) at (3.75, 0.75) {};
		\node [style=none] (41) at (2.75, 0.75) {};
		\node [style=none] (42) at (5.25, 2.25) {};
		\node [style=none] (43) at (2.75, -3.5) {};
		\node [style=disintegration, minimum height=2.1cm, minimum width=1.5cm] (44) at (4, -0.7) {};
		\node [style=none] (45) at (5.75, 2) {$Y$};
		\node [style=none] (46) at (2.25, -3.25) {$Z$};
		\node [style=none] (47) at (4.5, -3.5) {};
		\node [style=none] (48) at (5, -3.25) {$X$};
	\end{pgfonlayer}
	\begin{pgfonlayer}{edgelayer}
		\draw (3.center) to (2);
		\draw [in=-90, out=30, looseness=1.25] (2) to (4.center);
		\draw [bend left=90, looseness=1.75] (10.center) to (9.center);
		\draw (12.center) to (10.center);
		\draw (4.center) to (11.center);
		\draw (9.center) to (7);
		\draw [in=165, out=-90, looseness=1.25] (7) to (2);
		\draw (36.center) to (35);
		\draw [in=-90, out=30, looseness=1.25] (35) to (37.center);
		\draw [bend left=90, looseness=1.75] (41.center) to (40.center);
		\draw (43.center) to (41.center);
		\draw (37.center) to (42.center);
		\draw (40.center) to (39);
		\draw [in=165, out=-90, looseness=1.25] (39) to (35);
		\draw (34) to (47.center);
	\end{pgfonlayer}
\end{tikzpicture}}}
		\end{equation}
	\end{minipage}
        \hspace*{1.5em}
	\begin{minipage}{0.54\textwidth}
		\begin{equation}
			\label{eqn:dagger}
			\vcenter{\hbox{\quad\begin{tikzpicture}
	\begin{pgfonlayer}{nodelayer}
		\node [style=black dot] (0) at (2, -1) {};
		\node [style=none] (1) at (2.75, 0.25) {};
		\node [style=none] (2) at (1.25, 1.5) {};
		\node [style=none] (3) at (2.75, 1.5) {};
		\node [style=none] (4) at (3.75, -0.5) {$=$};
		\node [style=small box] (5) at (5.5, -2.5) {$\omega$};
		\node [style=small box] (6) at (5.5, -1) {$f$};
		\node [style=black dot] (7) at (5.5, 0) {};
		\node [style=small box] (8) at (2, -2) {$\omega$};
		\node [style=small box] (9) at (1.25, 0.25) {$f$};
		\node [style=small box] (10) at (6.25, 1.25) {$f^{\dag}_{\omega}$};
		\node [style=none] (11) at (4.75, 1.25) {};
		\node [style=none] (12) at (4.75, 2.5) {};
		\node [style=none] (13) at (6.25, 2.5) {};
		\node [style=small box] (14) at (10.5, -1.75) {$c$};
		\node [style=small box] (15) at (9.75, 0.5) {$f$};
		\node [style=black dot] (16) at (10.5, -0.75) {};
		\node [style=none] (17) at (10.5, -3.25) {};
		\node [style=none] (18) at (11.25, 0.5) {};
		\node [style=none] (19) at (9.75, 2) {};
		\node [style=none] (20) at (11.25, 2) {};
		\node [style=none] (21) at (12.25, -0.5) {$=$};
		\node [style=none] (22) at (14.25, -3.75) {};
		\node [style=small box] (23) at (14.25, -2) {$c$};
		\node [style=small box] (24) at (14.25, -0.5) {$f$};
		\node [style=black dot] (25) at (14.25, 0.5) {};
		\node [style=black dot] (26) at (14.25, -3.25) {};
		\node [style=medium box] (27) at (15.5, 1.75) {$f^{\dag}_{c}$};
		\node [style=none] (28) at (13.5, 1.25) {};
		\node [style=none] (29) at (15.75, 1.25) {};
		\node [style=none] (30) at (15.5, 2.75) {};
		\node [style=none] (31) at (15, 1.25) {};
		\node [style=none] (32) at (13.5, 2.75) {};
	\end{pgfonlayer}
	\begin{pgfonlayer}{edgelayer}
		\draw [in=-90, out=30, looseness=1.25] (0) to (1.center);
		\draw (1.center) to (3.center);
		\draw (8) to (0);
		\draw (5) to (6);
		\draw (6) to (7);
		\draw (9) to (2.center);
		\draw [in=-90, out=150, looseness=1.25] (0) to (9);
		\draw (11.center) to (12.center);
		\draw (10) to (13.center);
		\draw [in=-90, out=30, looseness=1.25] (7) to (10);
		\draw [in=-90, out=150, looseness=1.25] (7) to (11.center);
		\draw (20.center) to (18.center);
		\draw [in=30, out=-90] (18.center) to (16);
		\draw [in=-90, out=165] (16) to (15);
		\draw (15) to (19.center);
		\draw (16) to (14);
		\draw (14) to (17.center);
		\draw (27) to (30.center);
		\draw [in=30, out=-90] (31.center) to (25);
		\draw [in=270, out=150] (25) to (28.center);
		\draw [in=15, out=-90, looseness=0.75] (29.center) to (26);
		\draw (26) to (23);
		\draw (23) to (24);
		\draw (24) to (25);
		\draw (22.center) to (26);
		\draw (32.center) to (28.center);
	\end{pgfonlayer}
\end{tikzpicture}}}
		\end{equation}
	\end{minipage}
These daggers satisfy appropriate instantiations of the disintegration
equation~\eqref{eqn:disintegration}, which we make explicit above on
the right.
\end{definition}

We now come to our new result, for removal of shaded boxes. It
plays a crucial role in the compositional reasoning for disintegration
that is developed in this paper.

\begin{proposition}
	\label{prop:boxremoval}
	In a CD-category with normalisation and cancellative comparators, 
	a
	state $\omega$ and a channel $c$,
	both with full support, satisfy:
	\[ \vcenter{\hbox{\begin{tikzpicture}
	\begin{pgfonlayer}{nodelayer}
		\node [style=small box] (0) at (7.5, -2) {};
		\node [style=small box] (1) at (7.5, -2) {$c$};
		\node [style=black dot] (2) at (7.5, -1) {};
		\node [style=none] (3) at (7.5, -1.75) {};
		\node [style=none] (4) at (8, -0.25) {};
		\node [style=none] (8) at (6, -0.25) {};
		\node [style=none] (9) at (8, 1) {};
		\node [style=none] (10) at (6, -3.25) {};
		\node [style=disintegration, minimum height=1.6cm, minimum width=1.3cm] (11) at (7.05, -1.15) {};
		\node [style=none] (14) at (7.5, -3.25) {};
		\node [style=none] (16) at (7, -0.25) {};
		\node [style=none] (17) at (9.5, -0.75) {$=$};
		\node [style=none] (18) at (10.75, -2.5) {};
		\node [style=none] (19) at (11.75, -2.5) {};
		\node [style=none] (20) at (10.75, 0.75) {};
		\node [style=upground] (21) at (11.75, -0.5) {};
		\node [style=small box] (22) at (-3.5, -2) {};
		\node [style=small box] (23) at (-3.5, -2) {$\omega$};
		\node [style=black dot] (24) at (-3.5, -1) {};
		\node [style=none] (25) at (-3.5, -1.75) {};
		\node [style=none] (26) at (-3, -0.25) {};
		\node [style=none] (27) at (-5, -0.25) {};
		\node [style=none] (28) at (-3, 1) {};
		\node [style=none] (29) at (-5, -3.25) {};
		\node [style=disintegration, minimum height=1.6cm, minimum width=1.3cm] (30) at (-3.95, -1.15) {};
		\node [style=none] (32) at (-4, -0.25) {};
		\node [style=none] (33) at (-1.5, -0.75) {$=$};
		\node [style=none] (34) at (-0.25, -2.5) {};
		\node [style=none] (36) at (-0.25, 0.75) {};
		\node [style=none] (37) at (2.75, -0.75) {and};
	\end{pgfonlayer}
	\begin{pgfonlayer}{edgelayer}
		\draw (3.center) to (2);
		\draw [in=-90, out=30, looseness=1.25] (2) to (4.center);
		\draw (10.center) to (8.center);
		\draw (4.center) to (9.center);
		\draw (1) to (14.center);
		\draw [in=165, out=-90, looseness=1.25] (16.center) to (2);
		\draw [bend left=90, looseness=2.00] (8.center) to (16.center);
		\draw (20.center) to (18.center);
		\draw (21) to (19.center);
		\draw (25.center) to (24);
		\draw [in=-90, out=30, looseness=1.25] (24) to (26.center);
		\draw (29.center) to (27.center);
		\draw (26.center) to (28.center);
		\draw [in=165, out=-90, looseness=1.25] (32.center) to (24);
		\draw [bend left=90, looseness=2.00] (27.center) to (32.center);
		\draw (36.center) to (34.center);
	\end{pgfonlayer}
\end{tikzpicture}}} \]
\end{proposition}

\begin{myproof}
  The equation for a state $\omega$ on the left is a special case, so
  we do the proof for the more general case on the right. It uses
  Frobenius~\eqref{eqn:frobenius},
  Definition~\ref{def:normalisation}~\eqref{def:normalisation:comp},
  and the fact that $c$ has full support.
  \[ \vcenter{\hbox{\begin{tikzpicture}
	\begin{pgfonlayer}{nodelayer}
		\node [style=small box] (0) at (4.5, -2) {};
		\node [style=small box] (1) at (4.5, -2) {$c$};
		\node [style=black dot] (2) at (4.5, -1) {};
		\node [style=none] (3) at (4.5, -1.75) {};
		\node [style=none] (4) at (5, -0.25) {};
		\node [style=none] (8) at (3.25, -0.25) {};
		\node [style=none] (9) at (5, 1) {};
		\node [style=none] (10) at (3.25, -3.25) {};
		\node [style=disintegration, minimum height=1.6cm, minimum width=1.1cm] (11) at (4.15, -1.15) {};
		\node [style=none] (14) at (4.5, -3.25) {};
		\node [style=none] (16) at (4, -0.25) {};
		\node [style=none] (17) at (21.5, -1.25) {$=$};
		\node [style=none] (18) at (22.75, -2.75) {};
		\node [style=none] (19) at (23.75, -2.75) {};
		\node [style=none] (20) at (22.75, 0.5) {};
		\node [style=upground] (21) at (23.75, -0.75) {};
		\node [style=none] (22) at (6.5, -1.25) {$=$};
		\node [style=small box] (23) at (10, -1.75) {};
		\node [style=small box] (24) at (10, -1.75) {$c$};
		\node [style=black dot] (25) at (8.5, -2.25) {};
		\node [style=none] (26) at (8.5, -3.25) {};
		\node [style=none] (27) at (9, -1.5) {};
		\node [style=none] (29) at (8, 0.75) {};
		\node [style=disintegration, minimum height=1.3cm, minimum width=1.5cm] (31) at (9.3, -1.3) {};
		\node [style=none] (32) at (10, -3.25) {};
		\node [style=none] (33) at (8, -1.5) {};
		\node [style=none] (34) at (9, -0.75) {};
		\node [style=none] (35) at (10, -0.75) {};
		\node [style=none] (36) at (12, -1.25) {$=$};
		\node [style=small box] (38) at (15, -1) {$c$};
		\node [style=black dot] (39) at (13.5, -2.5) {};
		\node [style=none] (40) at (13.5, -3.25) {};
		\node [style=none] (41) at (14, -1.75) {};
		\node [style=none] (42) at (13, 0.75) {};
		\node [style=disintegration, minimum height=1.1cm, minimum width=1.0cm] (43) at (14.8, -0.65) {};
		\node [style=none] (44) at (15, -3.25) {};
		\node [style=none] (45) at (13, -1.75) {};
		\node [style=none] (46) at (14, -0.25) {};
		\node [style=none] (47) at (15, -0.25) {};
		\node [style=none] (48) at (17, -1.25) {$=$};
		\node [style=black dot] (49) at (18.75, -2.5) {};
		\node [style=none] (50) at (18.75, -3.25) {};
		\node [style=none] (51) at (19.25, -1.75) {};
		\node [style=none] (52) at (18.25, -1.75) {};
		\node [style=upground] (53) at (19.25, -1.25) {};
		\node [style=none] (54) at (18.25, 0.75) {};
		\node [style=none] (55) at (20, -3.25) {};
		\node [style=upground] (56) at (20, -0.75) {};
	\end{pgfonlayer}
	\begin{pgfonlayer}{edgelayer}
		\draw (3.center) to (2);
		\draw [in=-90, out=30, looseness=1.25] (2) to (4.center);
		\draw (10.center) to (8.center);
		\draw (4.center) to (9.center);
		\draw (1) to (14.center);
		\draw [in=165, out=-90, looseness=1.25] (16.center) to (2);
		\draw [bend left=90, looseness=2.00] (8.center) to (16.center);
		\draw (20.center) to (18.center);
		\draw (21) to (19.center);
		\draw (26.center) to (25);
		\draw [in=-90, out=30, looseness=1.25] (25) to (27.center);
		\draw (24) to (32.center);
		\draw [in=165, out=-90, looseness=1.25] (33.center) to (25);
		\draw [bend left=90, looseness=2.00] (34.center) to (35.center);
		\draw (29.center) to (33.center);
		\draw (35.center) to (24);
		\draw (34.center) to (27.center);
		\draw (40.center) to (39);
		\draw [in=-90, out=30, looseness=1.25] (39) to (41.center);
		\draw (38) to (44.center);
		\draw [in=165, out=-90, looseness=1.25] (45.center) to (39);
		\draw [bend left=90, looseness=2.00] (46.center) to (47.center);
		\draw (42.center) to (45.center);
		\draw (47.center) to (38);
		\draw (46.center) to (41.center);
		\draw (50.center) to (49);
		\draw [in=-90, out=30, looseness=1.25] (49) to (51.center);
		\draw [in=165, out=-90, looseness=1.25] (52.center) to (49);
		\draw (54.center) to (52.center);
		\draw (53) to (51.center);
		\draw (56) to (55.center);
	\end{pgfonlayer}
\end{tikzpicture}}} \eqno{\QEDbox} \]
\end{myproof}


\section{Applications to Bayesian networks}\label{sec:bayesiannetwork}

This section elaborates two examples of Bayesian networks in order to
illustrate how the disintegration rules from the previous two sections
can be applied in a compositional manner. These rules turn a Bayesian
network with outputs only into a network that also has input wires, so
that conditioning can be computed via composition, acting on the
evidence as input. There is a fairly straightforward translation of
traditional Bayesian network notation into string diagram, see
\textit{e.g.}~\cite{Fong12,JacobsZ21}: the main difference is that
copying is written explicitly as $\minicopy$ in string diagrams.

\subsection{Conditioning for fault trees}\label{subsec:faulttree}

 \begin{wrapfigure}{r}{0.4\textwidth}
 	\vspace{-5em}
 	\begin{equation}
 		\label{FaulttreeJoint}
 		\tau \quad=\qquad \vcenter{\hbox{\begin{tikzpicture}
	\begin{pgfonlayer}{nodelayer}
		\node [style=none] (0) at (3, 3.75) {};
		\node [style=none] (1) at (4, 3.75) {};
		\node [style=none] (2) at (3, 4.5) {};
		\node [style=none] (3) at (4, 4.5) {};
		\node [style=none] (4) at (3.25, 3.75) {};
		\node [style=none] (5) at (3.75, 3.75) {};
		\node [style=none] (6) at (3.5, 5) {};
		\node [style=none] (7) at (3.5, 5.5) {};
		\node [style=none] (8) at (2, 1.5) {};
		\node [style=none] (9) at (3, 1.5) {};
		\node [style=none] (10) at (2, 2.25) {};
		\node [style=none] (11) at (3, 2.25) {};
		\node [style=none] (12) at (2.25, 1.5) {};
		\node [style=none] (13) at (2.75, 1.5) {};
		\node [style=none] (14) at (2.5, 2.75) {};
		\node [style=white dot, minimum size=1em] (15) at (4.5, 2.5) {$\scriptstyle\nicefrac{1}{5}$};
		\node [style=white dot, minimum size=1em] (17) at (0.5, -2.25) {$\scriptstyle\nicefrac{1}{2}$};
		\node [style=white dot, minimum size=1em] (19) at (2.5, -2.25) {$\scriptstyle\nicefrac{1}{3}$};
		\node [style=white dot, minimum size=1em] (20) at (4.5, -2.25) {$\scriptstyle\nicefrac{1}{4}$};
		\node [style=black dot] (22) at (2.5, -1.5) {};
		\node [style=none] (23) at (1, -1) {};
		\node [style=none] (24) at (2, -1) {};
		\node [style=none] (25) at (1, -0.25) {};
		\node [style=none] (26) at (2, -0.25) {};
		\node [style=none] (27) at (1.25, -0.87) {};
		\node [style=none] (28) at (1.75, -0.87) {};
		\node [style=none] (29) at (1.5, 0.25) {};
		\node [style=none] (30) at (3, -1) {};
		\node [style=none] (31) at (4, -1) {};
		\node [style=none] (32) at (3, -0.25) {};
		\node [style=none] (33) at (4, -0.25) {};
		\node [style=none] (34) at (3.25, -0.87) {};
		\node [style=none] (35) at (3.75, -0.87) {};
		\node [style=none] (36) at (3.5, 0.25) {};
		\node [style=black dot] (37) at (0.5, -1.5) {};
		\node [style=black dot] (38) at (4.5, -1.5) {};
		\node [style=none] (39) at (6.75, 1.5) {};
		\node [style=none] (40) at (6.75, 5.5) {};
		\node [style=none] (41) at (0, 2) {};
		\node [style=none] (42) at (6, 2.5) {};
		\node [style=none] (43) at (0, 5.5) {};
		\node [style=none] (44) at (-0.75, 5.5) {};
		\node [style=none] (45) at (-0.75, 1.25) {};
		\node [style=none] (51) at (2.5, 0.75) {};
		\node [style=black dot] (52) at (1.5, 0.75) {};
		\node [style=black dot] (53) at (3.5, 0.75) {};
		\node [style=black dot] (54) at (4.5, 3.25) {};
		\node [style=black dot] (55) at (2.5, 3.25) {};
		\node [style=none] (56) at (1.75, 3.75) {};
		\node [style=none] (57) at (1, 1.75) {};
		\node [style=none] (58) at (1.75, 5.5) {};
		\node [style=none] (59) at (5.25, 3.75) {};
		\node [style=none] (60) at (5.25, 5.5) {};
		\node [style=none] (61) at (6, 5.5) {};
		\node [style=none] (62) at (1, 5.5) {};
	\end{pgfonlayer}
	\begin{pgfonlayer}{edgelayer}
		\draw (0.center) to (1.center);
		\draw (1.center) to (3.center);
		\draw (2.center) to (0.center);
		\draw [bend left=90, looseness=1.75] (2.center) to (3.center);
		\draw (7.center) to (6.center);
		\draw (8.center) to (9.center);
		\draw (9.center) to (11.center);
		\draw (10.center) to (8.center);
		\draw [bend left=90, looseness=1.75] (10.center) to (11.center);
		\draw [bend left, looseness=1.25] (23.center) to (24.center);
		\draw (24.center) to (26.center);
		\draw (25.center) to (23.center);
		\draw [in=-165, out=90, looseness=0.75] (25.center) to (29.center);
		\draw [in=90, out=-15, looseness=0.75] (29.center) to (26.center);
		\draw [bend left, looseness=1.25] (30.center) to (31.center);
		\draw (31.center) to (33.center);
		\draw (32.center) to (30.center);
		\draw [in=-165, out=90, looseness=0.75] (32.center) to (36.center);
		\draw [in=90, out=-15, looseness=0.75] (36.center) to (33.center);
		\draw (22) to (19);
		\draw [in=-90, out=165, looseness=1.25] (22) to (28.center);
		\draw [in=-90, out=15, looseness=1.25] (22) to (34.center);
		\draw (44.center) to (45.center);
		\draw [in=165, out=-90, looseness=1.25] (45.center) to (37);
		\draw [in=135, out=-90, looseness=1.25] (35.center) to (38);
		\draw [in=-90, out=15] (38) to (39.center);
		\draw (39.center) to (40.center);
		\draw [in=-105, out=15, looseness=1.25] (37) to (27.center);
		\draw (37) to (17);
		\draw (38) to (20);
		\draw (52) to (29.center);
		\draw (53) to (36.center);
		\draw (55) to (14.center);
		\draw (54) to (15);
		\draw [in=-90, out=165, looseness=1.25] (52) to (41.center);
		\draw [in=-90, out=15, looseness=1.25] (52) to (12.center);
		\draw [in=-90, out=150, looseness=1.25] (53) to (13.center);
		\draw [in=-90, out=30, looseness=0.75] (53) to (42.center);
		\draw (43.center) to (41.center);
		\draw [in=165, out=-90, looseness=1.25] (5.center) to (54);
		\draw (51.center) to (22);
		\draw [in=150, out=-90, looseness=1.25] (56.center) to (55);
		\draw [in=-90, out=30, looseness=1.25] (55) to (4.center);
		\draw [in=90, out=-90] (57.center) to (51.center);
		\draw (42.center) to (61.center);
		\draw (59.center) to (60.center);
		\draw [in=-90, out=15, looseness=1.25] (54) to (59.center);
		\draw (56.center) to (58.center);
		\draw (62.center) to (57.center);
	\end{pgfonlayer}
\end{tikzpicture}}}
 	\end{equation}
 	\vspace{-1em}
 \end{wrapfigure}

Fault trees form a graphical formalism that is used in safety and
reliability engineering, see \textit{e.g.}~\cite{JimenezHS21} for 
an overview (or also~\cite{PiedeleuTSZ24} for a string diagrammatic
approach). They can be seen as Bayesian networks, with logical
gates as conditional probability tables. We adapt the leading 
illustration from~\cite{LopuhaaZwakenberg24} and present it
as a joint distribution $\tau$, visualised on the right.

The encircled numbers $r$ at the leafs correspond to coin
distributions $\flip(r) = r\ket{1} + (1-r)\ket{0}$ on the set
$\finset{2} = \{0,1\}$. The above or-gates (with a curved bottom) and
and-gates (with a straight bottom) are deterministic channels
$\orchan, \andchan \colon \finset{2}\times\finset{2} \rightarrow
\Dst(\finset{2})$ given by $\orchan(x,y) = 1\bigket{x\vee y}$ and
$\andchan(x,y) = 1\bigket{x\wedge y}$. The joint distribution $\tau$
described in~\eqref{FaulttreeJoint} is a distribution on
$\finset{2}^{8} = \finset{2} \times \cdots \times \finset{2}$.

\begin{figure*}[th]
 \begin{center}
 	$\vcenter{\hbox{\input{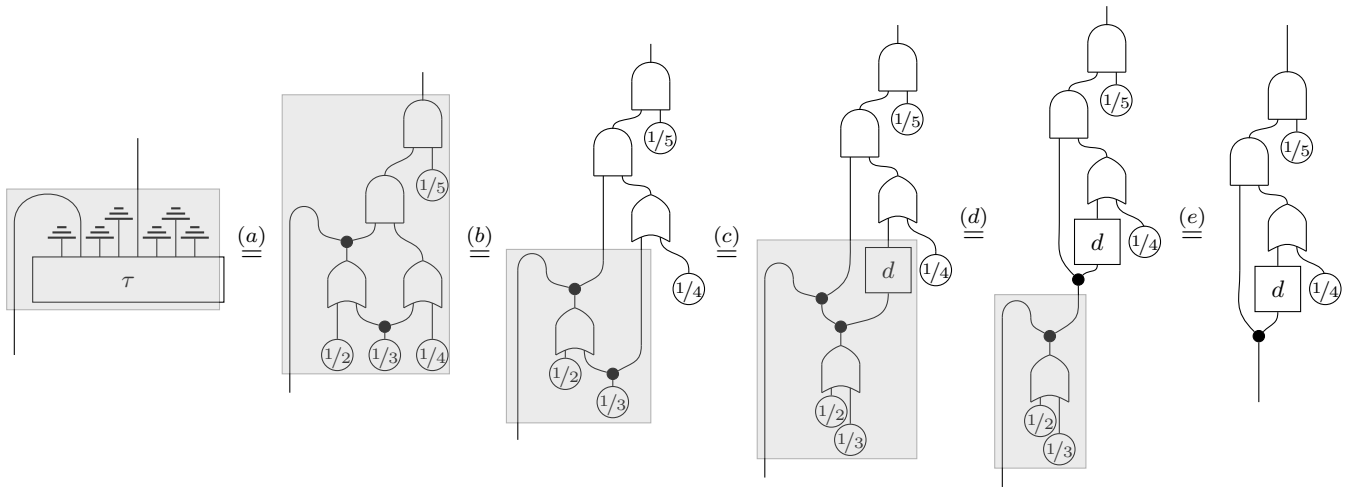}}}$
 \end{center}
 \caption{Disintegration of the Bayesian fault tree
 	network~\eqref{FaulttreeJoint}, via the introduction of a 
 	dagger channel $d$ and via eventual removal of the shaded box, 
 	see the beginning of Section~\ref{sec:bayesiannetwork} for 
 	details.}
 	\label{fig:faulttree}
 \end{figure*}

 Now suppose we wish to condition on the second wire from the left
 in~\eqref{FaulttreeJoint}, and wish to learn the effect on the fifth
 (top) wire. This means that we can discard all other
 wires. Figure~\ref{fig:faulttree} describes how disintegration for
 this Bayesian (fault tree) network works. We elaborate on what
 happens in a step-by-step manner. Equation~$(a)$ on the left involves
 unpacking the description in~\eqref{FaulttreeJoint} and removing the
 discarded wires. In step~$(b)$ channels are pulled out of the shaded
 box, using
 Definition~\ref{def:normalisation}~\eqref{def:normalisation:comp}.
 Subsequently, in step~$(c)$, the channel $d$ is introduced as dagger
 of $\orchan \after (\flip(\nicefrac{1}{2})\otimes\idmap) \colon
 \finset{2} \rightarrow \Dst(\finset{2})$, with
 $\flip(\nicefrac{1}{3})$ as prior, according to the left equation
 in~\eqref{eqn:dagger}. In step~$(d)$ the associativity of copying
 $\minicopy$ is used and the (dagger) channel $d$ is pulled out of the
 shaded box, again as in
 Definition~\ref{def:normalisation}~\eqref{def:normalisation:comp}.
 Finally, the shaded box is removed entirely in step~$(e)$, via
 Proposition~\ref{prop:boxremoval}.

\begin{wrapfigure}{r}{0.5\textwidth}\vspace*{-2em}
 	\[ \left\{\begin{array}{rcl}
 		d(1) 
 		& = &
 		\frac{1}{2}\ket{1} + \frac{1}{2}\ket{0}
 		\\
 		d(0)
 		& = &
 		1\ket{0}
 	\end{array}\right.
 	\hspace*{3em}
 	\left\{\begin{array}{rcl}
 		1 
 		& \mapsto &
 		\frac{1}{8}\ket{1} + \frac{7}{8}\ket{0}
 		\\
 		0
 		& \mapsto &
 		1\ket{0}.
 	\end{array}\right. \vspace*{-1em} \]
 \end{wrapfigure}
 At this stage we have turned the original
 network~\eqref{FaulttreeJoint} with outputs only into an `open'
 network with both and input and ouput, at the right of
 Figure~\ref{fig:faulttree}. It is not hard to actually compute what
 this network does, in a compositional manner. The dagger channel $d$
 is shown on the right, together with entire disintegrated network,
 both as channels $\finset{2} \rightarrow \Dst(\finset{2})$. This last
 outcome $0 \mapsto 1\ket{0}$ of the disintegrated network is not
 surprising: if one forces the point in the network where we condition
 to be~$0$, the two and-gates above it will produce~$0$ as output.

\subsection{Conditioning for the `Child' Bayesian 
network}\label{subsec:child}


We consider (part of) the `Child' Bayesian network, a famous example
from the literature~\cite{SpiegelhalterDLC93} that is part of a
standard repository of Bayesian networks\footnote{See
\href{https://www.bnlearn.com/bnrepository/discrete-medium.html\#child}{www.bnlearn.com/bnrepository/discrete-medium.html\#child}}.
It involves various medical possibilities and their statistical
relationships for the diagnosis of partical child diseases.

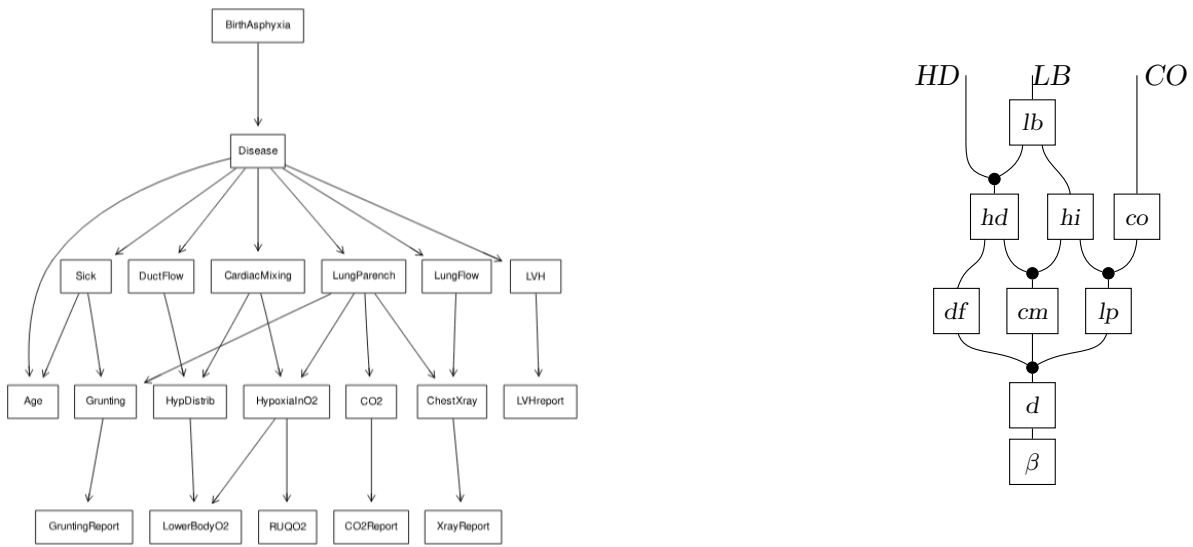
\begin{figure*}[th]
	\begin{center}
		$\vcenter{\hbox{\includegraphics[width=20em]{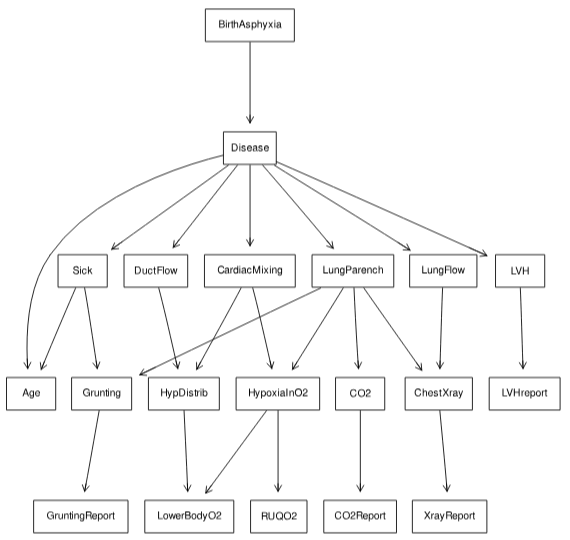}}}$
		\hspace*{10em}
		\begin{tabular}{c}
			$\vcenter{\hbox{\begin{tikzpicture}
	\begin{pgfonlayer}{nodelayer}
		\node [style=small box] (0) at (0, -3.5) {$\beta$};
		\node [style=small box] (1) at (0, -2) {$\mathsl{d}$};
		\node [style=black dot] (2) at (0, -1) {};
		\node [style=none] (3) at (-2, 0.25) {};
		\node [style=none] (4) at (0, 0.25) {};
		\node [style=none] (5) at (2, 0.25) {};
		\node [style=small box] (6) at (-2, 0.5) {$\mathsl{df}$};
		\node [style=small box] (7) at (2, 0.5) {$\mathsl{lp}$};
		\node [style=small box] (8) at (0, 0.5) {$\mathsl{cm}$};
		\node [style=black dot] (9) at (0, 1.5) {};
		\node [style=small box] (11) at (-1, 3) {$\mathsl{hd}$};
		\node [style=small box] (12) at (2.75, 3) {$\mathsl{co}$};
		\node [style=small box] (13) at (1, 3) {$\mathsl{hi}$};
		\node [style=none] (14) at (-2, 1) {};
		\node [style=none] (15) at (-0.75, 2.5) {};
		\node [style=none] (16) at (0.75, 2.5) {};
		\node [style=none] (19) at (-1.25, 2.5) {};
		\node [style=black dot] (20) at (2, 1.5) {};
		\node [style=none] (21) at (1.25, 2.5) {};
		\node [style=none] (22) at (2.75, 2.5) {};
		\node [style=black dot] (23) at (-1, 4) {};
		\node [style=small box] (24) at (0, 5.5) {$\mathsl{lb}$};
		\node [style=none] (25) at (-0.25, 5) {};
		\node [style=none] (26) at (0.25, 5) {};
		\node [style=none] (27) at (1, 3.5) {};
		\node [style=none] (28) at (-1.75, 5) {};
		\node [style=none] (29) at (0, 6.75) {};
		\node [style=none] (30) at (2.75, 6.75) {};
		\node [style=none] (31) at (-1.75, 6.75) {};
		\node [style=none] (32) at (-2.5, 6.75) {$\mathsl{HD}$};
		\node [style=none] (33) at (0.5, 6.75) {$\mathsl{LB}$};
		\node [style=none] (34) at (3.5, 6.75) {$\mathsl{CO}$};
	\end{pgfonlayer}
	\begin{pgfonlayer}{edgelayer}
		\draw (2) to (1);
		\draw (1) to (0);
		\draw (6) to (3.center);
		\draw (8) to (4.center);
		\draw (7) to (5.center);
		\draw (4.center) to (2);
		\draw [in=-90, out=30, looseness=1.25] (2) to (5.center);
		\draw [in=-90, out=150, looseness=1.25] (2) to (3.center);
		\draw (9) to (8);
		\draw (14.center) to (6);
		\draw [in=-90, out=165] (9) to (15.center);
		\draw [in=-90, out=15] (9) to (16.center);
		\draw [in=-90, out=75, looseness=1.50] (14.center) to (19.center);
		\draw [in=-90, out=165] (20) to (21.center);
		\draw [in=-90, out=15] (20) to (22.center);
		\draw (20) to (7);
		\draw (23) to (11);
		\draw (31.center) to (28.center);
		\draw (29.center) to (24);
		\draw (30.center) to (12);
		\draw (13) to (27.center);
		\draw [in=-90, out=90] (27.center) to (26.center);
		\draw [in=-90, out=30] (23) to (25.center);
		\draw [in=-90, out=165, looseness=1.25] (23) to (28.center);
	\end{pgfonlayer}
\end{tikzpicture}}}$
		\end{tabular}
	\end{center}
	\caption{The original child Bayesian network
		from~\cite{SpiegelhalterDLC93} on the left (from top to 
		bottom), and
		the relevant portion of it as string diagram on the right 
		(from
		bottom to top).}
	\label{fig:childjoint}
\end{figure*}

The entire network is given on the left in
Figure~\ref{fig:childjoint}.  We are interested in a particular
disintegration $\mathsl{HD} \times \mathsl{CO} \rightarrow
\Dst\big(\mathsl{LB}\big)$, see below, so we formalise only the
relevant part as a string diagram on the right.

The first type is for BirthAsphyxia $(\mathsl{BA})$, which can be
present or not. Further, there are sets Disease ($\mathsl{DI}$), with
six elements PFC, TGA, Fallot, PAIVS, TAPVD, Lung; DuctFlow
($\mathsl{DF}$) with three elements Lt\_to\_Rt, None, Rt\_to\_Lt;
CardiacMixing ($\mathsl{CM}$) containing None, Mild, Complete, Transp;
LungParench $(\mathsl{LP}$) with Normal, Congested, Abnormal; CO2
($\mathsl{CO}$) containing Normal, Low, High; HypDistrib
($\mathsl{HD}$) with Equal, Unequal; HypoxiaInO2 ($\mathsl{HI}$)
containing Mild, Moderate, Severe; and LowerBodyO2 ($\mathsl{LB}$)
containing elements `\textless5', `5-12', `12+'. We shall use 
abbreviations
for the elements of these sets of the form:
\[ \begin{array}{rclcrclcrcl}
	\mathsl{BA}
	& = &
	\big\{\ba, \no{\ba}\big\}
        & \qquad &
	\mathsl{CM}
	& = &
	\big\{\CMno, \mi, \co, \tr\big\}
        & \qquad &
	\mathsl{HD}
	& = &
	\big\{\eq, \no{\eq}\big\}
	\\
	\mathsl{DI}
	& = &
	\big\{\pfc, \tga, \flt, \pis, \tpd, \lng\big\}
        & &
	\mathsl{LP}
	& = &
	\big\{\nr, \cg, \ab\big\}
        & &
	\mathsl{HI}
	& = &
	\big\{\mi, \mo, \se\big\}
	\\
	\mathsl{DF}
	& = &
	\big\{\lt, \DFno, \rt\big\}
        & &
	\mathsl{CO}
	& = &
	\big\{\nr, \lo, \hi\big\}
        & &
	\mathsl{LB}
	& = &
	\big\{\ltfi, \fitw, \gttw\big\}.
\end{array} \]

\noindent The string diagram on the right in 
Figure~\ref{fig:childjoint}
involves a distribution $\beta = 0.1\ket{\ba} + 0.9\ket{\no{\ba}}$ and
channels:
\begin{equation}
	\label{diag:childchannels}
	\begin{array}{c}
		\mathsl{BA} \stackrel{\mathsl{d}}{\longrightarrow} \Dst\big(\mathsl{DI}\big)
		\qquad
		\mathsl{DI} \stackrel{\mathsl{df}}{\longrightarrow} \Dst\big(\mathsl{DF}\big)
		\qquad
		\mathsl{DI} \stackrel{\mathsl{cm}}{\longrightarrow} \Dst\big(\mathsl{CM}\big)
		\qquad
		\mathsl{DI} \stackrel{\mathsl{lp}}{\longrightarrow} \Dst\big(\mathsl{LP}\big)
		\qquad
		\mathsl{LP} \stackrel{\mathsl{co}}{\longrightarrow} \Dst\big(\mathsl{CO}\big)
		\\
		\mathsl{DF} \times \mathsl{CM} 
		\stackrel{\mathsl{hd}}{\longrightarrow} \Dst\big(\mathsl{HD}\big)
		\qquad
		\mathsl{CM} \times \mathsl{LP} 
		\stackrel{\mathsl{hi}}{\longrightarrow} \Dst\big(\mathsl{HI}\big)
		\qquad
		\mathsl{HD} \times \mathsl{HI} 
		\stackrel{\mathsl{lb}}{\longrightarrow} \Dst\big(\mathsl{LB}\big)
	\end{array}
\end{equation}

\begin{figure*}[th]
	\begin{center}
		$\vcenter{\hbox{\input{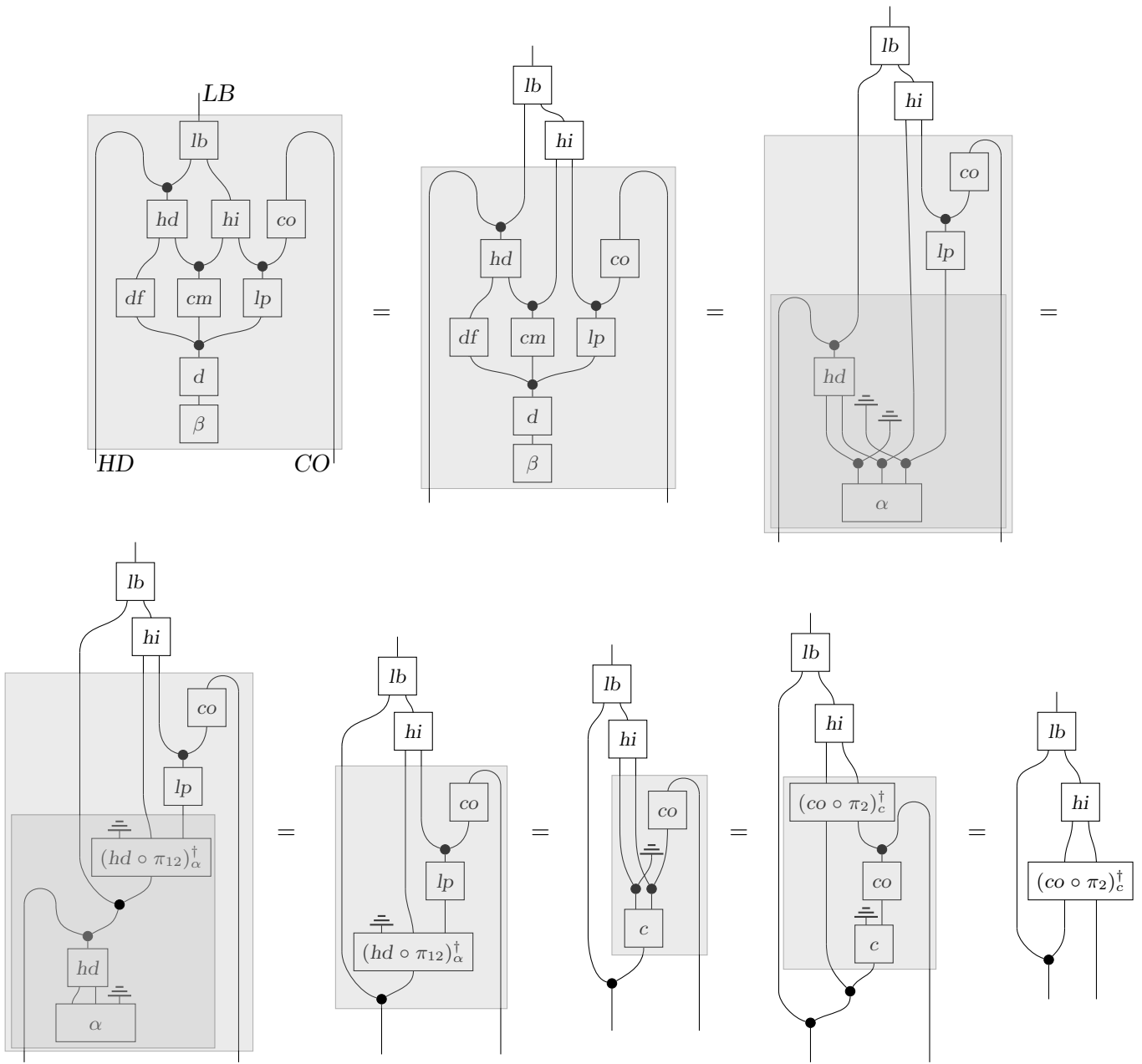}}}$
	\end{center}
	\caption{The disintegration steps for the child Bayesian
          network on the right in Figure~\ref{fig:childjoint}, using
          the abbreviations~\eqref{diag:childabbreviations}.}
	\label{fig:childdisintegration}
\end{figure*}

\noindent The definitions of these channels can be found in
Appendix~\ref{app:data}. Here we look only at how they are used in the
Bayesian network on the right of Figure~\ref{fig:childjoint}.  The
disintegration $\mathsl{HD} \times \mathsl{CO} \rightarrow
\Dst\big(\mathsl{LB}\big)$ that we are interested in is elaborated in
Figure~\ref{fig:childdisintegration}. It uses the following two
abbreviations.
\begin{equation}
	\label{diag:childabbreviations}
	\vcenter{\hbox{\begin{tikzpicture}
	\begin{pgfonlayer}{nodelayer}
		\node [style=small box] (0) at (-0.5, -3.5) {$\beta$};
		\node [style=small box] (1) at (-0.5, -2) {$\mathsl{d}$};
		\node [style=black dot] (2) at (-0.5, -1) {};
		\node [style=none] (3) at (-2, 0.25) {};
		\node [style=none] (4) at (-0.5, 0.25) {};
		\node [style=none] (5) at (1, 0.25) {};
		\node [style=small box] (6) at (-2, 0.5) {$\mathsl{df}$};
		\node [style=small box] (8) at (-0.5, 0.5) {$\mathsl{cm}$};
		\node [style=none] (14) at (-0.5, 1.75) {};
		\node [style=none] (16) at (-2, 1.75) {};
		\node [style=none] (19) at (1, 1.75) {};
		\node [style=none] (29) at (-2.625, 1.75) {$\mathsl{DF}$};
		\node [style=none] (30) at (0.2, 1.75) {$\mathsl{CM}$};
		\node [style=none] (31) at (1.625, 1.75) {$\mathsl{DI}$};
		\node [style=none] (32) at (-6.5, -1) {};
		\node [style=none] (33) at (-5.75, -1) {};
		\node [style=none] (34) at (-7.25, -1) {};
		\node [style=semilarge box] (35) at (-6.5, -1.5) {$\alpha$};
		\node [style=none] (36) at (-6.5, 0.25) {};
		\node [style=none] (37) at (-5.75, 0.25) {};
		\node [style=none] (38) at (-7.25, 0.25) {};
		\node [style=none] (39) at (-4, -1.25) {$\coloneqq$};
		\node [style=small box] (40) at (6.25, -1.5) {$c$};
		\node [style=none] (41) at (6.5, -1) {};
		\node [style=none] (42) at (6, -1) {};
		\node [style=none] (43) at (12, 0) {};
		\node [style=small box] (44) at (12, 0.25) {$\mathsl{lp}$};
		\node [style=none] (45) at (11, -1.5) {};
		\node [style=semilarge box] (47) at (11, -2) {$(\mathsl{hd} \after\pi_{12})^{\dag}_{\alpha}$};
		\node [style=none] (48) at (10, -1.5) {};
		\node [style=upground] (49) at (10, -1) {};
		\node [style=none] (51) at (12, -1.5) {};
		\node [style=none] (52) at (6.5, 0.5) {};
		\node [style=none] (53) at (6, 0.5) {};
		\node [style=none] (54) at (11, -3.5) {};
		\node [style=none] (55) at (6.25, -3.5) {};
		\node [style=none] (56) at (12, 1.5) {};
		\node [style=none] (57) at (11, 1.5) {};
		\node [style=none] (58) at (8, -1.25) {$\coloneqq$};
		\node [style=none] (59) at (10.25, 1.5) {$\mathsl{CM}$};
		\node [style=none] (60) at (12.625, 1.5) {$\mathsl{LP}$};
		\node [style=none] (61) at (10.25, -3.25) {$\mathsl{HD}$};
	\end{pgfonlayer}
	\begin{pgfonlayer}{edgelayer}
		\draw (2) to (1);
		\draw (1) to (0);
		\draw (6) to (3.center);
		\draw (8) to (4.center);
		\draw (4.center) to (2);
		\draw [in=-90, out=30, looseness=1.25] (2) to (5.center);
		\draw [in=-90, out=150, looseness=1.25] (2) to (3.center);
		\draw (16.center) to (6);
		\draw (14.center) to (8);
		\draw (19.center) to (5.center);
		\draw (38.center) to (34.center);
		\draw (36.center) to (32.center);
		\draw (37.center) to (33.center);
		\draw (44) to (43.center);
		\draw (49) to (48.center);
		\draw (43.center) to (51.center);
		\draw (53.center) to (42.center);
		\draw (52.center) to (41.center);
		\draw (40) to (55.center);
		\draw (56.center) to (44);
		\draw (47) to (54.center);
		\draw (57.center) to (45.center);
	\end{pgfonlayer}
\end{tikzpicture}}}
\end{equation}

\noindent We leave it to the interested reader to recognise the
various diagram transformations steps that are applied. We do point
out that this disintegration uses two daggers, both with respect to a
distribution~$\alpha$ and with respect to a channel~$c$, see
Definition~\ref{def:dagger} for the two versions.

Eventually, at the bottom-right of
Figure~\ref{fig:childdisintegration} we obtain a description of the
channel $\mathsl{HD} \times \mathsl{CO} \rightarrow
\Dst\big(\mathsl{LB}\big)$. Using the detailed channel descriptions
from the appendix it can be computed concretely as in 
(\ref{eqn:childdisintegration}).

\begin{wrapfigure}{r}{0.6\textwidth}
	\label{fig:child_results}
	\begin{equation}
		\label{eqn:childdisintegration}
		\begin{array}{rcl}
			(\eq, \nr)
			& \longmapsto &
			0.356\bigket{\ltfi} + 0.497\bigket{\fitw} + 
			0.147\bigket{\gttw}
			\\
			(\eq, \lo)
			& \longmapsto &
			0.356\bigket{\ltfi} + 0.496\bigket{\fitw} + 
			0.147\bigket{\gttw}
			\\
			(\eq, \hi)
			& \longmapsto &
			0.359\bigket{\ltfi} + 0.489\bigket{\fitw} + 
			0.152\bigket{\gttw}
			\\
			(\no{\eq}, \nr)
			& \longmapsto &
			0.512\bigket{\ltfi} + 0.427\bigket{\fitw} + 
			0.061\bigket{\gttw}
			\\
			(\no{\eq}, \lo)
			& \longmapsto &
			0.512\bigket{\ltfi} + 0.427\bigket{\fitw} + 
			0.0611\bigket{\gttw}
			\\
			(\no{\eq}, \hi)
			& \longmapsto &
			0.505\bigket{\ltfi} + 0.432\bigket{\fitw} + 
			0.0631\bigket{\gttw}.
		\end{array}
	\end{equation}
\end{wrapfigure}

\noindent The precise medical relevance of these distributions is not
our focus. What we do like to point out is that our approach yields
the entire disintegration channel, for all inputs. In contrast, in
conventional inference in Bayesian networks~\cite{Pearl88} one obtains
outcomes for each of the above lines separately\footnote{We indeed
	checked these channel outcomes~\eqref{eqn:childdisintegration} 
	with
	six separate inference queries using the \emph{pgmpy} Python 
	library
	from \href{https://pgmpy.org}{pgmpy.org}.}. For instance, the six
lines~\eqref{eqn:childdisintegration} of the extracted channel quickly
show that the outcome is determined by the first input element ($\eq$
or $\no{\eq}$) and not by the second input ($\nr$, $\hi$, $\lo$).

Moreover, the channel that we obtain in this way can be pre-computed,
so that actual inferences of this kind can be done quickly. Further,
if we have an evidence distribution on $\mathsl{HD} \times
\mathsl{CO}$ we can apply the above
channel~\eqref{eqn:childdisintegration} to it via pushforward (Kleisli
extension). In this way we perform updating in the style of Jeffrey,
see~\cite{Jacobs19c,Jacobs21c} for more information.

\section{Applications to conditional 
independence}\label{sec:independence}

Disintegration is a technique that shows up in conditional 
independence
arguments. This section will give an illustration, based on an example
from the literature. In general, we say that for a map $h \colon Z
\rightarrow X\otimes Y$ conditional independence of $X,Y$ given $Z$
holds, often written as $X \coprod Y \mid Z$, when:

\begin{equation}
	\label{diag:independence}
	\vcenter{\hbox{\begin{tikzpicture}
	\begin{pgfonlayer}{nodelayer}
		\node [style=medium box] (0) at (-3.75, 0) {$h$};
		\node [style=small box] (1) at (-0.25, 0.25) {$f$};
		\node [style=small box] (2) at (1.75, 0.25) {$g$};
		\node [style=black dot] (3) at (0.75, -0.75) {};
		\node [style=none] (4) at (-3.75, -1.5) {};
		\node [style=none] (5) at (-4.25, 0.25) {};
		\node [style=none] (6) at (-3.25, 0.25) {};
		\node [style=none] (7) at (-3.25, 1.5) {};
		\node [style=none] (8) at (-4.25, 1.5) {};
		\node [style=none] (9) at (-0.25, 1.5) {};
		\node [style=none] (10) at (1.75, 1.5) {};
		\node [style=none] (11) at (0.75, -1.5) {};
		\node [style=none] (12) at (-2.75, 1.5) {$Y$};
		\node [style=none] (13) at (-4.75, 1.5) {$X$};
		\node [style=none] (14) at (-4.25, -1.5) {$Z$};
		\node [style=none] (15) at (-0.75, 1.5) {$X$};
		\node [style=none] (16) at (2.25, 1.5) {$Y$};
		\node [style=none] (17) at (0.25, -1.5) {$Z$};
		\node [style=none] (18) at (-2, 0) {$=$};
	\end{pgfonlayer}
	\begin{pgfonlayer}{edgelayer}
		\draw (8.center) to (5.center);
		\draw (6.center) to (7.center);
		\draw (0) to (4.center);
		\draw (3) to (11.center);
		\draw [in=165, out=-90, looseness=1.25] (1) to (3);
		\draw [in=-90, out=15, looseness=1.25] (3) to (2);
		\draw (2) to (10.center);
		\draw (1) to (9.center);
	\end{pgfonlayer}
\end{tikzpicture}}} 
	\qquad\mbox{for some $f,g$.}
\end{equation}

In~\cite[({\sc CommonCause})]{LiAH23}, based
on~\cite[Fig.~6(a)]{BaoDHS21}, a particular joint distribution
$\sigma$ on $\finset{2}^5$ is defined in a basic probabilistic
programming language. We reproduce the code below on the left.  It 
uses $\leftarrow$ for sampling and $\|$ for disjunction. 
The distribution defined by the program forms a string diagram of the 
form on the right below.

\hspace*{-4em}\begin{minipage}{0.3\textwidth}
	\[ \vcenter{\hbox{\begin{minipage}[t]{0.22\textwidth}
				\begin{tabular}{l}
					\texttt{Z <- flip(1/2)} \\[-.2em]
					\texttt{X <- flip(1/2)} \\[-.2em]
					\texttt{Y <- flip(1/2)} \\[-.2em]
					\texttt{A = X || Z} \\[-.2em]
					\texttt{B = Z || Y} \\ [-.2em]
					\texttt{return (Z, X, Y, A, B)}
				\end{tabular}
	\end{minipage}}}\quad\quad\quad \]	
\end{minipage}
\hspace*{6em}
\begin{minipage}{0.6\textwidth}
	\[ \sigma \;\;=\;\;\vcenter{\hbox{\begin{tikzpicture}
	\begin{pgfonlayer}{nodelayer}
		\node [style=small box] (0) at (-5.25, -2.25) {$\gamma$};
		\node [style=small box] (1) at (-3.25, -2.25) {$\gamma$};
		\node [style=small box] (2) at (-1.25, -2.25) {$\gamma$};
		\node [style=black dot] (3) at (-5.25, -0.75) {};
		\node [style=black dot] (4) at (-3.25, -1.25) {};
		\node [style=black dot] (5) at (-1.25, -1.25) {};
		\node [style=black dot] (6) at (-1.25, 0) {};
		\node [style=none] (7) at (-0.75, 0.5) {};
		\node [style=none] (8) at (0.25, 0.5) {};
		\node [style=none] (9) at (-0.75, 1.25) {};
		\node [style=none] (10) at (0.25, 1.25) {};
		\node [style=none] (11) at (-0.5, 0.63) {};
		\node [style=none] (12) at (0, 0.63) {};
		\node [style=none] (13) at (-0.25, 1.75) {};
		\node [style=none] (14) at (-2.75, 0.5) {};
		\node [style=none] (15) at (-1.75, 0.5) {};
		\node [style=none] (16) at (-2.75, 1.25) {};
		\node [style=none] (17) at (-1.75, 1.25) {};
		\node [style=none] (18) at (-2, 0.63) {};
		\node [style=none] (19) at (-2.5, 0.63) {};
		\node [style=none] (20) at (-2.25, 1.75) {};
		\node [style=none] (22) at (-6.75, 2.5) {};
		\node [style=none] (23) at (-5.5, 2.5) {};
		\node [style=none] (24) at (-4.25, 2.5) {};
		\node [style=none] (25) at (-0.25, 2.5) {};
		\node [style=none] (26) at (-2.25, 2.5) {};
		\node [style=none] (27) at (-7.25, 2.5) {$Z$};
		\node [style=none] (28) at (-6, 2.5) {$X$};
		\node [style=none] (29) at (-4.75, 2.5) {$Y$};
		\node [style=none] (30) at (-1.75, 2.5) {$A$};
		\node [style=none] (31) at (0.25, 2.5) {$B$};
	\end{pgfonlayer}
	\begin{pgfonlayer}{edgelayer}
		\draw [bend left, looseness=1.25] (7.center) to (8.center);
		\draw (8.center) to (10.center);
		\draw (9.center) to (7.center);
		\draw [in=-165, out=90, looseness=0.75] (9.center) to (13.center);
		\draw [in=90, out=-15, looseness=0.75] (13.center) to (10.center);
		\draw [bend left, looseness=1.25] (14.center) to (15.center);
		\draw (15.center) to (17.center);
		\draw (16.center) to (14.center);
		\draw [in=-165, out=90, looseness=0.75] (16.center) to (20.center);
		\draw [in=90, out=-15, looseness=0.75] (20.center) to (17.center);
		\draw (3) to (0);
		\draw (4) to (1);
		\draw (5) to (2);
		\draw [in=-90, out=15, looseness=0.75] (3) to (6);
		\draw [in=-90, out=180] (4) to (23.center);
		\draw [in=-90, out=180, looseness=1.25] (5) to (24.center);
		\draw [in=165, out=-90, looseness=0.75] (22.center) to (3);
		\draw [in=-90, out=30, looseness=0.75] (5) to (12.center);
		\draw [in=-105, out=15] (6) to (11.center);
		\draw (13.center) to (25.center);
		\draw (20.center) to (26.center);
		\draw [in=-90, out=15, looseness=0.75] (4) to (19.center);
		\draw [in=-90, out=165, looseness=1.25] (6) to (18.center);
	\end{pgfonlayer}
\end{tikzpicture}}}
	\qquad\mbox{where $\gamma=\flip(\frac{1}{2})$.} \]
\end{minipage}

\medskip

\noindent It may be seen as a Bayesian network. The claim made
in~\cite{BaoDHS21,LiAH23} is that $A \coprod B \mid Z$ via
disintegration. A graphical proof looks as follows. 
\[ \vcenter{\hbox{\begin{tikzpicture}
	\begin{pgfonlayer}{nodelayer}
		\node [style=small box] (0) at (-3.25, -1.75) {$\gamma$};
		\node [style=small box] (1) at (-1.75, -1.75) {$\gamma$};
		\node [style=small box] (2) at (0.25, -1.75) {$\gamma$};
		\node [style=black dot] (3) at (-3.25, -0.75) {};
		\node [style=black dot] (6) at (-0.75, -0.25) {};
		\node [style=none] (7) at (-0.5, 0.25) {};
		\node [style=none] (8) at (0.5, 0.25) {};
		\node [style=none] (9) at (-0.5, 1) {};
		\node [style=none] (10) at (0.5, 1) {};
		\node [style=none] (11) at (-0.25, 0.38) {};
		\node [style=none] (12) at (0.25, 0.38) {};
		\node [style=none] (13) at (0, 1.5) {};
		\node [style=none] (14) at (-2, 0.25) {};
		\node [style=none] (15) at (-1, 0.25) {};
		\node [style=none] (16) at (-2, 1) {};
		\node [style=none] (17) at (-1, 1) {};
		\node [style=none] (18) at (-1.25, 0.38) {};
		\node [style=none] (19) at (-1.75, 0.38) {};
		\node [style=none] (20) at (-1.5, 1.5) {};
		\node [style=none] (21) at (-3.75, -0.25) {};
		\node [style=none] (24) at (0, 2.25) {};
		\node [style=none] (25) at (-1.5, 2.25) {};
		\node [style=medium box] (31) at (-9, -1) {\hspace*{0.4cm}$\sigma$\hspace*{0.4cm}};
		\node [style=none] (32) at (-9, -0.5) {};
		\node [style=none] (64) at (-8.5, -0.5) {};
		\node [style=none] (65) at (-8, -0.5) {};
		\node [style=none] (66) at (-9.5, -0.5) {};
		\node [style=none] (67) at (-10, -0.5) {};
		\node [style=none] (68) at (-10, 0.25) {};
		\node [style=none] (69) at (-8.5, 2) {};
		\node [style=none] (70) at (-8, 2) {};
		\node [style=none] (71) at (-10.75, 0.25) {};
		\node [style=none] (72) at (-10.75, -2.5) {};
		\node [style=upground] (73) at (-9.5, 0.25) {};
		\node [style=upground] (74) at (-9, 0.75) {};
		\node [style=disintegration, minimum height=1.4cm, minimum width=1.65cm] (75) at (-9.3, -0.35) {};
		\node [style=none] (76) at (-6.25, -0.25) {$=$};
		\node [style=none] (77) at (-4.5, -0.25) {};
		\node [style=none] (78) at (-4.5, -3) {};
		\node [style=disintegration, minimum height=2.1cm, minimum width=2.85cm] (79) at (-1.85, -0.4) {};
		\node [style=none] (80) at (2.5, -0.25) {$=$};
		\node [style=small box] (81) at (5.75, -3.5) {$\gamma$};
		\node [style=small box] (82) at (4.75, -0.5) {$\gamma$};
		\node [style=small box] (83) at (7.75, -0.5) {$\gamma$};
		\node [style=black dot] (84) at (5.75, -2.5) {};
		\node [style=black dot] (85) at (6.25, -0.25) {};
		\node [style=none] (86) at (6.5, 0.5) {};
		\node [style=none] (87) at (7.5, 0.5) {};
		\node [style=none] (88) at (6.5, 1.25) {};
		\node [style=none] (89) at (7.5, 1.25) {};
		\node [style=none] (90) at (6.75, 0.63) {};
		\node [style=none] (91) at (7.25, 0.63) {};
		\node [style=none] (92) at (7, 1.75) {};
		\node [style=none] (93) at (5, 0.5) {};
		\node [style=none] (94) at (6, 0.5) {};
		\node [style=none] (95) at (5, 1.25) {};
		\node [style=none] (96) at (6, 1.25) {};
		\node [style=none] (97) at (5.75, 0.63) {};
		\node [style=none] (98) at (5.25, 0.63) {};
		\node [style=none] (99) at (5.5, 1.75) {};
		\node [style=none] (100) at (5.25, -2) {};
		\node [style=none] (101) at (7, 2.5) {};
		\node [style=none] (102) at (5.5, 2.5) {};
		\node [style=none] (103) at (4.5, -2) {};
		\node [style=none] (104) at (4.5, -4.75) {};
		\node [style=disintegration, minimum height=1.4cm, minimum width=1.1cm] (105) at (5.45, -2.85) {};
		\node [style=small box] (107) at (11.75, -1) {$\gamma$};
		\node [style=small box] (108) at (14.75, -1) {$\gamma$};
		\node [style=black dot] (110) at (13.25, -3) {};
		\node [style=none] (111) at (13.5, 0) {};
		\node [style=none] (112) at (14.5, 0) {};
		\node [style=none] (113) at (13.5, 0.75) {};
		\node [style=none] (114) at (14.5, 0.75) {};
		\node [style=none] (115) at (13.75, 0.13) {};
		\node [style=none] (116) at (14.25, 0.13) {};
		\node [style=none] (117) at (14, 1.25) {};
		\node [style=none] (118) at (12, 0) {};
		\node [style=none] (119) at (13, 0) {};
		\node [style=none] (120) at (12, 0.75) {};
		\node [style=none] (121) at (13, 0.75) {};
		\node [style=none] (122) at (12.75, 0.13) {};
		\node [style=none] (123) at (12.25, 0.13) {};
		\node [style=none] (124) at (12.5, 1.25) {};
		\node [style=none] (126) at (14, 2) {};
		\node [style=none] (127) at (12.5, 2) {};
		\node [style=none] (128) at (9.75, -0.25) {$=$};
		\node [style=none] (129) at (13.25, -4.25) {};
	\end{pgfonlayer}
	\begin{pgfonlayer}{edgelayer}
		\draw [bend left, looseness=1.25] (7.center) to (8.center);
		\draw (8.center) to (10.center);
		\draw (9.center) to (7.center);
		\draw [in=-165, out=90, looseness=0.75] (9.center) to (13.center);
		\draw [in=90, out=-15, looseness=0.75] (13.center) to (10.center);
		\draw [bend left, looseness=1.25] (14.center) to (15.center);
		\draw (15.center) to (17.center);
		\draw (16.center) to (14.center);
		\draw [in=-165, out=90, looseness=0.75] (16.center) to (20.center);
		\draw [in=90, out=-15, looseness=0.75] (20.center) to (17.center);
		\draw (3) to (0);
		\draw [in=-90, out=15, looseness=0.75] (3) to (6);
		\draw [in=165, out=-90, looseness=0.75] (21.center) to (3);
		\draw [in=-90, out=15, looseness=1.25] (6) to (11.center);
		\draw (13.center) to (24.center);
		\draw (20.center) to (25.center);
		\draw [in=-90, out=165, looseness=1.25] (6) to (18.center);
		\draw (67.center) to (68.center);
		\draw [bend left=270, looseness=2.25] (68.center) to (71.center);
		\draw (71.center) to (72.center);
		\draw (66.center) to (73);
		\draw (32.center) to (74);
		\draw (64.center) to (69.center);
		\draw (65.center) to (70.center);
		\draw (19.center) to (1);
		\draw (77.center) to (78.center);
		\draw [bend left=90, looseness=2.00] (77.center) to (21.center);
		\draw (12.center) to (2);
		\draw [bend left, looseness=1.25] (86.center) to (87.center);
		\draw (87.center) to (89.center);
		\draw (88.center) to (86.center);
		\draw [in=-165, out=90, looseness=0.75] (88.center) to (92.center);
		\draw [in=90, out=-15, looseness=0.75] (92.center) to (89.center);
		\draw [bend left, looseness=1.25] (93.center) to (94.center);
		\draw (94.center) to (96.center);
		\draw (95.center) to (93.center);
		\draw [in=-165, out=90, looseness=0.75] (95.center) to (99.center);
		\draw [in=90, out=-15, looseness=0.75] (99.center) to (96.center);
		\draw (84) to (81);
		\draw [in=-90, out=15, looseness=0.75] (84) to (85);
		\draw [in=165, out=-90, looseness=0.75] (100.center) to (84);
		\draw [in=-90, out=30, looseness=1.25] (85) to (90.center);
		\draw (92.center) to (101.center);
		\draw (99.center) to (102.center);
		\draw [in=-90, out=165, looseness=1.25] (85) to (97.center);
		\draw [in=90, out=-90, looseness=1.25] (98.center) to (82);
		\draw (103.center) to (104.center);
		\draw [bend left=90, looseness=2.00] (103.center) to (100.center);
		\draw [in=90, out=-90, looseness=1.25] (91.center) to (83);
		\draw [bend left, looseness=1.25] (111.center) to (112.center);
		\draw (112.center) to (114.center);
		\draw (113.center) to (111.center);
		\draw [in=-165, out=90, looseness=0.75] (113.center) to (117.center);
		\draw [in=90, out=-15, looseness=0.75] (117.center) to (114.center);
		\draw [bend left, looseness=1.25] (118.center) to (119.center);
		\draw (119.center) to (121.center);
		\draw (120.center) to (118.center);
		\draw [in=-165, out=90, looseness=0.75] (120.center) to (124.center);
		\draw [in=90, out=-15, looseness=0.75] (124.center) to (121.center);
		\draw [in=-90, out=60, looseness=0.75] (110) to (115.center);
		\draw (117.center) to (126.center);
		\draw (124.center) to (127.center);
		\draw [in=-90, out=120, looseness=0.75] (110) to (122.center);
		\draw [in=90, out=-90, looseness=1.25] (123.center) to (107);
		\draw [in=90, out=-90, looseness=1.25] (116.center) to (108);
		\draw (110) to (129.center);
	\end{pgfonlayer}
\end{tikzpicture}}} \]

\noindent It uses
Definition~\ref{def:normalisation}~\eqref{def:normalisation:comp} and
Proposition~\ref{prop:boxremoval}. The final diagram above is clearly 
an instance of the diagram on the right in~\eqref{diag:independence}. 
The proof in~\cite{LiAH23} is much more concrete, and distinguishes 
whether the first coin flip (with type $Z$) produces~$1$ or~$0$.

\section{Applications to Probabilistic 
Programming}\label{sec:programming}

We have informally used a probabilistic program in the last section to
describe a joint distribution over five variables, and have written
$\leftarrow$ for sampling. A fully-fledged Bayesian probabilistic
programming language
(\textit{e.g.}~\cite{GoodmanS14,CusumanoSLM19,BinghamCJOPKSSHG19}) 
adds to
this the capacity to condition on data (\lstinline|observe|), and
perform inference (\lstinline|normalize|). We refer
to~\cite{MeentPYW21} for an introduction to the subject.

\begin{wrapfigure}{l}{0.5\textwidth}
	\vspace{-1em}
	\[ \vcenter{\hbox{\begin{minipage}[t]{0.22\textwidth}
				\scalebox{0.83}{
					\begin{tabular}{l}
						\texttt{alice n = normalize $\$$ do} 
						\\[-.2em] 
						\texttt{  loc <- location} \\[-.2em]
						\texttt{  prediction <- bob(n-1)} \\[-.2em]
						\texttt{  condition} \\[-.2em]
						\texttt{      (loc == prediction)} \\[-.2em]
						\texttt{  return loc} 
				\end{tabular}}
	\end{minipage}}}
	\hspace*{2em}
	\vcenter{\hbox{\begin{minipage}[t]{0.22\textwidth}
				\scalebox{0.83}{
					\begin{tabular}{l}
						\texttt{bob 0 = location}\\[-.2em] 
						\texttt{bob n = normalize $\$$ do} \\[-.2em] 
						\texttt{  loc <- location}\\[-.2em] 
						\texttt{  prediction <- alice n}\\[-.2em] 
						\texttt{  condition}\\[-.2em] 
						\texttt{    (loc == prediction)}\\[-.2em] 
						\texttt{  return loc}
				\end{tabular}}
	\end{minipage}}} \]
	\caption{Pseudocode for the coordination game}
	\label{fig:alice_bob_pseudocode}
\end{wrapfigure}

From a semantic point of view, probabilistic programs and string
diagrams are tightly related and often inter-convertible. Observe can
be interpreted as the comparator, and normalize as the shaded
box. Conversely, a programming language can be seen as an internal
language or type-theoretic calculus for statistical
problems~\cite{LavoreJR24,Stein22}. Equations such as in
Proposition~\ref{prop:boxremoval} can then be understood as admissible
program transformations, which preserve the meaning of the program and
may cause faster execution.

A striking feature of probabilistic programming is that
\lstinline|normalize| commands can be nested. Such \emph{nested
	queries} can model agents making inferences about each other's
cognitive processes. This kind of reasoning about reasoning is central
to communication theory and to the theory of
mind~\cite{StuhlmullerG14}. Equations for boxes thereby become tools
for such ``reasoning about reasoning'' in the sense
of~\cite{ZhangA22}.

\begin{wrapfigure}{r}{0.2\textwidth}
  \label{eq:square} 
  \vspace{-2.8em}
  \[ \vcenter{\hbox{\begin{tikzpicture}
	\begin{pgfonlayer}{nodelayer}
		\node [style=small box] (0) at (0, 0.25) {$\omega$};
		\node [style=small box] (1) at (1.5, 0.25) {$\omega$};
		\node [style=black dot] (2) at (0, 1.25) {};
		\node [style=none] (3) at (1.5, 2) {};
		\node [style=none] (4) at (-0.75, 2) {};
		\node [style=none] (6) at (0.75, 2) {};
		\node [style=none] (7) at (-2.25, 0.75) {$\coloneqq$};
		\node [style=none] (8) at (-3.5, 0.75) {$\omega^{2}$};
		\node [style=none] (9) at (-0.75, 2.75) {};
		\node [style=disintegration, minimum height=1.5cm, minimum width=1.6cm] (10) at (0.7, 0.95) {};
	\end{pgfonlayer}
	\begin{pgfonlayer}{edgelayer}
		\draw (3.center) to (1);
		\draw (2) to (0);
		\draw [in=150, out=-90] (4.center) to (2);
		\draw [in=-90, out=30] (2) to (6.center);
		\draw [bend left=90, looseness=1.50] (6.center) to (3.center);
		\draw (9.center) to (4.center);
	\end{pgfonlayer}
\end{tikzpicture}}} \]
\end{wrapfigure}
First, for an arbitrary distribution $\omega\in\Dst(X)$ we can
similarly define $\omega^{2} \in \Dst(X)$ as on the right.  This is
the normalisation of the subdistribution $\sum_{x}
\omega(x)^{2}\ket{x}$. An iterated version $\omega^{n}$ will be used 
below. When $n$ grows, this $\omega^{n}$ quickly becomes $1\ket{x}$, 
where $x\in\supp(\omega)$ has the highest probability (if there is 
one).

We reproduce a simple example from~\cite[Section~3.1]{StuhlmullerG14},
of a \emph{coordination game}, originally due
to~\cite{Schelling80}. Two agents (Alice and Bob) want to meet up at
one of two locations but did not agree which one and can not 
communicate
any more. So they are forced to guess where the other person would go
--- which must recursively take their own behavior into account.

There is a prior distribution `$\location$' with popularity weights
for places to meet. Each agent then simulates the thought process of
the other, up to a specified recursion depth~$n$. On Figure 
\ref{fig:alice_bob_pseudocode}, we translate the
problem into Haskell code with a probabilistic programming framework
(such as \emph{MonadBayes}~\cite{ScibiorGG15}).

				

\begin{wrapfigure}{r}{0.5\textwidth}\vspace*{-1em}
	\[ \vcenter{\hbox{\scalebox{0.9}{{\begin{tikzpicture}
	\begin{pgfonlayer}{nodelayer}
		\node [style=small box] (34) at (-1.35, -0.5) {$\alice{n}$};
		\node [style=none] (35) at (0.25, -0.25) {$\coloneqq$};
		\node [style=small box] (36) at (2.5, -1) {$\location$};
		\node [style=small box] (37) at (5.5, -1) {$\bob{n-1}$};
		\node [style=none] (38) at (5.5, 1) {};
		\node [style=black dot] (39) at (2.5, 0) {};
		\node [style=none] (40) at (3.25, 1) {};
		\node [style=none] (41) at (1.75, 0.75) {};
		\node [style=none] (42) at (1.75, 2.5) {};
		\node [style=none] (43) at (3.25, 1) {};
		\node [style=disintegration, minimum height=2.1cm, minimum width=3cm] (44) at (3.95, 0.15) {};
		\node [style=none] (45) at (-1.35, 1.5) {};
		\node [style=small box] (46) at (9, -0.5) {$\bob{n}$};
		\node [style=none] (47) at (10.5, -0.25) {$\coloneqq$};
		\node [style=small box] (48) at (13, -1) {$\location$};
		\node [style=small box] (49) at (15.75, -1) {$\alice{n}$};
		\node [style=none] (50) at (15.75, 1) {};
		\node [style=black dot] (51) at (13, 0) {};
		\node [style=none] (52) at (13.75, 1) {};
		\node [style=none] (53) at (12.25, 0.75) {};
		\node [style=none] (54) at (12.25, 2.5) {};
		\node [style=none] (55) at (13.75, 1) {};
		\node [style=disintegration, minimum height=2cm, minimum width=2.8cm] (56) at (14.15, 0.15) {};
		\node [style=none] (57) at (9, 1.5) {};
	\end{pgfonlayer}
	\begin{pgfonlayer}{edgelayer}
		\draw (39) to (36);
		\draw [in=-90, out=150, looseness=1.25] (39) to (41.center);
		\draw (41.center) to (42.center);
		\draw [in=-90, out=30, looseness=1.25] (39) to (40.center);
		\draw (40.center) to (43.center);
		\draw [bend left=90, looseness=1.75] (43.center) to (38.center);
		\draw (38.center) to (37);
		\draw (45.center) to (34);
		\draw (51) to (48);
		\draw [in=-90, out=150, looseness=1.25] (51) to (53.center);
		\draw (53.center) to (54.center);
		\draw [in=-90, out=30, looseness=1.25] (51) to (52.center);
		\draw (52.center) to (55.center);
		\draw [bend left=90, looseness=1.75] (55.center) to (50.center);
		\draw (50.center) to (49);
		\draw [in=90, out=-90] (57.center) to (46);
	\end{pgfonlayer}
\end{tikzpicture}}}}} 
        \vspace*{-1em} \]
\end{wrapfigure}
The mutually recursive functions \lstinline|alice| and \lstinline|bob|
perform nested normalisation and inference. Mathematically, we capture
the meaning of the program in the string diagrams on the right, where
$\bob{0} = \location$. This can be evaluated in terms of the powers
from Diagram~\eqref{eq:square}. Via Lemma~\ref{lem:nestedbox}, we can
eliminate (inner) nested boxes, see below, so that $\bob{n}$ is equal
to the normalisation of $\location^{2n+1}$, where $\location^{2n+1}$
is the $(2n+1)$-fold power (comparison) of the $\location$
distribution with itself.
\[ \hspace*{-1em}\vcenter{\hbox{\scalebox{0.85}{\begin{tikzpicture}
	\begin{pgfonlayer}{nodelayer}
		\node [style=small box] (36) at (0.75, -1.25) {$\location$};
		\node [style=small box] (37) at (3.5, -1.25) {$\bob{n}$};
		\node [style=none] (38) at (3.5, 0.75) {};
		\node [style=black dot] (39) at (0.75, -0.25) {};
		\node [style=none] (40) at (1.5, 0.75) {};
		\node [style=none] (41) at (0, 0.5) {};
		\node [style=none] (42) at (0, 2.5) {};
		\node [style=none] (43) at (1.5, 0.75) {};
		\node [style=disintegration, minimum height=1.8cm, minimum width=2.7cm] (44) at (1.95, -0.2) {};
		\node [style=none] (47) at (-4.5, 0.5) {$=$};
		\node [style=small box] (48) at (-2.25, 1) {$\location$};
		\node [style=none] (50) at (0, 3) {};
		\node [style=black dot] (51) at (-2.25, 2) {};
		\node [style=none] (52) at (-1.5, 3) {};
		\node [style=none] (53) at (-3, 2.75) {};
		\node [style=none] (54) at (-3, 4.5) {};
		\node [style=none] (55) at (-1.5, 3) {};
		\node [style=disintegration, minimum height=3.0cm, minimum width=4.3cm] (56) at (0.55, 0.9) {};
		\node [style=small box] (58) at (11, -1.25) {$\location$};
		\node [style=small box] (59) at (14.75, -1.25) {$\location^{2n+1}$};
		\node [style=none] (60) at (14.75, 0.75) {};
		\node [style=black dot] (61) at (11, -0.25) {};
		\node [style=none] (62) at (11.75, 0.75) {};
		\node [style=none] (63) at (10.25, 0.5) {};
		\node [style=none] (64) at (10.25, 2.5) {};
		\node [style=none] (65) at (11.75, 0.75) {};
		\node [style=disintegration, minimum height=0.8cm, minimum width=2.2cm] (66) at (14.75, -1.25) {};
		\node [style=none] (67) at (5.5, 0.5) {$=$};
		\node [style=small box] (68) at (8, 1) {$\location$};
		\node [style=none] (69) at (10.25, 3) {};
		\node [style=black dot] (70) at (8, 2) {};
		\node [style=none] (71) at (8.75, 3) {};
		\node [style=none] (72) at (7.25, 2.75) {};
		\node [style=none] (73) at (7.25, 4.5) {};
		\node [style=none] (74) at (8.75, 3) {};
		\node [style=disintegration, minimum height=3.05cm, minimum width=5.4cm] (75) at (11.7, 0.85) {};
		\node [style=small box] (76) at (23.5, -1.25) {$\location$};
		\node [style=small box] (77) at (27.25, -1.25) {$\location^{2n+1}$};
		\node [style=none] (78) at (27.25, 0.75) {};
		\node [style=black dot] (79) at (23.5, -0.25) {};
		\node [style=none] (80) at (24.5, 0.75) {};
		\node [style=none] (81) at (22.25, 0.5) {};
		\node [style=none] (83) at (24.5, 0.75) {};
		\node [style=none] (85) at (18, 0.5) {$=$};
		\node [style=small box] (86) at (20.25, -1.25) {$\location$};
		\node [style=none] (87) at (22.25, 0.5) {};
		\node [style=black dot] (88) at (20.25, -0.25) {};
		\node [style=none] (89) at (21, 0.75) {};
		\node [style=none] (90) at (19.5, 0.5) {};
		\node [style=none] (91) at (19.5, 2.75) {};
		\node [style=none] (92) at (21, 0.75) {};
		\node [style=disintegration, minimum height=2.1cm, minimum width=5.4cm] (93) at (24, 0.05) {};
		\node [style=small box] (94) at (33, 0) {$\location^{2n+3}$};
		\node [style=disintegration, minimum height=0.8cm, minimum width=2.2cm] (95) at (33, 0) {};
		\node [style=none] (96) at (30.25, 0.75) {$=$};
		\node [style=none] (97) at (33, 3) {};
		\node [style=small box] (98) at (-6.5, 0) {$\bob{n+1}$};
		\node [style=none] (99) at (-6.5, 2.5) {};
	\end{pgfonlayer}
	\begin{pgfonlayer}{edgelayer}
		\draw (39) to (36);
		\draw [in=-90, out=150, looseness=1.25] (39) to (41.center);
		\draw (41.center) to (42.center);
		\draw [in=-90, out=30, looseness=1.25] (39) to (40.center);
		\draw (40.center) to (43.center);
		\draw [bend left=90, looseness=1.25] (43.center) to (38.center);
		\draw (38.center) to (37);
		\draw (51) to (48);
		\draw [in=-90, out=150, looseness=1.25] (51) to (53.center);
		\draw (53.center) to (54.center);
		\draw [in=-90, out=30, looseness=1.25] (51) to (52.center);
		\draw (52.center) to (55.center);
		\draw [bend left=90, looseness=1.75] (55.center) to (50.center);
		\draw (42.center) to (50.center);
		\draw (61) to (58);
		\draw [in=-90, out=150, looseness=1.25] (61) to (63.center);
		\draw (63.center) to (64.center);
		\draw [in=-90, out=30, looseness=1.25] (61) to (62.center);
		\draw (62.center) to (65.center);
		\draw [bend left=90, looseness=1.25] (65.center) to (60.center);
		\draw (60.center) to (59);
		\draw (70) to (68);
		\draw [in=-90, out=150, looseness=1.25] (70) to (72.center);
		\draw (72.center) to (73.center);
		\draw [in=-90, out=30, looseness=1.25] (70) to (71.center);
		\draw (71.center) to (74.center);
		\draw [bend left=90, looseness=1.75] (74.center) to (69.center);
		\draw (64.center) to (69.center);
		\draw (79) to (76);
		\draw [in=-90, out=-180] (79) to (81.center);
		\draw [in=-90, out=30, looseness=1.25] (79) to (80.center);
		\draw (80.center) to (83.center);
		\draw [bend left=90] (83.center) to (78.center);
		\draw (78.center) to (77);
		\draw (88) to (86);
		\draw [in=-90, out=150, looseness=1.25] (88) to (90.center);
		\draw (90.center) to (91.center);
		\draw [in=-90, out=30, looseness=1.25] (88) to (89.center);
		\draw (89.center) to (92.center);
		\draw [bend left=90, looseness=1.75] (92.center) to (87.center);
		\draw (97.center) to (94);
		\draw (99.center) to (98);
	\end{pgfonlayer}
\end{tikzpicture}}}} \]

\noindent As a result, $\alice{n}$ is the normalisation of the
$2n$-power of $\location$. In the limit $n \to \infty$, both Alice and
Bob will end up going to the most popular place, that is, to
$s\in\supp(\location)$ with the highest probability.


\section{Applications to causality}\label{sec:causality}

\begin{wrapfigure}{r}{0.5\textwidth}
	\label{fig:causalmodel}
	\begin{equation}
		\label{diag:smokingjoint}
		\begin{array}{rcl}
			\sigma
			& = &
			\frac{1}{5}\ket{s,t,c}
			\\[+0.2em]
			& & \;\; +\,
			\frac{1}{50}\ket{s,t,\no{c}}
			\\[+0.2em]
			& & \;\; +\,
			\frac{1}{20}\ket{s,\no{t},c}
			\\[+0.2em]
			& & \;\; +\,
			\frac{1}{10}\ket{s,\no{t},\no{c}}
			\\[+0.2em]
			& & \;\; +\,
			\frac{1}{50}\ket{\no{s},t,c}
			\\[+0.2em]
			& & \;\; +\,
			\frac{1}{100}\ket{\no{s},t,\no{c}}
			\\[+0.2em]
			& & \;\; +\,
			\frac{1}{10}\ket{\no{s},\no{t},c}
			\\[+0.2em]
			& & \;\; +\,
			\frac{1}{2}\ket{\no{s},\no{t},\no{c}}.
		\end{array}
		\hspace*{5em}
		\vcenter{\hbox{\begin{tikzpicture}
	\begin{pgfonlayer}{nodelayer}
		\node [style=small box] (0) at (0, -2) {$\gamma$};
		\node [style=black dot] (1) at (0, -1) {};
		\node [style=small box] (2) at (0, 0) {$f$};
		\node [style=black dot] (3) at (0, 1) {};
		\node [style=medium box] (4) at (0.5, 4) {$h$};
		\node [style=none] (6) at (1, 3.75) {};
		\node [style=none] (7) at (1, 0) {};
		\node [style=none] (8) at (-1.75, 3) {};
		\node [style=none] (10) at (0.5, 5) {};
		\node [style=none] (11) at (-1.75, 5) {};
		\node [style=none] (13) at (1, 5) {$C$};
		\node [style=none] (14) at (-0.5, 5) {$T$};
		\node [style=none] (15) at (-2.25, 5) {$S$};
		\node [style=small box] (16) at (0, 2) {$g$};
		\node [style=black dot] (17) at (0, 3) {};
		\node [style=none] (18) at (0, 3.75) {};
		\node [style=none] (20) at (-1, 5) {};
		\node [style=none] (21) at (-1, 4) {};
		\node [style=none] (22) at (1.25, -0.75) {$G$};
	\end{pgfonlayer}
	\begin{pgfonlayer}{edgelayer}
		\draw (8.center) to (11.center);
		\draw (4) to (10.center);
		\draw (7.center) to (6.center);
		\draw (2) to (3);
		\draw (1) to (2);
		\draw (0) to (1);
		\draw [in=-90, out=0] (1) to (7.center);
		\draw [in=-90, out=180] (3) to (8.center);
		\draw (16) to (17);
		\draw (17) to (18.center);
		\draw (3) to (16);
		\draw (21.center) to (20.center);
		\draw [in=-90, out=-180] (17) to (21.center);
	\end{pgfonlayer}
\end{tikzpicture}}}
	\end{equation}
\end{wrapfigure}

We elaborate a leading example from the causality literature in the
current framework, from~\cite{Pearl09}, see also
\textit{e.g.}~\cite{Nielsen12}. We build on the approach
from~\cite{JacobsKZ21} using string-diagrammatic surgery for
intervention. Although the example is well-known and extensively
discussed elsewhere, our treatment does have an original element,
namely the introduction of the dagger channel, see immediately
after~\eqref{diag:smokingdo} for details.

The example is about a possible causal relation between smoking $S$, 
cancer $C$, and tar $T$, with a possible confounding role played by
genetic consitution $G$. It starts with a joint distribution
$\sigma\in \Dst\big(S\times T\times C\big)$, involving sets $S =
\{s,\no{s}\}$ for smoking and non-smoking, and similarly for cancer
and tar: $C = \{c,\no{c}\}$ and $T = \{t, \no{t}\}$.  We \emph{know}
what this $\sigma$ is, for instance from some empirical study, see on 
the left above, and we \emph{assume} that it has the shape described 
on the right.

\noindent We see the confounding role via the assumed genetic
distribution $\gamma\in\Dst(G)$. The assumed channels $f,g,h$ and the
distribution $\gamma$ can not all be obtained from the joint
distribution $\sigma$, but we can get something. For instance, the
composite $f \after \gamma$ becomes available when we marginalise out
$T$ and $C$.  Similarly, the channel $g$ can be extracted by
disintegrating from $S$ to $T$, see below for details.

We start with the causal surgery procedure, by punching a hole below
the $S$ copy bullet, as in:

\[ \vcenter{\hbox{\begin{tikzpicture}
	\begin{pgfonlayer}{nodelayer}
		\node [style=small box] (0) at (0, -2) {$\gamma$};
		\node [style=black dot] (1) at (0, -1) {};
		\node [style=small box] (2) at (0, 0) {$f$};
		\node [style=black dot] (3) at (0, 1) {};
		\node [style=medium box] (4) at (0.5, 4) {$h$};
		\node [style=none] (6) at (1, 3.75) {};
		\node [style=none] (7) at (1, 0) {};
		\node [style=none] (8) at (-1.75, 3) {};
		\node [style=none] (10) at (0.5, 5) {};
		\node [style=none] (11) at (-1.75, 5) {};
		\node [style=none] (13) at (1, 5) {$C$};
		\node [style=none] (14) at (-0.5, 5) {$T$};
		\node [style=none] (15) at (-2.25, 5) {$S$};
		\node [style=small box] (16) at (0, 2) {$g$};
		\node [style=black dot] (17) at (0, 3) {};
		\node [style=none] (18) at (0, 3.75) {};
		\node [style=none] (20) at (-1, 5) {};
		\node [style=none] (21) at (-1, 4) {};
		\node [style=none] (22) at (1.25, -0.75) {$G$};
	\end{pgfonlayer}
	\begin{pgfonlayer}{edgelayer}
		\draw (8.center) to (11.center);
		\draw (4) to (10.center);
		\draw (7.center) to (6.center);
		\draw (2) to (3);
		\draw (1) to (2);
		\draw (0) to (1);
		\draw [in=-90, out=0] (1) to (7.center);
		\draw [in=-90, out=180] (3) to (8.center);
		\draw (16) to (17);
		\draw (17) to (18.center);
		\draw (3) to (16);
		\draw (21.center) to (20.center);
		\draw [in=-90, out=-180] (17) to (21.center);
	\end{pgfonlayer}
\end{tikzpicture}}}
\xymatrix{\mbox{}\ar@{|->}[r]^-{\text{\footnotesize punch}}
	& \mbox{}}
\vcenter{\hbox{\begin{tikzpicture}
	\begin{pgfonlayer}{nodelayer}
		\node [style=small box] (0) at (-2.5, -3.75) {$\gamma$};
		\node [style=black dot] (1) at (-2.5, -2.75) {};
		\node [style=black dot] (3) at (-2.5, -0.75) {};
		\node [style=medium box] (4) at (-2, 2.25) {$h$};
		\node [style=none] (6) at (-1.5, 2) {};
		\node [style=none] (7) at (-1.5, -1.75) {};
		\node [style=none] (8) at (-4.25, 1.25) {};
		\node [style=none] (10) at (-2, 3.25) {};
		\node [style=none] (11) at (-4.25, 3.25) {};
		\node [style=none] (13) at (-1.5, 3.25) {$C$};
		\node [style=none] (14) at (-3, 3.25) {$T$};
		\node [style=none] (15) at (-4.75, 3.25) {$S$};
		\node [style=small box] (16) at (-2.5, 0.25) {$g$};
		\node [style=black dot] (17) at (-2.5, 1.25) {};
		\node [style=none] (18) at (-2.5, 2) {};
		\node [style=none] (20) at (-3.5, 3.25) {};
		\node [style=none] (21) at (-3.5, 2.25) {};
		\node [style=upground] (22) at (-2.5, -2.25) {};
		\node [style=downground] (23) at (-2.5, -1.25) {};
		\node [style=small box] (24) at (4, 0.25) {$\gamma$};
		\node [style=black dot] (26) at (2.5, -0.75) {};
		\node [style=medium box] (27) at (3, 2.25) {$h$};
		\node [style=none] (28) at (3.5, 2) {};
		\node [style=none] (29) at (4, 0.5) {};
		\node [style=none] (30) at (0.75, 1.25) {};
		\node [style=none] (31) at (3, 3.25) {};
		\node [style=none] (32) at (0.75, 3.25) {};
		\node [style=none] (33) at (3.5, 3.25) {$C$};
		\node [style=none] (34) at (2, 3.25) {$T$};
		\node [style=none] (35) at (0.25, 3.25) {$S$};
		\node [style=small box] (36) at (2.5, 0.25) {$g$};
		\node [style=black dot] (37) at (2.5, 1.25) {};
		\node [style=none] (38) at (2.5, 2) {};
		\node [style=none] (39) at (1.5, 3.25) {};
		\node [style=none] (40) at (1.5, 2.25) {};
		\node [style=downground] (42) at (2.5, -1.25) {};
		\node [style=none] (43) at (-0.25, 0) {$=$};
	\end{pgfonlayer}
	\begin{pgfonlayer}{edgelayer}
		\draw (8.center) to (11.center);
		\draw (4) to (10.center);
		\draw (7.center) to (6.center);
		\draw (0) to (1);
		\draw [in=-90, out=0] (1) to (7.center);
		\draw [in=-90, out=180] (3) to (8.center);
		\draw (16) to (17);
		\draw (17) to (18.center);
		\draw (3) to (16);
		\draw (21.center) to (20.center);
		\draw [in=-90, out=-180] (17) to (21.center);
		\draw (23) to (3);
		\draw (1) to (22);
		\draw (30.center) to (32.center);
		\draw (27) to (31.center);
		\draw [in=-90, out=180] (26) to (30.center);
		\draw (36) to (37);
		\draw (37) to (38.center);
		\draw (26) to (36);
		\draw (40.center) to (39.center);
		\draw [in=-90, out=-180] (37) to (40.center);
		\draw (42) to (26);
		\draw [in=-90, out=90, looseness=1.50] (29.center) to (28.center);
	\end{pgfonlayer}
\end{tikzpicture}}} \]

\noindent The leaf $\unif$ is used for a uniform distribution.

It is good to be aware that this punching happens in a
hypothetical string diagram. But it is a way to get the required
causal `do' channel by discarding and disintegration, applied to the
diagram above on the right. We calculate using the disintegration 
rules, especially the box removal rule from 
Proposition~\ref{prop:boxremoval}.
\begin{equation}
	\label{diag:smokingpunched}
	\vcenter{\hbox{\begin{tikzpicture}
	\begin{pgfonlayer}{nodelayer}
		\node [style=small box] (0) at (-2.5, 0.5) {$\gamma$};
		\node [style=black dot] (1) at (-4, -1.75) {};
		\node [style=medium box] (2) at (-3.5, 2.5) {$h$};
		\node [style=none] (3) at (-3, 2.25) {};
		\node [style=none] (4) at (-2.5, 0.75) {};
		\node [style=none] (5) at (-5, -1) {};
		\node [style=none] (6) at (-3.5, 4) {};
		\node [style=none] (7) at (-6, -1) {};
		\node [style=none] (8) at (-3, 4) {$C$};
		\node [style=none] (9) at (-6.5, -3.25) {$S$};
		\node [style=small box] (10) at (-4, 0.5) {$g$};
		\node [style=downground] (12) at (-4, -2.25) {};
		\node [style=small box] (13) at (2, -0.75) {$\gamma$};
		\node [style=medium box] (15) at (1, 1) {$h$};
		\node [style=none] (16) at (1.5, 0.75) {};
		\node [style=none] (17) at (2, -0.5) {};
		\node [style=none] (18) at (0.5, -2.75) {};
		\node [style=none] (19) at (1, 3.25) {};
		\node [style=none] (21) at (1.5, 3.25) {$C$};
		\node [style=none] (22) at (0, -2.75) {$S$};
		\node [style=small box] (23) at (0.5, -0.75) {$g$};
		\node [style=none] (24) at (0.5, 0.75) {};
		\node [style=none] (26) at (-6, -3.25) {};
		\node [style=none] (28) at (-1, 0) {$=$};
		\node [style=none] (48) at (-4, 2.25) {};
		\node [style=small box] (52) at (-8.5, 0) {$\gamma$};
		\node [style=black dot] (53) at (-10, -1.25) {};
		\node [style=medium box] (54) at (-9.5, 2) {$h$};
		\node [style=none] (55) at (-9, 1.75) {};
		\node [style=none] (56) at (-8.5, 0.25) {};
		\node [style=none] (57) at (-11, -0.25) {};
		\node [style=none] (58) at (-9.5, 3.5) {};
		\node [style=none] (59) at (-12, -0.25) {};
		\node [style=none] (60) at (-9, 3.5) {$C$};
		\node [style=none] (61) at (-12.5, -2.75) {$S$};
		\node [style=small box] (62) at (-10, 0) {$g$};
		\node [style=downground] (63) at (-10, -1.75) {};
		\node [style=none] (64) at (-12, -2.75) {};
		\node [style=disintegration, minimum height=2.5cm, minimum width=2.25cm] (65) at (-9.95, 0.3) {};
		\node [style=none] (66) at (-7, 0) {$=$};
		\node [style=black dot] (67) at (-10, 1) {};
		\node [style=none] (68) at (-10, 1.75) {};
		\node [style=none] (69) at (-11, 2) {};
		\node [style=upground] (70) at (-11, 3.25) {};
		\node [style=disintegration, minimum height=1.1cm, minimum width=1.3cm] (71) at (-4.9, -1.4) {};
	\end{pgfonlayer}
	\begin{pgfonlayer}{edgelayer}
		\draw [bend left=270, looseness=2.00] (5.center) to (7.center);
		\draw (2) to (6.center);
		\draw [in=-90, out=180] (1) to (5.center);
		\draw (1) to (10);
		\draw (12) to (1);
		\draw [in=-90, out=90, looseness=1.50] (4.center) to (3.center);
		\draw (15) to (19.center);
		\draw [in=-90, out=90, looseness=1.50] (17.center) to (16.center);
		\draw (23) to (24.center);
		\draw (26.center) to (7.center);
		\draw (18.center) to (23);
		\draw [bend left=270, looseness=2.00] (57.center) to (59.center);
		\draw (54) to (58.center);
		\draw [in=-90, out=180] (53) to (57.center);
		\draw (53) to (62);
		\draw (63) to (53);
		\draw [in=-90, out=90, looseness=1.50] (56.center) to (55.center);
		\draw (64.center) to (59.center);
		\draw (67) to (68.center);
		\draw [in=-90, out=-180] (67) to (69.center);
		\draw (70) to (69.center);
		\draw (67) to (62);
		\draw (48.center) to (10);
	\end{pgfonlayer}
\end{tikzpicture}}}
\end{equation}

\noindent At this stage the situation is: we may not be able to
extract the ingredients of the original joint distribution $\sigma$
in~\eqref{diag:smokingjoint}, but we may be able to obtain the
ingredients for the causal channel on the right
in~\eqref{diag:smokingpunched}. This works, as we shall demonstrate,
in several steps.  

We first discard the second two wires of the joint
distribution~\eqref{diag:smokingjoint}, giving the marginal on the
left below. Next we extract the channel $g\colon S \rightarrow T$ from
$\sigma$ in, using again Proposition~\ref{prop:boxremoval}:
	\[ \vcenter{\hbox{\begin{tikzpicture}
	\begin{pgfonlayer}{nodelayer}
		\node [style=medium box] (20) at (2.25, -0.25) {$\sigma$};
		\node [style=none] (21) at (1.75, 0.25) {};
		\node [style=none] (22) at (2.75, 0.25) {};
		\node [style=none] (23) at (2.25, 0.25) {};
		\node [style=none] (25) at (1.75, 2) {};
		\node [style=none] (28) at (1.25, 2) {$S$};
		\node [style=upground] (29) at (2.75, 1) {};
		\node [style=upground] (30) at (2.25, 1.5) {};
		\node [style=small box] (50) at (6.5, -0.25) {$\gamma$};
		\node [style=small box] (52) at (6.5, 1.25) {$f$};
		\node [style=none] (53) at (6, 2.5) {$S$};
		\node [style=none] (54) at (6.5, 2.5) {};
		\node [style=none] (55) at (4.5, 1) {$=$};
	\end{pgfonlayer}
	\begin{pgfonlayer}{edgelayer}
		\draw (21.center) to (25.center);
		\draw (22.center) to (29);
		\draw (30) to (23.center);
		\draw (54.center) to (52);
		\draw (52) to (50);
	\end{pgfonlayer}
\end{tikzpicture}}}
        \hspace*{5em}
        \vcenter{\hbox{\begin{tikzpicture}
	\begin{pgfonlayer}{nodelayer}
		\node [style=none] (26) at (4, 1.5) {$=$};
		\node [style=small box] (27) at (7.25, -1.5) {$\gamma$};
		\node [style=black dot] (28) at (7.25, -0.5) {};
		\node [style=small box] (29) at (7.25, 0.5) {$f$};
		\node [style=black dot] (30) at (7.25, 1.5) {};
		\node [style=none] (32) at (8.25, 4.25) {};
		\node [style=none] (33) at (8.25, 0.5) {};
		\node [style=none] (34) at (6, 3.25) {};
		\node [style=none] (35) at (5.25, 3.25) {};
		\node [style=none] (36) at (6.75, 5.5) {$T$};
		\node [style=none] (37) at (4.75, -2.75) {$S$};
		\node [style=small box] (38) at (7.25, 2.5) {$g$};
		\node [style=black dot] (39) at (7.25, 3.5) {};
		\node [style=none] (40) at (7.25, 4.25) {};
		\node [style=none] (41) at (6.25, 5.5) {};
		\node [style=none] (42) at (6.25, 4.5) {};
		\node [style=upground] (44) at (7.25, 4.25) {};
		\node [style=none] (45) at (5.25, -3) {};
		\node [style=upground] (47) at (8.25, 4.25) {};
		\node [style=none] (48) at (9.75, 1.5) {$=$};
		\node [style=small box] (49) at (12.25, -0.75) {$\gamma$};
		\node [style=small box] (51) at (12.25, 0.75) {$f$};
		\node [style=black dot] (52) at (12.25, 1.75) {};
		\node [style=none] (55) at (11.75, 2.5) {};
		\node [style=none] (56) at (11, 2.5) {};
		\node [style=none] (57) at (12.75, 5.25) {$T$};
		\node [style=none] (58) at (10.5, -2.25) {$S$};
		\node [style=small box] (59) at (12.25, 4) {$g$};
		\node [style=none] (62) at (12.25, 5.25) {};
		\node [style=none] (66) at (11, -2.5) {};
		\node [style=disintegration, minimum height=3.5cm, minimum width=1.9cm] (67) at (6.85, 1.15) {};
		\node [style=disintegration, minimum height=2.4cm, minimum width=1.2cm] (68) at (11.9, 0.8) {};
		\node [style=none] (69) at (16, 4.5) {$T$};
		\node [style=small box] (70) at (15.5, 1.5) {$g$};
		\node [style=none] (71) at (15.5, 4.5) {};
		\node [style=none] (72) at (14, 1.5) {$=$};
		\node [style=none] (73) at (15, -1.75) {$S$};
		\node [style=none] (74) at (15.5, -2) {};
		\node [style=medium box] (76) at (2, 1) {$\sigma$};
		\node [style=none] (77) at (2, 1.5) {};
		\node [style=none] (78) at (2.5, 1.5) {};
		\node [style=none] (79) at (1.5, 1.5) {};
		\node [style=none] (80) at (0.75, -0.75) {};
		\node [style=none] (81) at (2, 4.25) {};
		\node [style=none] (82) at (1.5, 2.5) {};
		\node [style=none] (83) at (0.75, 2.5) {};
		\node [style=none] (84) at (0.25, -0.5) {$S$};
		\node [style=upground] (85) at (2.5, 2.25) {};
		\node [style=none] (86) at (2.5, 4.25) {$T$};
		\node [style=disintegration, minimum height=1.4cm, minimum width=1.3cm] (87) at (1.8, 1.65) {};
	\end{pgfonlayer}
	\begin{pgfonlayer}{edgelayer}
		\draw [bend left=270, looseness=2.00] (34.center) to (35.center);
		\draw (33.center) to (32.center);
		\draw (29) to (30);
		\draw (28) to (29);
		\draw (27) to (28);
		\draw [in=-90, out=0] (28) to (33.center);
		\draw [in=-90, out=180] (30) to (34.center);
		\draw (38) to (39);
		\draw (39) to (40.center);
		\draw (30) to (38);
		\draw (42.center) to (41.center);
		\draw [in=-90, out=-180] (39) to (42.center);
		\draw (35.center) to (45.center);
		\draw [bend left=270, looseness=2.00] (55.center) to (56.center);
		\draw (51) to (52);
		\draw [in=-90, out=180] (52) to (55.center);
		\draw (52) to (59);
		\draw (56.center) to (66.center);
		\draw (62.center) to (59);
		\draw (51) to (49);
		\draw (71.center) to (70);
		\draw (70) to (74.center);
		\draw [bend left=270, looseness=2.00] (82.center) to (83.center);
		\draw (77.center) to (81.center);
		\draw (78.center) to (85);
		\draw (79.center) to (82.center);
		\draw (83.center) to (80.center);
	\end{pgfonlayer}
\end{tikzpicture}}} \]
	
\noindent Finally, we extract the channel $S\times T \rightarrow
\Dst(C)$ via disintegration:
\[ \vcenter{\hbox{\begin{tikzpicture}
	\begin{pgfonlayer}{nodelayer}
		\node [style=small box] (0) at (2.25, -3.25) {$\gamma$};
		\node [style=black dot] (1) at (2.25, -2.25) {};
		\node [style=small box] (2) at (2.25, -1.25) {$f$};
		\node [style=black dot] (3) at (2.25, -0.25) {};
		\node [style=medium box] (4) at (2.75, 2.75) {$h$};
		\node [style=none] (5) at (3.25, 2.5) {};
		\node [style=none] (6) at (3.25, -1.25) {};
		\node [style=none] (7) at (1, 0.75) {};
		\node [style=none] (8) at (0.25, 0.75) {};
		\node [style=none] (10) at (0.75, -4.75) {$S$};
		\node [style=small box] (11) at (2.25, 0.75) {$g$};
		\node [style=black dot] (12) at (2.25, 1.75) {};
		\node [style=none] (13) at (2.25, 2.5) {};
		\node [style=none] (15) at (1.25, 2.5) {};
		\node [style=none] (17) at (0.25, -4.75) {};
		\node [style=disintegration, minimum height=3.8cm, minimum width=2.1cm] (18) at (1.65, -0.25) {};
		\node [style=none] (19) at (-1.25, -0.25) {$=$};
		\node [style=medium box] (20) at (-3, -0.75) {$\sigma$};
		\node [style=none] (21) at (-3, -0.25) {};
		\node [style=none] (22) at (-2.5, -0.25) {};
		\node [style=none] (23) at (-3.5, -0.25) {};
		\node [style=none] (24) at (-4.25, -2.5) {};
		\node [style=none] (26) at (-3.5, 0.75) {};
		\node [style=none] (27) at (-4.25, 0.75) {};
		\node [style=none] (28) at (-3.75, -2.5) {$S$};
		\node [style=none] (30) at (-5.25, -2.5) {$T$};
		\node [style=disintegration, minimum height=1.7cm, minimum width=1.5cm] (31) at (-3.5, 0.1) {};
		\node [style=none] (32) at (-3, 0.75) {};
		\node [style=none] (33) at (-2.5, 3.25) {};
		\node [style=none] (34) at (-4.75, 0.75) {};
		\node [style=none] (35) at (-4.75, -2.5) {};
		\node [style=none] (36) at (-3, 3.25) {$C$};
		\node [style=none] (37) at (-0.75, -4.75) {$T$};
		\node [style=none] (38) at (-0.25, 2.5) {};
		\node [style=none] (39) at (-0.25, -4.75) {};
		\node [style=none] (40) at (2.75, 4.25) {};
		\node [style=none] (41) at (2.25, 4.25) {$C$};
		\node [style=none] (42) at (4.5, -0.25) {$=$};
		\node [style=small box] (43) at (8.25, -3.75) {$\gamma$};
		\node [style=black dot] (44) at (8.25, -1.25) {};
		\node [style=black dot] (46) at (8.25, -0.25) {};
		\node [style=medium box] (47) at (8.75, 2.75) {$h$};
		\node [style=none] (48) at (9.25, 2.5) {};
		\node [style=none] (49) at (9.75, -0.25) {};
		\node [style=none] (50) at (7, 0.75) {};
		\node [style=none] (51) at (6.25, 0.75) {};
		\node [style=none] (52) at (6.75, -5.25) {$S$};
		\node [style=small box] (53) at (8.25, 0.75) {$g$};
		\node [style=black dot] (54) at (8.25, 1.75) {};
		\node [style=none] (55) at (8.25, 2.5) {};
		\node [style=none] (56) at (7.25, 2.5) {};
		\node [style=none] (57) at (6.25, -5.25) {};
		\node [style=none] (59) at (5.25, -5.25) {$T$};
		\node [style=none] (60) at (5.75, 2.5) {};
		\node [style=none] (61) at (5.75, -5.25) {};
		\node [style=none] (62) at (8.75, 4.25) {};
		\node [style=none] (63) at (8.25, 4.25) {$C$};
		\node [style=none] (64) at (11.5, -0.25) {$=$};
		\node [style=small box] (65) at (8.25, -2.25) {$f$};
		\node [style=small box] (66) at (9.75, 0.75) {$f^{\dag}_{\gamma}$};
		\node [style=disintegration, minimum height=4.0cm, minimum width=2.55cm] (67) at (8, -0.5) {};
		\node [style=disintegration, minimum height=3.0cm, minimum width=2.25cm] (68) at (8.3, -1.5) {};
		\node [style=small box] (69) at (15.25, -3.75) {$\gamma$};
		\node [style=black dot] (70) at (15.25, -1.25) {};
		\node [style=black dot] (71) at (15.25, 0.25) {};
		\node [style=medium box] (72) at (15.75, 3.25) {$h$};
		\node [style=none] (73) at (16.25, 3) {};
		\node [style=none] (75) at (14, -0.75) {};
		\node [style=none] (76) at (13.25, -0.75) {};
		\node [style=none] (77) at (13.75, -5.25) {$S$};
		\node [style=small box] (78) at (15.25, 1.25) {$g$};
		\node [style=black dot] (79) at (15.25, 2.25) {};
		\node [style=none] (80) at (15.25, 3) {};
		\node [style=none] (81) at (14.25, 3) {};
		\node [style=none] (82) at (13.25, -5.25) {};
		\node [style=none] (83) at (12.25, -5.25) {$T$};
		\node [style=none] (84) at (12.75, 3) {};
		\node [style=none] (85) at (12.75, -5.25) {};
		\node [style=none] (86) at (15.75, 4.75) {};
		\node [style=none] (87) at (15.25, 4.75) {$C$};
		\node [style=none] (88) at (18.25, -0.25) {$=$};
		\node [style=small box] (89) at (15.25, -2.25) {$f$};
		\node [style=small box] (90) at (16.75, 1.25) {$f^{\dag}_{\gamma}$};
		\node [style=disintegration, minimum height=4.3cm, minimum width=2.55cm] (91) at (15, -0.2) {};
		\node [style=disintegration, minimum height=2.2cm, minimum width=2.25cm] (92) at (15.3, -2.3) {};
		\node [style=black dot] (95) at (21.75, -4) {};
		\node [style=medium box] (96) at (22.25, 2.75) {$h$};
		\node [style=none] (97) at (22.75, 2.5) {};
		\node [style=none] (100) at (22.25, -5.25) {$S$};
		\node [style=small box] (101) at (21.75, -2.75) {$g$};
		\node [style=black dot] (102) at (21.75, -1.75) {};
		\node [style=none] (103) at (21.75, 2.5) {};
		\node [style=none] (104) at (20.75, -1) {};
		\node [style=none] (105) at (21.75, -5.25) {};
		\node [style=none] (106) at (19.25, -5.25) {$T$};
		\node [style=none] (107) at (19.75, -1) {};
		\node [style=none] (108) at (19.75, -5.25) {};
		\node [style=none] (109) at (22.25, 4.25) {};
		\node [style=none] (110) at (21.75, 4.25) {$C$};
		\node [style=none] (111) at (24.5, -0.25) {$=$};
		\node [style=small box] (113) at (22.75, 0.75) {$f^{\dag}_{\gamma}$};
		\node [style=disintegration, minimum height=2.0cm, minimum width=1.8cm] (114) at (21.25, -2.4) {};
		\node [style=none] (115) at (22.75, -2.75) {};
		\node [style=medium box] (116) at (26.75, 0.25) {$h$};
		\node [style=none] (117) at (27.25, 0) {};
		\node [style=none] (118) at (26.25, 0) {};
		\node [style=none] (119) at (26.75, 3) {};
		\node [style=none] (120) at (26.25, 3) {$C$};
		\node [style=small box] (121) at (27.25, -2) {$f^{\dag}_{\gamma}$};
		\node [style=none] (122) at (27.75, -4.5) {$S$};
		\node [style=none] (123) at (27.25, -4.5) {};
		\node [style=none] (124) at (25.75, -4.5) {$T$};
		\node [style=none] (125) at (26.25, -4.5) {};
	\end{pgfonlayer}
	\begin{pgfonlayer}{edgelayer}
		\draw [bend left=270, looseness=2.00] (7.center) to (8.center);
		\draw (6.center) to (5.center);
		\draw (2) to (3);
		\draw (1) to (2);
		\draw (0) to (1);
		\draw [in=-90, out=0] (1) to (6.center);
		\draw [in=-90, out=180] (3) to (7.center);
		\draw (11) to (12);
		\draw (12) to (13.center);
		\draw (3) to (11);
		\draw [in=-90, out=-180] (12) to (15.center);
		\draw (8.center) to (17.center);
		\draw [bend left=270, looseness=2.00] (26.center) to (27.center);
		\draw (23.center) to (26.center);
		\draw (27.center) to (24.center);
		\draw [bend left=90, looseness=1.75] (34.center) to (32.center);
		\draw (34.center) to (35.center);
		\draw (22.center) to (33.center);
		\draw (38.center) to (39.center);
		\draw [bend left=90, looseness=1.50] (38.center) to (15.center);
		\draw (4) to (40.center);
		\draw (32.center) to (21.center);
		\draw [bend left=270, looseness=2.00] (50.center) to (51.center);
		\draw [in=-90, out=0] (44) to (49.center);
		\draw [in=-90, out=180] (46) to (50.center);
		\draw (53) to (54);
		\draw (54) to (55.center);
		\draw (46) to (53);
		\draw [in=-90, out=-180] (54) to (56.center);
		\draw (51.center) to (57.center);
		\draw (60.center) to (61.center);
		\draw [bend left=90, looseness=1.50] (60.center) to (56.center);
		\draw (47) to (62.center);
		\draw (66) to (49.center);
		\draw [in=-90, out=90, looseness=1.75] (66) to (48.center);
		\draw (44) to (65);
		\draw (44) to (46);
		\draw (43) to (65);
		\draw [bend left=270, looseness=2.00] (75.center) to (76.center);
		\draw (78) to (79);
		\draw (79) to (80.center);
		\draw (71) to (78);
		\draw [in=-90, out=-180] (79) to (81.center);
		\draw (76.center) to (82.center);
		\draw (84.center) to (85.center);
		\draw [bend left=90, looseness=1.50] (84.center) to (81.center);
		\draw (72) to (86.center);
		\draw [in=-90, out=90, looseness=1.75] (90) to (73.center);
		\draw (70) to (89);
		\draw (70) to (71);
		\draw (69) to (89);
		\draw [in=180, out=-90] (75.center) to (70);
		\draw [in=-90, out=15, looseness=1.25] (71) to (90);
		\draw (101) to (102);
		\draw (102) to (103.center);
		\draw (95) to (101);
		\draw [in=-90, out=-180] (102) to (104.center);
		\draw (107.center) to (108.center);
		\draw [bend left=90, looseness=1.50] (107.center) to (104.center);
		\draw (96) to (109.center);
		\draw [in=-90, out=90, looseness=1.75] (113) to (97.center);
		\draw [in=-90, out=30] (95) to (115.center);
		\draw (115.center) to (113);
		\draw (105.center) to (95);
		\draw (116) to (119.center);
		\draw [in=-90, out=90, looseness=1.75] (121) to (117.center);
		\draw (121) to (123.center);
		\draw (125.center) to (118.center);
	\end{pgfonlayer}
\end{tikzpicture}}} \]

\noindent The most complex step is the second one, which involves both
(a)~the change from one shaded box to two nested boxes, as in
Lemma~\ref{lem:nestedbox}, and (b)~the introduction of the dagger
$f^{\dag}_{\gamma}$ of the channel $f$, with the distribution $\gamma$
as prior. The final equation also combined two steps, namely
(a)~pulling the copier out, as in
Definition~\ref{def:normalisation}~\eqref{def:normalisation:comp}, and
(b)~removing the channel $g$ via Proposition~\ref{prop:boxremoval}.


We are now in a position to assemble the different pieces and obtain
the causal smoking-to-cancer channel $\mathsl{do} \colon S \rightarrow
C$ from the distribution $\sigma$, in the following manner.
\begin{equation}
	\label{diag:smokingdo}
	\vcenter{\hbox{\begin{tikzpicture}
	\begin{pgfonlayer}{nodelayer}
		\node [style=small box] (0) at (3.25, -0.75) {$\gamma$};
		\node [style=medium box] (1) at (2.25, 1) {$h$};
		\node [style=none] (2) at (2.75, 0.75) {};
		\node [style=none] (3) at (3.25, -0.5) {};
		\node [style=none] (4) at (1.75, -2.75) {};
		\node [style=none] (5) at (2.25, 3.25) {};
		\node [style=none] (6) at (2.75, 3) {$C$};
		\node [style=none] (7) at (1.25, -2.5) {$S$};
		\node [style=small box] (8) at (1.75, -0.75) {$g$};
		\node [style=none] (9) at (1.75, 0.75) {};
		\node [style=none] (10) at (-0.25, 0) {$\coloneqq$};
		\node [style=none] (11) at (-1.75, 0) {$\mathsl{do}$};
		\node [style=small box] (12) at (8.5, -2.25) {$\gamma$};
		\node [style=medium box] (13) at (7.5, 3.25) {$h$};
		\node [style=none] (14) at (8, 3) {};
		\node [style=none] (15) at (8.5, 1.5) {};
		\node [style=none] (16) at (7, -3.75) {};
		\node [style=none] (17) at (7.5, 4.75) {};
		\node [style=none] (18) at (8, 4.5) {$C$};
		\node [style=none] (19) at (6.5, -3.5) {$S$};
		\node [style=small box] (20) at (7, -2.25) {$g$};
		\node [style=none] (21) at (7, 3) {};
		\node [style=none] (22) at (5.25, 0) {$=$};
		\node [style=small box] (23) at (8.5, 1.5) {$f^{\dag}_{\gamma}$};
		\node [style=small box] (24) at (8.5, -0.75) {$f$};
		\node [style=none] (25) at (10.75, 0) {$=$};
		\node [style=medium box] (26) at (13.5, -2.5) {$\sigma$};
		\node [style=none] (27) at (14.25, 0.75) {};
		\node [style=none] (28) at (14, -2) {};
		\node [style=none] (29) at (13, -2) {};
		\node [style=none] (30) at (12.25, -4) {};
		\node [style=none] (31) at (13, -1) {};
		\node [style=none] (32) at (12.25, -1) {};
		\node [style=none] (33) at (11.75, -3.75) {$S$};
		\node [style=upground] (34) at (14, -1.25) {};
		\node [style=disintegration, minimum height=1.4cm, minimum width=1.3cm] (35) at (13.3, -1.85) {};
		\node [style=medium box] (36) at (16, -2.5) {$\sigma$};
		\node [style=none] (37) at (14.75, 0.75) {};
		\node [style=none] (38) at (16.5, -2) {};
		\node [style=none] (39) at (16, -2) {};
		\node [style=upground] (40) at (16.5, -1.25) {};
		\node [style=upground] (41) at (16, -0.75) {};
		\node [style=medium box] (42) at (16, 1.75) {$\sigma$};
		\node [style=none] (43) at (16, 2.25) {};
		\node [style=none] (44) at (16.5, 2.25) {};
		\node [style=none] (45) at (15.5, 2.25) {};
		\node [style=none] (46) at (15.5, 3.25) {};
		\node [style=none] (47) at (14.75, 3.25) {};
		\node [style=disintegration, minimum height=1.7cm, minimum width=1.5cm] (48) at (15.5, 2.6) {};
		\node [style=none] (49) at (16, 3.25) {};
		\node [style=none] (50) at (16.5, 5) {};
		\node [style=none] (51) at (14.25, 3.25) {};
		\node [style=none] (52) at (16, 4.75) {$C$};
		\node [style=none] (53) at (15.5, -2) {};
		\node [style=none] (54) at (15.5, -0.25) {};
		\node [style=none] (55) at (13.5, -2) {};
		\node [style=none] (56) at (13.5, -0.25) {};
	\end{pgfonlayer}
	\begin{pgfonlayer}{edgelayer}
		\draw (1) to (5.center);
		\draw [in=-90, out=90, looseness=1.50] (3.center) to (2.center);
		\draw (8) to (9.center);
		\draw (4.center) to (8);
		\draw (13) to (17.center);
		\draw [in=-90, out=90, looseness=1.50] (15.center) to (14.center);
		\draw (20) to (21.center);
		\draw (16.center) to (20);
		\draw (23) to (24);
		\draw (24) to (12);
		\draw [bend left=270, looseness=2.00] (31.center) to (32.center);
		\draw (28.center) to (34);
		\draw (29.center) to (31.center);
		\draw (32.center) to (30.center);
		\draw (38.center) to (40);
		\draw (41) to (39.center);
		\draw [bend left=270, looseness=2.00] (46.center) to (47.center);
		\draw (45.center) to (46.center);
		\draw [bend left=90, looseness=1.75] (51.center) to (49.center);
		\draw (44.center) to (50.center);
		\draw (49.center) to (43.center);
		\draw (27.center) to (51.center);
		\draw (47.center) to (37.center);
		\draw (54.center) to (53.center);
		\draw (56.center) to (55.center);
		\draw [in=90, out=-90, looseness=1.25] (37.center) to (54.center);
		\draw [in=-90, out=90] (56.center) to (27.center);
	\end{pgfonlayer}
\end{tikzpicture}}}
\end{equation}

\noindent This assembling makes use of one crucial trick, namely the
composite $f^{\dag}_{\gamma} \after f$ in the second equation.  This
allowed since $\gamma = f^{\dag}_{\gamma} \after f \after \gamma$.
This follows from the equation on the left in~\eqref{eqn:dagger}, by
discarding the first wire on both sides. This do channel $\mathsl{do}
\colon S \rightarrow \Dst(C)$ in~\eqref{diag:smokingdo} yields the
same outcomes as in~\cite{JacobsKZ21}:
\[ \begin{array}{rclcrcl}
	\mathsl{do}(s)
	& = &
	0.5423\ket{c} + 0.4577\ket{\no{c}}
	& \qquad &
	\mathsl{do}(\no{s})
	& = &
	0.2535\ket{c} + 0.7465\ket{\no{c}}.
\end{array} \]

In the end, let us reconsider what happened. We started from a joint
distribution $\sigma\in\Dst\big(S\times T\times C\big)$
in~\eqref{diag:smokingjoint} and assumed a certain (inaccessible)
internal structure. Via an intervention punch we removed the
confounding influence and determined what the resulting intervention
map from smoking to cancer would be. It turned out that we can
identify this intervention map via purely external manipulations of
the original joint distribution $\sigma$. The result is a composition,
on the right in~\eqref{diag:smokingdo}, of these manipulated portions
of $\sigma$, obtained via disintegration and marginalisation. The
resulting intervention channel was then actually computed, via these
separate portions. 

Finally, we observe that the choice of the mediating channel $g$
matters. One could try to choose the identity channel $g = \idmap$ to
get a diagram of the shape (\ref{diag:smokingjoint}). In that case,
the channel $S\times T \rightarrow \Dst(C)$ could not be extracted,
since the identity channel lacks full support. This shows that the
notion of full support is crucial when considering causal
identifiability.

\section{Applications to counterfactuals}\label{sec:counterfactual}

Counterfactual reasoning is quite common and follows a `had' pattern
that is easy to recognise, like in: had we left earlier, we would have
avoided the traffic jam. Such phrases contain an antecedent (about
leaving early) and a conclusion (ending up in a traffic jam), where
the antecedent is factually false --- we did not leave early --- but
is still used to draw a conclusion.  Such reasoning patterns form a
challenge in logic. They change a small part of reality, but leave
many other things the same, and then come to a conclusion. In a
probabilistic setting, this small change --- the `had' part of the
counterfactual --- can be modelled as an intervention, of the kind
that we have seen in the previous section on causality. We shall
describe counterfactual reasoning by duplicating the situation at
hand, into a factual and counterfactual world, as proposed
in~\cite{BalkeP94} (see also~\cite{Pearl09}), in such a way that the
probabilistic parts are shared, while the (deterministic) mechanisms
in both worlds are duplicated but the same. This involves decomposing
a probabilistic channel in terms of a deterministic channel together
with an `exogenous' distribution, in which all probability is
concentrated. This separation is a general phenomenon that is
described as randomness pushback in~\cite{Fritz20}. The exogenous
distributions capture the unobserved background factors that are left
unexplained and are simply accepted as they are. This approach is
illustrated below in terms of string diagrams, where we shall, once
again, exploit that we can simplify string diagrams after
intervention, via the discard rules for $\ground$ and also that we can
apply disintegration simplifications via rewriting.

A commonly used example in the (probabilistic) literature on
counterfactuals involves people going to a party, with a question: had
Bob not gone, would a scuffle not have happened, see
\textit{e.g.}~\cite{BalkeP94,JacobsKZ21}. We include a different, less
familiar illustration, taken from~\cite[Ex.~27.2]{BareinboimCII22}. It
involves a counterfactual of the form: had the patient been treated,
would they have survived? In~\cite{BareinboimCII22} the relevant
channels are already decomposed into a deterministic channel and an
exogenous distribution. In the represenation as string diagram below,
there are four such distributions $\flip(r) = r\ket{1} + (1 -
r)\ket{0} \in \Dst(\finset{2})$ with (roughly) the original names.
\[ \vcenter{\hbox{\scalebox{0.9}{\begin{tikzpicture}
	\begin{pgfonlayer}{nodelayer}
		\node [style=small box] (0) at (1.5, 0) {$U_z$};
		\node [style=small box] (1) at (3, 0) {$U_r$};
		\node [style=small box] (2) at (4.5, 0) {$U_x$};
		\node [style=small box] (3) at (6, 0) {$U_y$};
		\node [style=black dot] (4) at (3, 1) {};
		\node [style=none] (5) at (1.5, 1.75) {};
		\node [style=none] (6) at (2.5, 1.75) {};
		\node [style=black dot] (7) at (2, 3.25) {};
		\node [style=medium box] (8) at (2, 2.25) {$f_z$};
		\node [style=none] (9) at (2.5, 4) {};
		\node [style=none] (10) at (3.5, 4) {};
		\node [style=black dot] (11) at (3, 5.5) {};
		\node [style=medium box] (12) at (3, 4.5) {$f_x$};
		\node [style=none] (13) at (1.5, 4) {};
		\node [style=none] (14) at (1.5, 8) {};
		\node [style=none] (15) at (2.25, 6.25) {};
		\node [style=none] (16) at (3.75, 6.25) {};
		\node [style=none] (17) at (4.25, 6.25) {};
		\node [style=none] (18) at (4.75, 6.25) {};
		\node [style=medium box] (19) at (4.25, 6.75) {$f_y$};
		\node [style=none] (20) at (4.25, 8) {};
		\node [style=none] (21) at (2.25, 8) {};
		\node [style=none] (22) at (4.75, 8) {$Y$};
		\node [style=none] (23) at (2.75, 8) {$X$};
		\node [style=none] (24) at (1, 8) {$Z$};
	\end{pgfonlayer}
	\begin{pgfonlayer}{edgelayer}
		\draw (4) to (1);
		\draw [in=90, out=-90, looseness=1.25] (5.center) to (0);
		\draw [in=165, out=-90, looseness=1.25] (6.center) to (4);
		\draw (7) to (8);
		\draw (11) to (12);
		\draw [in=-90, out=15] (7) to (9.center);
		\draw [in=-90, out=165] (7) to (13.center);
		\draw (13.center) to (14.center);
		\draw [in=-90, out=180, looseness=1.25] (11) to (15.center);
		\draw [in=90, out=-90, looseness=1.25] (10.center) to (2);
		\draw [in=-90, out=15, looseness=0.75] (4) to (17.center);
		\draw [in=-90, out=90, looseness=1.25] (3) to (18.center);
		\draw (21.center) to (15.center);
		\draw (20.center) to (19);
		\draw [in=-90, out=15] (11) to (16.center);
	\end{pgfonlayer}
\end{tikzpicture}}}}
\hspace*{5em}
\begin{array}{rclcrclcrclcrcl}
	U_{r}
	& = &
	\flip\big(\frac{1}{4}\big)
        \\[+0.5em]
	U_{z}
	& = &
	\flip\big(\frac{19}{20}\big)
        \\[+0.5em]
	U_{x}
	& = &
	\flip\big(\frac{9}{10}\big)
        \\[+0.5em]
        U_{y}
	& = &
	\flip\big(\frac{7}{10}\big)
\end{array}
\hspace*{5em}
\begin{array}{rcl}
	f_{z}(a,b)
	& = &
	\begin{cases}
		1\bigket{1} & \mbox{if }a=b=1
		\\
		1\bigket{0} & \mbox{otherwise}
	\end{cases}
	\\
	f_{x}(a,b)
	& = &
	\begin{cases}
		1\bigket{1} & {\begin{array}[t]{l}
				\mbox{if } a=b=1 \\[-0.5em] \mbox{or 
				}a=b=0\end{array}}
		\\
		1\bigket{0} & \mbox{otherwise}
	\end{cases}
	\\
	f_{y}(a,b,c)
	& = &
	\begin{cases}
		1\bigket{1} & {\begin{array}[t]{l}
				\mbox{if } a=b=1 \\[-0.5em]
				\mbox{or }a=0, \, b=c=1  \\[-0.5em]
				\mbox{or }a=b=c=0\end{array}}
		\\
		1\bigket{0} & \mbox{otherwise}
	\end{cases}
\end{array} \]

\noindent There are three associated deterministic channels $f_{z},
f_{x} \colon \finset{2} \times \finset{2} \rightarrow
\Dst\big(\finset{2}\big)$ and $f_{z} \colon \finset{2} \times
\finset{2} \times \finset{2} \rightarrow
\Dst\big(\finset{2}\big)$. They are defined on the left below, and
used in the string diagram on the right. The labels / types $X,Y,Z$
are all equal to $\finset{2}$, but with different names: $Z$ is for
symptoms, $X$ is for treatment, and $Y$ is for survival.

First we are interested in a causal connection between
treatment $X$ and survival $Y$, where we need to take the confounding
influence via $U_r$ into account. We thus remove $f_X$ via a punch,
discard $Z$, then disintegrate the result to get a direct connection 
from
$X$ to $Y$.  This gives the following situation.
\[ \vcenter{\hbox{\begin{tikzpicture}
	\begin{pgfonlayer}{nodelayer}
		\node [style=small box] (0) at (5, -4) {$U_z$};
		\node [style=small box] (1) at (6.75, -4) {$U_r$};
		\node [style=small box] (2) at (8.25, -4) {$U_x$};
		\node [style=small box] (3) at (9.75, -4) {$U_y$};
		\node [style=black dot] (4) at (6.75, -3) {};
		\node [style=none] (5) at (5, -2.25) {};
		\node [style=none] (6) at (6, -2.25) {};
		\node [style=black dot] (7) at (5.5, -0.75) {};
		\node [style=medium box] (8) at (5.5, -1.75) {$f_z$};
		\node [style=none] (9) at (6.25, 0) {};
		\node [style=none] (10) at (7.25, 0) {};
		\node [style=black dot] (11) at (6.75, 1.5) {};
		\node [style=none] (12) at (4.75, 0) {};
		\node [style=none] (13) at (4.75, 1) {};
		\node [style=none] (14) at (6, 2.25) {};
		\node [style=none] (15) at (7.5, 2.25) {};
		\node [style=none] (16) at (8, 2.25) {};
		\node [style=none] (17) at (8.5, 2.25) {};
		\node [style=medium box] (18) at (8, 2.75) {$f_y$};
		\node [style=none] (19) at (8, 4.5) {};
		\node [style=none] (20) at (3.75, 2.25) {};
		\node [style=none] (21) at (8.5, 4.5) {$Y$};
		\node [style=none] (22) at (3.25, -5.5) {$X$};
		\node [style=downground] (23) at (6.75, 1) {};
		\node [style=upground] (24) at (7.25, 0.15) {};
		\node [style=upground] (25) at (6.25, 0.15) {};
		\node [style=upground] (26) at (4.75, 1.15) {};
		\node [style=none] (27) at (3.75, -5.5) {};
		\node [style=disintegration, minimum height=4.15cm, minimum width=3.5cm] (28) at (7.05, -0.6) {};
		\node [style=none] (29) at (11.75, -0.75) {$=$};
		\node [style=none] (30) at (13.25, 0.5) {};
		\node [style=none] (31) at (13.75, 0.5) {};
		\node [style=none] (32) at (14.25, 0.5) {};
		\node [style=medium box] (33) at (13.75, 1) {$f_y$};
		\node [style=none] (34) at (13.75, 2.75) {};
		\node [style=none] (35) at (14.25, 2.75) {$Y$};
		\node [style=small box] (36) at (15.75, -1.75) {$U_y$};
		\node [style=small box] (37) at (14.25, -1.75) {$U_r$};
		\node [style=none] (38) at (12.75, -4) {$X$};
		\node [style=none] (39) at (13.25, -4) {};
		\node [style=none] (40) at (2.25, -0.75) {$\coloneqq$};
		\node [style=none] (41) at (1, -0.75) {$\mathsl{do}$};
	\end{pgfonlayer}
	\begin{pgfonlayer}{edgelayer}
		\draw (4) to (1);
		\draw [in=90, out=-90, looseness=1.25] (5.center) to (0);
		\draw [in=165, out=-90, looseness=1.25] (6.center) to (4);
		\draw (7) to (8);
		\draw [in=-90, out=15] (7) to (9.center);
		\draw [in=-90, out=165] (7) to (12.center);
		\draw (12.center) to (13.center);
		\draw [in=-90, out=180, looseness=1.25] (11) to (14.center);
		\draw [in=90, out=-90, looseness=1.25] (10.center) to (2);
		\draw [in=-90, out=15, looseness=0.75] (4) to (16.center);
		\draw [in=-90, out=90, looseness=1.25] (3) to (17.center);
		\draw [bend left=90, looseness=1.50] (20.center) to (14.center);
		\draw (19.center) to (18);
		\draw [in=-90, out=15] (11) to (15.center);
		\draw (11) to (23);
		\draw (20.center) to (27.center);
		\draw (34.center) to (33);
		\draw [in=90, out=-90, looseness=1.25] (32.center) to (36);
		\draw [in=90, out=-90, looseness=1.25] (31.center) to (37);
		\draw (39.center) to (30.center);
	\end{pgfonlayer}
\end{tikzpicture}}} \]

\noindent The equation on the right is obtained via some obvious
re-wiring and via the by now familiar application of
Definition~\ref{def:normalisation}~\eqref{def:normalisation:comp} and
Proposition~\ref{prop:boxremoval}. This `do' channel
$\mathsl{do}\colon X \rightarrow Y$ can now be computed, as
$\mathsl{do}(1) = \frac{1}{4}\ket{1} + \frac{3}{4}\ket{0}$ and
$\mathsl{do}(0) = \frac{2}{5}\ket{1} + \frac{3}{5}\ket{0}$.  This
outcome is as in~\cite{BareinboimCII22}. It shows that the treatment
(input~$1$ for $\mathsl{do}$) does not really have effect of survival
(output~$1$).

Next, we look at the counterfactual question. There is a patient who
did not get treatment and died. Thus we can ask, had the patient been
treated, would they have survived? We form a twinned diagram,
coupling the factual and the counterfactual situation via the
exogenous distributions:
\[ \vcenter{\hbox{\begin{tikzpicture}
	\begin{pgfonlayer}{nodelayer}
		\node [style=small box] (0) at (-4.75, 0) {$U_z$};
		\node [style=small box] (1) at (2.5, 0) {$U_r$};
		\node [style=small box] (2) at (-2.5, 2.25) {$U_x$};
		\node [style=small box] (3) at (4.5, 0) {$U_y$};
		\node [style=black dot] (4) at (2.5, 1) {};
		\node [style=none] (6) at (-4.75, 1.75) {};
		\node [style=none] (7) at (-3.75, 1.75) {};
		\node [style=black dot] (8) at (-4.25, 3.25) {};
		\node [style=medium box] (9) at (-4.25, 2.25) {$f_z$};
		\node [style=none] (10) at (-3.5, 4) {};
		\node [style=none] (11) at (-2.5, 4) {};
		\node [style=black dot] (12) at (-3, 5.5) {};
		\node [style=medium box] (13) at (-3, 4.5) {$f_x$};
		\node [style=none] (14) at (-5, 4) {};
		\node [style=none] (16) at (-3.75, 6.25) {};
		\node [style=none] (17) at (-2.25, 6.25) {};
		\node [style=none] (18) at (-1.75, 6.25) {};
		\node [style=none] (19) at (-1.25, 6.25) {};
		\node [style=medium box] (20) at (-1.75, 6.75) {$f_y$};
		\node [style=none] (21) at (-1.75, 8) {};
		\node [style=none] (22) at (-3.75, 8) {};
		\node [style=none] (23) at (-1.25, 8) {$Y$};
		\node [style=none] (24) at (-3.25, 8) {$X$};
		\node [style=none] (43) at (8, 6.25) {};
		\node [style=none] (44) at (8.5, 6.25) {};
		\node [style=none] (45) at (9, 6.25) {};
		\node [style=medium box] (46) at (8.5, 6.75) {$f_y$};
		\node [style=none] (47) at (8.5, 8) {};
		\node [style=none] (49) at (9, 8) {$Y$};
		\node [style=small box] (50) at (7, 5) {$1$};
		\node [style=black dot] (51) at (4.5, 1) {};
		\node [style=upground] (52) at (-5, 5) {};
		\node [style=none] (53) at (-2.75, 8.75) {condition on $X=0$, $Y=0$};
		\node [style=none] (54) at (-2.75, -1.25) {factual};
		\node [style=none] (55) at (8, -1.25) {counterfactual};
		\node [shape=rectangle, draw=darkgray, dashed, minimum width=42mm, minimum height=50mm, yshift=0.5mm] (56) at (-2.8, 4.1) {};
		\node [shape=rectangle, draw=darkgray, dashed, minimum width=22mm, minimum height=50mm, yshift=0.5mm] (57) at (8, 4.1) {};
		\node [style=none] (58) at (8.5, 8.75) {export $Y$};
	\end{pgfonlayer}
	\begin{pgfonlayer}{edgelayer}
		\draw (4) to (1);
		\draw [in=90, out=-90, looseness=1.25] (6.center) to (0);
		\draw [in=180, out=-90, looseness=0.75] (7.center) to (4);
		\draw (8) to (9);
		\draw (12) to (13);
		\draw [in=-90, out=15] (8) to (10.center);
		\draw [in=-90, out=165] (8) to (14.center);
		\draw [in=-90, out=180, looseness=1.25] (12) to (16.center);
		\draw [in=90, out=-90, looseness=1.25] (11.center) to (2);
		\draw [in=-90, out=180] (4) to (18.center);
		\draw (22.center) to (16.center);
		\draw (21.center) to (20);
		\draw [in=-90, out=15] (12) to (17.center);
		\draw (47.center) to (46);
		\draw [in=-90, out=90, looseness=1.50] (50) to (43.center);
		\draw [in=-90, out=15] (4) to (44.center);
		\draw (51) to (3);
		\draw [in=-90, out=15] (51) to (45.center);
		\draw [in=-90, out=165, looseness=0.50] (51) to (19.center);
		\draw (52) to (14.center);
	\end{pgfonlayer}
\end{tikzpicture}}} \]

\noindent The factual side on the left describes the patient's
situation with no treatment ($X=0$) and no survival ($Y=0$). On the
counterfactual side on the right, treatment is enforced, via the
input~$1$ to $f_y$ and we like to learn about (export) the survival
probability via the $Y$-wire. In order to increase readability we will
use the following abbreviations.
\vspace{-1em}
\[ \vcenter{\hbox{\begin{tikzpicture}
	\begin{pgfonlayer}{nodelayer}
		\node [style=small box] (0) at (-4.75, -1) {$U_z$};
		\node [style=small box] (1) at (-2.75, -1) {$U_x$};
		\node [style=none] (2) at (-4.75, 0) {};
		\node [style=none] (3) at (-3.75, 0) {};
		\node [style=medium box] (5) at (-4.25, 0.5) {$f_z$};
		\node [style=none] (6) at (-4.25, 1.5) {};
		\node [style=none] (7) at (-3.25, 1.5) {};
		\node [style=medium box] (9) at (-3.75, 2) {$f_x$};
		\node [style=none] (13) at (-3.75, -2.25) {};
		\node [style=none] (14) at (-3.75, 3.25) {};
		\node [style=none] (15) at (-6.75, 0.5) {$\coloneqq$};
		\node [style=none] (16) at (-7.75, 0.5) {$g$};
		\node [style=none] (17) at (2, 0.5) {$\coloneqq$};
		\node [style=none] (18) at (1, 0.5) {$h$};
		\node [style=medium box] (19) at (4, 0.5) {$g^{\dag}_{U_r}$};
		\node [style=none] (20) at (4, -1) {};
		\node [style=none] (21) at (4, 2) {};
		\node [style=none] (22) at (5.75, 2) {};
		\node [style=small box] (23) at (5.75, 0.5) {$U_y$};
	\end{pgfonlayer}
	\begin{pgfonlayer}{edgelayer}
		\draw [in=90, out=-90, looseness=1.25] (2.center) to (0);
		\draw [in=90, out=-90, looseness=1.25] (7.center) to (1);
		\draw [in=-90, out=90, looseness=1.25] (5) to (6.center);
		\draw (14.center) to (9);
		\draw (3.center) to (13.center);
		\draw (21.center) to (19);
		\draw (19) to (20.center);
		\draw (22.center) to (23);
	\end{pgfonlayer}
\end{tikzpicture}}} \]

\noindent The disintegration of the combination of the above two
dashed diagrams, for factual and counterfactual, happens in

\[ 
\vcenter{\hbox{\begin{tikzpicture}
	\begin{pgfonlayer}{nodelayer}
		\node [style=small box] (1) at (-5.5, -5) {$U_r$};
		\node [style=small box] (3) at (-4, -5) {$U_y$};
		\node [style=black dot] (4) at (-5.5, -4) {};
		\node [style=none] (6) at (-6.5, -3.5) {};
		\node [style=black dot] (11) at (-6.5, -2) {};
		\node [style=medium box] (12) at (-6.5, -3) {$g$};
		\node [style=none] (14) at (-7, -1.25) {};
		\node [style=none] (15) at (-6, -1.25) {};
		\node [style=none] (16) at (-5.5, -1.25) {};
		\node [style=none] (17) at (-5, -1.25) {};
		\node [style=medium box] (18) at (-5.5, -0.75) {$f_y$};
		\node [style=none] (19) at (-5.5, -0.25) {};
		\node [style=none] (20) at (-7.75, -1.25) {};
		\node [style=none] (21) at (-9, -6.5) {$Y$};
		\node [style=none] (22) at (-7.25, -6.5) {$X$};
		\node [style=none] (23) at (-3.25, 0.5) {};
		\node [style=none] (24) at (-2.75, 0.5) {};
		\node [style=none] (25) at (-2.25, 0.5) {};
		\node [style=medium box] (26) at (-2.75, 1) {$f_y$};
		\node [style=none] (27) at (-2.75, 2.25) {};
		\node [style=none] (28) at (-2.25, 2.25) {$Y$};
		\node [style=small box] (29) at (-3.75, -0.75) {$1$};
		\node [style=black dot] (30) at (-4, -4) {};
		\node [style=none] (31) at (-8.5, -0.25) {};
		\node [style=none] (32) at (-7.75, -6.25) {};
		\node [style=none] (33) at (-8.5, -6.25) {};
		\node [style=disintegration, minimum height=3.8cm, minimum width=3.5cm] (34) at (-5.25, -2) {};
		\node [style=none] (35) at (-0.5, -2) {$=$};
		\node [style=small box] (36) at (3.5, -4.75) {$g$};
		\node [style=small box] (37) at (6, -1.75) {$U_y$};
		\node [style=black dot] (38) at (4.25, -0.75) {};
		\node [style=medium box] (41) at (4.25, -1.75) {$g^{\dag}_{U_r}$};
		\node [style=none] (42) at (2.75, -3) {};
		\node [style=none] (43) at (3.25, 0.25) {};
		\node [style=none] (44) at (3.75, 0.25) {};
		\node [style=none] (45) at (4.25, 0.25) {};
		\node [style=medium box] (46) at (3.75, 0.75) {$f_y$};
		\node [style=none] (47) at (3.75, 1.25) {};
		\node [style=none] (48) at (2, -3) {};
		\node [style=none] (49) at (0.75, -7.75) {$Y$};
		\node [style=none] (50) at (2.5, -7.75) {$X$};
		\node [style=none] (51) at (6, 2) {};
		\node [style=none] (52) at (6.5, 2) {};
		\node [style=none] (53) at (7, 2) {};
		\node [style=medium box] (54) at (6.5, 2.5) {$f_y$};
		\node [style=none] (55) at (6.5, 3.75) {};
		\node [style=none] (56) at (7, 3.75) {$Y$};
		\node [style=small box] (57) at (5.5, 0.75) {$1$};
		\node [style=black dot] (58) at (6, -0.75) {};
		\node [style=none] (59) at (1.25, 1.25) {};
		\node [style=none] (60) at (2, -7.5) {};
		\node [style=none] (61) at (1.25, -7.5) {};
		\node [style=black dot] (62) at (3.5, -3.75) {};
		\node [style=black dot] (63) at (4.25, -3) {};
		\node [style=none] (64) at (-10, -2) {$\coloneqq$};
		\node [style=none] (65) at (-11, -2) {$c$};
		\node [style=none] (66) at (8.75, -2) {$=$};
		\node [style=small box] (67) at (3.5, -6.25) {$U_r$};
		\node [style=disintegration, minimum height=5.1cm, minimum width=3.3cm] (68) at (4.25, -1.9) {};
		\node [style=black dot] (70) at (12.75, -4.25) {};
		\node [style=none] (73) at (11.5, -3) {};
		\node [style=none] (74) at (12, -3) {};
		\node [style=none] (75) at (12.5, -3) {};
		\node [style=medium box] (76) at (12, -2.5) {$f_y$};
		\node [style=none] (77) at (12, -1.75) {};
		\node [style=none] (79) at (10, -7.75) {$Y$};
		\node [style=none] (80) at (12.75, -7.75) {$X$};
		\node [style=none] (81) at (14.25, 2) {};
		\node [style=none] (82) at (14.75, 2) {};
		\node [style=none] (83) at (15.25, 2) {};
		\node [style=medium box] (84) at (14.75, 2.5) {$f_y$};
		\node [style=none] (85) at (14.75, 3.75) {};
		\node [style=none] (86) at (15.25, 3.75) {$Y$};
		\node [style=small box] (87) at (13.75, 0.75) {$1$};
		\node [style=black dot] (88) at (13.75, -4.25) {};
		\node [style=none] (89) at (10.5, -1.75) {};
		\node [style=none] (90) at (12.25, -7.5) {};
		\node [style=none] (91) at (10.5, -7.5) {};
		\node [style=medium box] (92) at (13.25, -5.5) {$h$};
		\node [style=none] (93) at (12.75, -5) {};
		\node [style=none] (94) at (13.75, -5) {};
		\node [style=black dot] (95) at (12.25, -6.75) {};
		\node [style=none] (96) at (11.5, -5.5) {};
		\node [style=disintegration, minimum height=2.6cm, minimum width=2.7cm] (97) at (12.9, -3.75) {};
		\node [style=none] (98) at (17, -2) {$=$};
		\node [style=black dot] (99) at (11.5, -4.25) {};
		\node [style=none] (100) at (13.75, -2.25) {};
		\node [style=upground] (101) at (13.75, -0.65) {};
		\node [style=none] (102) at (19.75, 0.75) {};
		\node [style=none] (103) at (20.25, 0.75) {};
		\node [style=none] (104) at (20.75, 0.75) {};
		\node [style=medium box] (105) at (20.25, 1.25) {$f_y$};
		\node [style=none] (106) at (20.25, 2.5) {};
		\node [style=none] (107) at (20.75, 2.5) {$Y$};
		\node [style=small box] (108) at (19.25, -0.5) {$1$};
		\node [style=medium box] (109) at (20.25, -3.5) {$\big(f_{y}\big)^{\dag}_{\idmap\otimes h}$};
		\node [style=none] (110) at (19.5, -3) {};
		\node [style=none] (111) at (20.25, -3) {};
		\node [style=none] (112) at (21, -3) {};
		\node [style=none] (113) at (19.5, -4) {};
		\node [style=none] (114) at (20.25, -4) {};
		\node [style=none] (115) at (21, -4) {};
		\node [style=none] (116) at (19.5, -2.25) {};
		\node [style=upground] (117) at (19.5, -2.15) {};
		\node [style=none] (118) at (21.25, -6) {$X$};
		\node [style=none] (119) at (20.75, -5.75) {};
		\node [style=black dot] (120) at (20.75, -5) {};
		\node [style=none] (121) at (19, -6) {$Y$};
		\node [style=none] (122) at (19.5, -5.75) {};
	\end{pgfonlayer}
	\begin{pgfonlayer}{edgelayer}
		\draw (4) to (1);
		\draw [in=165, out=-90, looseness=0.75] (6.center) to (4);
		\draw (11) to (12);
		\draw [in=-90, out=180, looseness=1.25] (11) to (14.center);
		\draw [in=-90, out=90, looseness=0.50] (4) to (16.center);
		\draw [bend left=90, looseness=1.75] (20.center) to (14.center);
		\draw (19.center) to (18);
		\draw [in=-90, out=15] (11) to (15.center);
		\draw (27.center) to (26);
		\draw [in=-90, out=90, looseness=1.50] (29) to (23.center);
		\draw [in=-90, out=15] (4) to (24.center);
		\draw (30) to (3);
		\draw [in=-90, out=15] (30) to (25.center);
		\draw [in=-90, out=150] (30) to (17.center);
		\draw [bend left=90] (31.center) to (19.center);
		\draw (31.center) to (33.center);
		\draw (20.center) to (32.center);
		\draw [in=-90, out=135] (38) to (44.center);
		\draw [bend left=90, looseness=1.75] (48.center) to (42.center);
		\draw (47.center) to (46);
		\draw (55.center) to (54);
		\draw [in=-90, out=90, looseness=1.50] (57) to (51.center);
		\draw [in=-90, out=15, looseness=1.50] (38) to (52.center);
		\draw (58) to (37);
		\draw [in=-90, out=15] (58) to (53.center);
		\draw [in=-90, out=150] (58) to (45.center);
		\draw [bend left=90] (59.center) to (47.center);
		\draw (59.center) to (61.center);
		\draw (48.center) to (60.center);
		\draw (62) to (36);
		\draw (38) to (41);
		\draw [in=-90, out=150] (62) to (42.center);
		\draw (41) to (63);
		\draw [in=-90, out=30] (62) to (63);
		\draw [in=-105, out=150, looseness=1.75] (63) to (43.center);
		\draw [in=-90, out=135] (70) to (74.center);
		\draw (77.center) to (76);
		\draw (85.center) to (84);
		\draw [in=-90, out=90, looseness=1.50] (87) to (81.center);
		\draw [in=-90, out=30] (70) to (82.center);
		\draw [in=-90, out=15, looseness=0.75] (88) to (83.center);
		\draw [in=-90, out=150] (88) to (75.center);
		\draw [bend left=90] (89.center) to (77.center);
		\draw (89.center) to (91.center);
		\draw [in=150, out=-90, looseness=1.25] (96.center) to (95);
		\draw [in=-90, out=30, looseness=1.25] (95) to (92);
		\draw (93.center) to (70);
		\draw (94.center) to (88);
		\draw (90.center) to (95);
		\draw (99) to (96.center);
		\draw (99) to (73.center);
		\draw [in=-90, out=30] (99) to (100.center);
		\draw (106.center) to (105);
		\draw [in=-90, out=90, looseness=1.50] (108) to (102.center);
		\draw (103.center) to (111.center);
		\draw [in=90, out=-90, looseness=1.25] (104.center) to (112.center);
		\draw (119.center) to (120);
		\draw [in=135, out=-90] (114.center) to (120);
		\draw [in=-90, out=60, looseness=1.50] (120) to (115.center);
		\draw [in=-90, out=90] (122.center) to (113.center);
		\draw (110.center) to (116.center);
		\draw (36) to (67);
		\draw (101) to (100.center);
	\end{pgfonlayer}
\end{tikzpicture}}} \] 

\noindent The abbreviation $g$ helps to simplify the situation in the first
shaded box. The dagger of this $g$ is introduced in the second shaded
box. In the next step, box removal is applied, formally after using
Lemma~\ref{lem:nestedbox}. At the same time, the abbreviation for $h$
is introduced and the copy is pulled out at the bottom, using
Definition~\ref{def:normalisation}~\eqref{def:normalisation:comp}.
Now we have a situation where a dagger with respect to a channel can
be introduced, as in~\eqref{eqn:daggerdef} on the right, namely the
dagger of $f_{y}$ with respect to the channel $\idmap\otimes h$.


Concretely, the counterfactual channel $c\colon X\times Y \rightarrow
\Dst(Y)$, with $X = Y = \finset{2}$, is:
\[ \begin{array}{rclcrclcrclcrcl}
	c(1,1)
	& = &
	1\bigket{1}
	& \qquad &
	c(0,1)
	& = &
	\frac{49}{454}\bigket{1} + \frac{405}{454}\bigket{0}
	& \qquad &
	c(1,0)
	& = &
	1\bigket{0}
	& \qquad &
	c(0,0)
	& = &
	\frac{1}{46}\bigket{1} + \frac{45}{46}\bigket{0}.
\end{array} \]


\noindent In~\cite{BareinboimCII22} only the latter outcome
$\frac{1}{46} \approx 0.0217$ occurs. Our disintegration approach
yields the entire channel, with all possible inputs. The low output at
$(0,0)$ of about $2\%$ survival chance, had the treatment been given,
confirms the earlier finding that the treatment has little positive
effect. We conclude with an interpretation of the different outcomes
of the channel $c$. The first two columns below describe the factual
situation.
\begin{center}
	\begin{tabular}{c|c|c}
		\begin{tabular}{c}\textbf{factual} \\[-0.4em]
			\textbf{treatment} \end{tabular}
		& 
		\begin{tabular}{c}\textbf{factual} \\[-0.4em]
			\textbf{survival } \end{tabular}
		&
		\begin{tabular}{c}\textbf{survival chance had} \\[-0.4em]
			\textbf{treatment been given} \end{tabular}
		\\
		\hline\hline
		1 & 1 & 1
		\\
		1 & 0 & 0
		\\
		0 & 1 & $\frac{49}{454}$
		\\
		0 & 0 & $\frac{1}{46}$
	\end{tabular}
\end{center}

\noindent The third line is intriguing. It describes the situation
where the patient survived without treatment, and where the survival
probability had the treatment been given is only $\frac{49}{454}
\approx 11\%$. Apparently, in this situation, the treatment has a
negative effect, lowering the chance of survival.

\section{Concluding remarks}\label{sec:conclusion}

This paper argues that inference in Bayesian networks, via 
conditioning / disintegration, can be done in a 
compositional manner. This is achieved by adding Proposition 
\ref{prop:boxremoval} to the calculus introduced in \cite{LorenzT23}, 
allowing one to directly compute with open models. This extends the 
range of applications of graphical techniques for probabilistic 
reasoning. It may become an ingredient of explainable AI, following 
the lines of~\cite{TullLCKC24}.  All our applications involve 
discrete probability, but the approach is sufficiently abstract to 
allow generalisation to suitable categories
for continuous probability. This is not trivial, for instance because
the cancellativity of caps that we use in
Definition~\ref{def:comparator} does not hold for Borel spaces, so a
different approach is needed. A possible approach would be to 
directly develop a theory of least disintegrations without relying on 
least normalisations and cancellative comparators. Partial Markov 
categories may provide a good setting for this, following 
\cite{LavoreR23,LavoreRSSz25}.

One further line of work is to develop automation for the rewriting of
string diagrams that happens in this paper. Another line of work is
to integrate the theory of partial Markov categories with (partial)
effectuses~\cite{ChoJWW15b,Cho15a,Jacobs15d}, where there is more
logical structure.

\bibliographystyle{entics}
\bibliography{bib}

\appendix


\section{Channel definitions in the child example}\label{app:data}

For completeness, we include the details of the channels /
boxes~\eqref{diag:childchannels} in the child Bayesian network example
in Section~\ref{sec:bayesiannetwork}. These details are used to 
calculate the actual disintegration~\eqref{eqn:childdisintegration}.


\[ \mathsl{LP} \stackrel{\mathsl{co}}{\longrightarrow} \Dst\big(\mathsl{CO}\big)
   \qquad
   \mathsl{DF} \times \mathsl{CM} \stackrel{\mathsl{hd}}{\longrightarrow} 
   \Dst\big(\mathsl{HD}\big)
   \qquad
   \mathsl{CM} \times \mathsl{LP} \stackrel{\mathsl{hi}}{\longrightarrow} 
   \Dst\big(\mathsl{HI}\big)
   \qquad
   \mathsl{HD} \times \mathsl{HI} \stackrel{\mathsl{lb}}{\longrightarrow} 
   \Dst\big(\mathsl{LB}\big). \]

\noindent First, the channel $\mathsl{d} \colon \mathsl{BA} \rightarrow
\Dst\big(\mathsl{DI}\big)$ is given by the following two equations.
\[ \begin{array}{rcl}
	\mathsl{d}(\ba) 
	& = &
	0.2\bigket{\pfc} + 0.3\bigket{\tga} + 0.25\bigket{\flt}
	+ 0.15\bigket{\pis} + 0.05\bigket{\tpd} + 
	0.05\bigket{\lng}
	\\
	\mathsl{d}(\no{\ba}) 
	& = & 
	0.03\bigket{\pfc} + 0.34\bigket{\tga} + 0.3\bigket{\flt}
	+  0.23\bigket{\pis} 
	+ 0.05\bigket{\tpd} + 0.05\bigket{\lng}
\end{array} \]

\noindent Next, we have three channels with $\mathsl{DI}$ as
domain. First, $\mathsl{df} \colon \mathsl{DI} \rightarrow
\Dst\big(\mathsl{DF}\big)$ is determined by:
\[ \begin{array}{rcl}
	\mathsl{df}(\pfc) & = &
	0.15\bigket{\lt} + 0.05\bigket{\DFno} + 0.8\bigket{\rt}
	\\
	\mathsl{df}(\tga) 
	& = & 
	0.1\bigket{\lt} + 0.8\bigket{\DFno} + 0.1\bigket{\rt}
	\\
	\mathsl{df}(\flt) & = & 
	0.8\bigket{\lt} + 0.2\bigket{\DFno} 
	\\
	\mathsl{df}(\pis) & = &
	1\bigket{\lt} \\
	\mathsl{df}(\tpd) & = & 
	0.33\bigket{\lt} + 0.33\bigket{\DFno} + 0.34\bigket{\rt}
	\\
	\mathsl{df}(\lng) & = & 
	0.2\bigket{\lt} + 0.4\bigket{\DFno} + 0.4\bigket{\rt}.
\end{array} \]

\noindent Next there is $\mathsl{cm} \colon \mathsl{DI} \rightarrow
\Dst\big(\mathsl{CM}\big)$ of the form:
\[ \begin{array}{rcl}
	\mathsl{cm}(\pfc) 
	& = & 
	0.4\bigket{\CMno} + 0.43\bigket{\mi} + 0.15\bigket{\co}
	+ 0.02\bigket{\tr}
	\\
	\mathsl{cm}(\tga) 
	& = &
	0.02\bigket{\CMno} + 0.09\bigket{\mi} + 0.09\bigket{\co}
	+ 0.8\bigket{\tr}
	\\
	\mathsl{cm}(\flt) 
	& = & 
	0.02\bigket{\CMno} + 0.16\bigket{\mi} + 0.8\bigket{\co}
	+ 0.02\bigket{\tr}
	\\
	\mathsl{cm}(\pis) 
	& = & 
	0.01\bigket{\CMno} + 0.02\bigket{\mi} + 0.95\bigket{\co}
	+ 0.02\bigket{\tr}
	\\
	\mathsl{cm}(\tpd) 
	& = & 0.01\bigket{\CMno} + 0.03\bigket{\mi} + 0.95\bigket{\co}
	+ 0.01\bigket{\tr}
	\\
	\mathsl{cm}(\lng) 
	& = & 
	0.4\bigket{\CMno} + 0.53\bigket{\mi} + 0.05\bigket{\co} 
	+ 0.02\bigket{\tr}.
\end{array} \]

\noindent And we also have $\mathsl{lp} \colon \mathsl{DI} \rightarrow
\Dst\big(\mathsl{LP}\big)$.
\[ \begin{array}{rcl}
	\mathsl{lp}(\pfc) 
	& = & 
	0.6\bigket{\nr} + 0.1\bigket{\cg}
	+ 0.3\bigket{\ab}
	\\
	\mathsl{lp}(\tga) 
	& = & 
	0.8\bigket{\nr} + 0.05\bigket{\cg}
	+ 0.15\bigket{\ab}
	\\
	\mathsl{lp}(\flt) 
	& = & 
	0.8\bigket{\nr} + 0.05\bigket{\cg}
	+ 0.15\bigket{\ab}
	\\
	\mathsl{lp}(\pis) 
	& = & 
	0.8\bigket{\nr} + 0.05\bigket{\cg}
	+ 0.15\bigket{\ab}
	\\
	\mathsl{lp}(\tpd) 
	& = & 0.1\bigket{\nr} + 0.6\bigket{\cg}
	+ 0.3\bigket{\ab}
	\\
	\mathsl{lp}(\lng) 
	& = & 
	0.03\bigket{\nr} + 0.25\bigket{\cg}
	+ 0.72\bigket{\ab}
\end{array} \]

\noindent The next channel is $\mathsl{co} \colon \mathsl{LP} \rightarrow
\Dst\big(\mathsl{CO}\big)$ consisting of three distributions:
\[ \begin{array}{rcl}
	\mathsl{co}(\nr) 
	& = & 0.8\bigket{\nr} + 0.1\bigket{\lo} + 0.1\bigket{\hi}
	\\
	\mathsl{co}(\cg) 
	& = & 
	0.65\bigket{\nr} + 0.05\bigket{\lo} + 0.3\bigket{\hi}
	\\
	\mathsl{co}(\ab) 
	& = & 
	0.45\bigket{\nr} + 0.05\bigket{\lo} + 0.5\bigket{\hi}
	\\
\end{array}\]

\noindent The firnal three channels with a product set as domain.
First, $\mathsl{hd} \colon\mathsl{DF} \times \mathsl{CM} \rightarrow
\Dst\big(\mathsl{HD}\big)$ is given by:
\[\begin{array}{rcl}
	\mathsl{hd}(\lt,\CMno) 
	& = & 
	0.95\bigket{\eq} + 0.05\bigket{\no{\eq}} 
	\\
	\mathsl{hd}(\lt,\mi) 
	& = & 
	0.95\bigket{\eq} + 0.05\bigket{\no{\eq}} 
	\\
	\mathsl{hd}(\lt,\co) 
	& = & 
	0.95\bigket{\eq} + 0.05\bigket{\no{\eq}} 
	\\
	\mathsl{hd}(\lt,\tr) 
	& = & 
	0.95\bigket{\eq} + 0.05\bigket{\no{\eq}} 
	\\
	\mathsl{hd}(\DFno,\CMno) 
	& = & 
	0.95\bigket{\eq} + 0.05\bigket{\no{\eq}} 
	\\
	\mathsl{hd}(\DFno,\mi) & = & 
	0.95\bigket{\eq} + 0.05\bigket{\no{\eq}} 
	\\
	\mathsl{hd}(\DFno,\co) 
	& = & 
	0.95\bigket{\eq} + 0.05\bigket{\no{\eq}} 
	\\
	\mathsl{hd}(\DFno,\tr) 
	& = & 
	0.95\bigket{\eq} + 0.05\bigket{\no{\eq}} 
	\\
	\mathsl{hd}(\rt,\CMno) 
	& = & 
	0.05\bigket{\eq} + 0.95\bigket{\no{\eq}} 
	\\
	\mathsl{hd}(\rt,\mi) 
	& = & 0.5\bigket{\eq} + 0.5\bigket{\no{\eq}} 
	\\
	\mathsl{hd}(\rt,\co) 
	& = &
	0.95\bigket{\eq} + 0.05\bigket{\no{\eq}} 
	\\
	\mathsl{hd}(\rt,\tr) 
	& = & 
	0.5\bigket{\eq} + 0.5\bigket{\no{\eq}}
\end{array} \]

\noindent The next one is $\mathsl{hi} \colon \mathsl{CM} \times
\mathsl{LP} \rightarrow \Dst\big(\mathsl{HI}\big)$.
\[ \begin{array}{rcl}
	\mathsl{hi}(\CMno,\nr) 
	& = & 
	0.93\bigket{\mi} + 0.05\bigket{\mo} + 0.02\bigket{\se}
	\\
	\mathsl{hi}(\CMno,\cg) 
	& = & 
	0.15\bigket{\mi} + 0.8\bigket{\mo} + 0.05\bigket{\se}
	\\
	\mathsl{hi}(\CMno,\ab) 
	& = & 0.7\bigket{\mi} + 0.2\bigket{\mo} + 0.1\bigket{\se}
	\\
	\mathsl{hi}(\mi,\nr) 
	& = & 0.1\bigket{\mi} + 0.8\bigket{\mo} + 0.1\bigket{\se}
	\\
	\mathsl{hi}(\mi,\cg) 
	& = & 
	0.1\bigket{\mi} + 0.75\bigket{\mo} + 0.15\bigket{\se}
	\\
	\mathsl{hi}(\mi,\ab) 
	& = & 
	0.1\bigket{\mi} + 0.65\bigket{\mo} + 0.25\bigket{\se}
	\\
	\mathsl{hi}(\co,\nr) 
	& = & 
	0.1\bigket{\mi} + 0.7\bigket{\mo} + 0.2\bigket{\se}
	\\
	\mathsl{hi}(\co,\cg) 
	& = & 0.05\bigket{\mi} + 0.65\bigket{\mo} + 0.3\bigket{\se}
	\\
	\mathsl{hi}(\co,\ab) 
	& = & 
	0.1\bigket{\mi} + 0.5\bigket{\mo} + 0.4\bigket{\se}
	\\
	\mathsl{hi}(\tr,\nr) 
	& = & 
	0.02\bigket{\mi} + 0.18\bigket{\mo} + 0.8\bigket{\se}
	\\
	\mathsl{hi}(\tr,\cg) 
	& = & 0.1\bigket{\mi} + 0.3\bigket{\mo} 
	+ 0.6\bigket{\se}
	\\
	\mathsl{hi}(\tr,\ab) 
	& = & 0.02\bigket{\mi} + 0.18\bigket{\mo} + 0.8\bigket{\se}.
\end{array} \]

\noindent As last channel we have $\mathsl{lb} \colon \mathsl{HD}
\times \mathsl{HI} \rightarrow \Dst\big(\mathsl{LB}\big)$.
\[ \begin{array}{rcl}
	\mathsl{lb}(\eq,\mi) 
	& = &
	0.1\bigket{\ltfi} + 0.3\bigket{\fitw} + 0.6\bigket{\gttw}
	\\
	\mathsl{lb}(\eq,\mo) 
	& = & 
	0.3\bigket{\ltfi} + 0.6\bigket{\fitw} + 0.1\bigket{\gttw}
	\\
	\mathsl{lb}(\eq,\se) 
	& = & 
	0.5\bigket{\ltfi} + 0.4\bigket{\fitw} + 0.1\bigket{\gttw}
	\\
	\mathsl{lb}(\no{\eq},\mi) 
	& = & 
	0.4\bigket{\ltfi} + 0.5\bigket{\fitw} + 0.1\bigket{\gttw}
	\\
	\mathsl{lb}(\no{\eq},\mo) 
	& = & 
	0.5\bigket{\ltfi} + 0.45\bigket{\fitw} + 0.05\bigket{\gttw}
	\\
	\mathsl{lb}(\no{\eq},\se) 
	& = & 
	0.6\bigket{\ltfi} + 0.35\bigket{\fitw} + 0.05\ketsl{12+}.
\end{array}\]

\end{document}